\documentclass[prx,floatfix,superscriptaddress,longbibliography,notitlepage, twocolumn, 10.5pt]{revtex4-2}
\usepackage[utf8]{inputenc}
\usepackage{graphicx}
\usepackage[dvipsnames]{xcolor}
\usepackage{float}
\usepackage[citecolor=RawSienna, colorlinks=true, linkcolor=MidnightBlue, urlcolor=MidnightBlue, linktocpage=true]{hyperref}
\usepackage[T1]{fontenc}
\usepackage{bbm}
\usepackage{amsmath}
\usepackage{lmodern}
\usepackage{amsfonts}
\usepackage{physics}
\usepackage{mathtools}
\usepackage{tikz}
\usepackage{mathtools}
\usepackage{appendix}
\usetikzlibrary{arrows.meta}
\usepackage{dsfont}
\usepackage{caption}
\usepackage{placeins}
\usepackage{subcaption}
\definecolor{myred}{RGB}{238, 102, 119}
\definecolor{myblue}{RGB}{68, 119, 170}
\definecolor{mygreen}{RGB}{34, 136, 51}

\begin{document}
\newtheorem{assumption}{Assumption}
\title{Continuous-time quantum optimisation without the adiabatic principle}
\author{Robert J.~Banks}
\email{robert.banks.20@ucl.ac.uk}
\affiliation{London Centre for Nanotechnology, UCL, London WC1H 0AH, UK}
\author{Georgios S.~Raftis}
\affiliation{London Centre for Nanotechnology, UCL, London WC1H 0AH, UK}
\author{Dan E.~Browne}
\affiliation{Department of Physics and Astronomy, UCL, London WC1E 6BT, UK}
\author{P.~A.~Warburton}
\affiliation{London Centre for Nanotechnology, UCL, London WC1H 0AH, UK}
\affiliation{Department of Electronic \& Electrical Engineering, UCL, London WC1E 7JE, UK}
\date{\today}

\begin{abstract}
Continuous-time quantum algorithms for combinatorial optimisation problems, such as quantum annealing, have previously been motivated by the adiabatic principle. A number of continuous-time approaches exploit dynamics, however, and therefore are no longer physically motivated by the adiabatic principle. In this work, we take Planck's principle as the underlying physical motivation for continuous-time quantum algorithms. Planck's principle states that the energy of an isolated system cannot decrease as the result of a cyclic process. We use this principle to justify monotonic schedules in quantum annealing, which are not adiabatic. This approach also highlights the limitations of reverse quantum annealing in an isolated system.
\end{abstract}

\maketitle

\section{Introduction}
Extensive work has gone into how quantum technologies might be able to solve combinatorial optimisation problems faster than classical techniques. Continuous-time quantum optimisation (CTQO) broadly refers to using analogue quantum systems to tackle combinatorial optimisation problems. Much of the early work focused on exploiting the adiabatic principle \cite{Far00, Alb18}. In adiabatic quantum optimisation (AQO) a quantum system is initialised in the ground state of some easy to prepare Hamiltonian. The initial Hamiltonian is referred to as the driver Hamiltonian. The combinatorial optimisation problem is encoded into the energy levels of a Hamiltonian, termed the problem Hamiltonian, with lower energy states corresponding to better solutions. The system in AQO is then evolved under a time-dependent Hamiltonian that interpolates between the driver and problem Hamiltonian, such that the system always remains in its instantaneous ground state. This approach is typically limited by the minimal spectral gap between the ground state and the first excited state. In many hard problems, this gap vanishes exponentially with the problem size. Consequently, AQO must have exponentially increasing run-times  \cite{Alb18}.  

In an attempt to circumvent the vanishing spectral gap, the adiabatic requirement may be relaxed. Quantum annealing (QA) more broadly refers to interpolating smoothly (not necessarily adiabatically) between two Hamiltonians to find the solution \cite{Kad98,Hau20}. The active decision to cause transitions is sometimes referred to as diabatic quantum annealing \cite{Cro21}. Other approaches such as continuous-time quantum walks (CTQWs) \cite{Cal19} and multi-stage quantum walks (MSQWs) \cite{Cal21} embrace large-scale dynamics not seen in AQO. Cyclic approaches such as reverse quantum annealing (RQA) \cite{Cha17,Wan22, Zha24} use non-adiabatic, non-monotonic schedules, to search locally. As approaches move away from AQO, the adiabatic theorem is no longer a reasonable motivation for their performance. Small system scaling provides compelling evidence for many of these approaches, but reasoning from an underlying physical principle is lacking.

Statistical physics has had large success in explaining phenomena where exactly solving a system is not feasible \cite{Blu09}. In a quantum system, exact diagonalisation is often prohibitively difficult, therefore we need other tools to reason about the behaviour of these large quantum systems. In this paper, we apply pure-state statistical physics \cite{Gog10,Dal16} to provide an analysis of continuous-time quantum approaches for optimisation, where the adiabatic principle is insufficient. 

This paper sets out to demonstrate how Planck's principle, an idea from statistical physics, can provide intuition into CTQO with a range of schedules, where the adiabatic theorem cannot be applied. Two types of schedules are considered: monotonically increasing and cyclic schedules. To gain intuition, this paper considers passive states, for these states it has been demonstrated that Planck's principle holds \cite{Kou21}. Planck's principle is used to show that the average quality of solutions increases for monotonic schedules and decreases for cyclic approaches compared to its initial value. This is then combined with the idea that local observables in isolated systems can be well approximated by a passive state in many circumstances. This is then used to provide intuition into CTQO more generally. This is further confirmed by numerical experiments.

Sec.~\ref{sec:CTQO} provides the background to the CTQO approaches considered in this paper. Sec.~\ref{sec:ass} introduces the statistical physics framework used in the rest of the paper in more detail. Sec.~\ref{sec:for} applies the statistical framework to forward QA approaches, specifically MSQWs. Sec.~\ref{sec:cyc} explores cyclic approaches, specifically RQA in an isolated quantum system. Throughout this paper we take $\hbar=1$ and $k_b=1$. The Pauli matrices are denoted by $X$, $Y$ and $Z$. The numerical simulations are carried out using the Python package QuTip \cite{Joh12,Joh13}. 

\section{Continuous-time quantum optimisation}
\label{sec:CTQO}
The aim of QA is to find (approximate) solutions to a combinatorial optimisation problem \cite{Hau20}. The problem Hamiltonian, $H_p$ is typically an Ising Hamiltonian, consisting of few-body (typically 2-body) terms. A driver Hamiltonian $H_d$, again consisting of few-body terms (typically 1-body terms) drives transitions between eigenstates of $H_p$ to try to build up amplitude in the low-energy eigenstates of $H_p$. In QA, the system is initialised in the ground state of the driver Hamiltonian. The system is then evolved under the Hamiltonian:
\begin{equation}
    H_{QA}=A(t)H_d+B(t)H_p,
\end{equation}
where $t$ denotes time. The time-dependent schedules, $A(t)$ and $B(t)$, are defined on the interval between $t=0$ and $t=t_f$. Typically, $A(t)$ is chosen to decrease monotonically with time, while $B(t)$ monotonically increases with time. Ideally, the schedule is chosen to minimise $\langle H_p (t_f) \rangle$. Often QA is justified by its apparent proximity to AQO, for which proofs of speed-up exist \cite{Rol02,Has21, aha05}.

\begin{figure}
    \centering
    \includegraphics[width=0.48\textwidth]{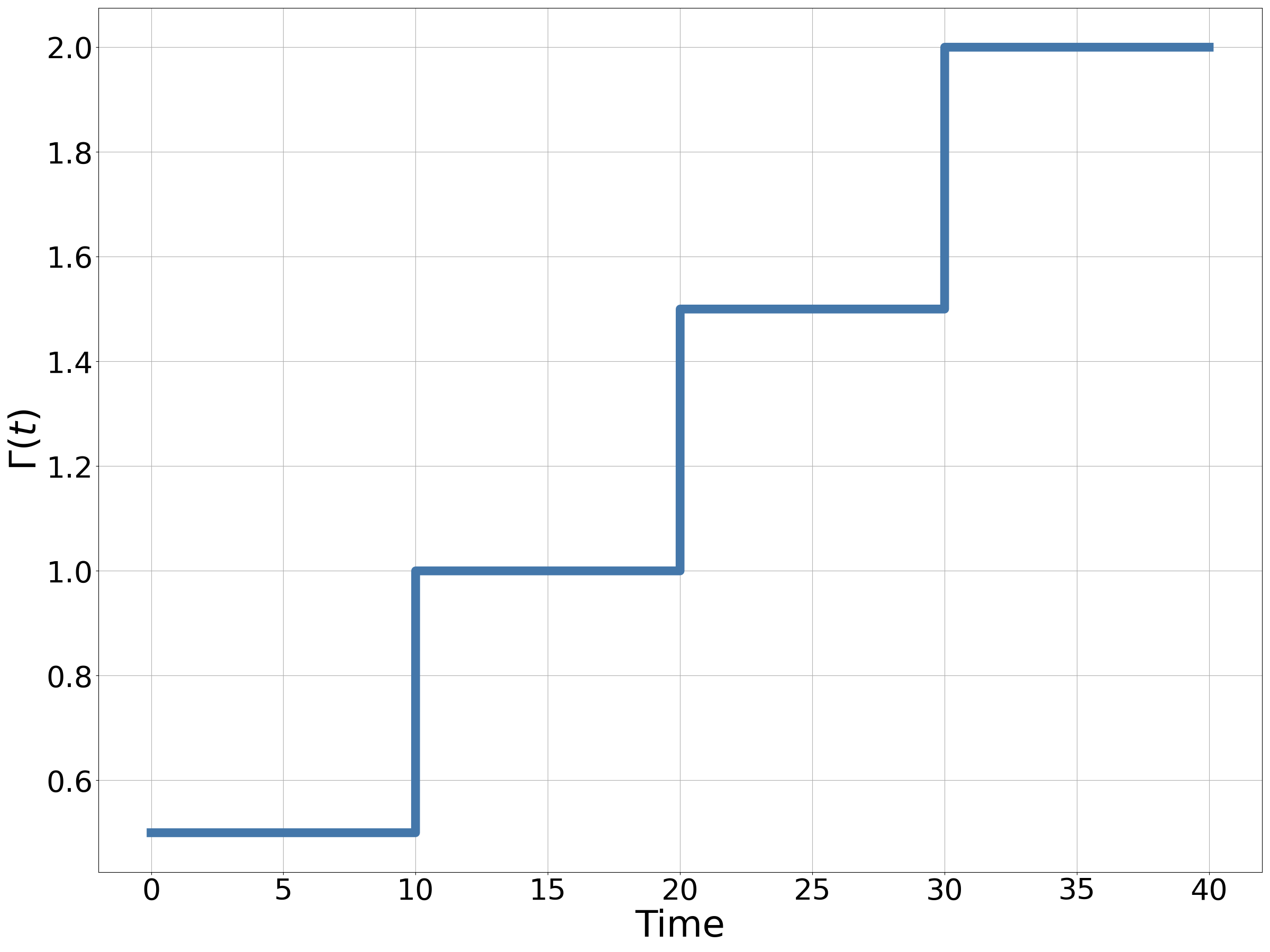}
    \caption{The schedule used in the MSQW example in Fig.~\ref{fig:msqw_ex}.}
    \label{fig:ex_sch}
\end{figure}

Recently, multi-stage quantum walks (MSQWs) have been proposed as a method for tackling combinatorial optimisation problems \cite{Cal21,Ban24}. These approaches consist of repeatedly quenching the system. The Hamiltonian is given by
\begin{equation}
    H_{QA}=H_d+\Gamma(t)H_p,
\end{equation}
where $\Gamma(t)$ is a piece-wise constant, non-decreasing function. Fig.~\ref{fig:ex_sch} shows an  example schedule. A stage refers to a period where the Hamiltonian is held constant. The initial state is the ground state of $H_d$. It has been numerically observed that the expectation of $H_p$ decreases (corresponding to better solutions on average) as $\Gamma(t)$ is increased \cite{Cal21,Ban24}. It is insufficient to appeal to the adiabatic theorem to explain the performance, and there are limited theoretical motivations for this approach. From energy conservation mechanisms, it has been shown that this approach can always do better than random guessing \cite{Cal21}. As this protocol is piece-wise constant with the time dependence entering through sudden quenches, it makes the approach more amenable to analytic investigation. For that reason, we focus on this approach in this paper. A single-stage MSQW is referred to as a continuous-time quantum walk (CTQW) \cite{Cal19, Cal21, Ban24}. In this approach, the Hamiltonian has no time dependence. CTQWs need careful parameter setting \cite{Cal19, Ban24}.

\begin{figure}
    \centering
    \includegraphics[width=0.48\textwidth]{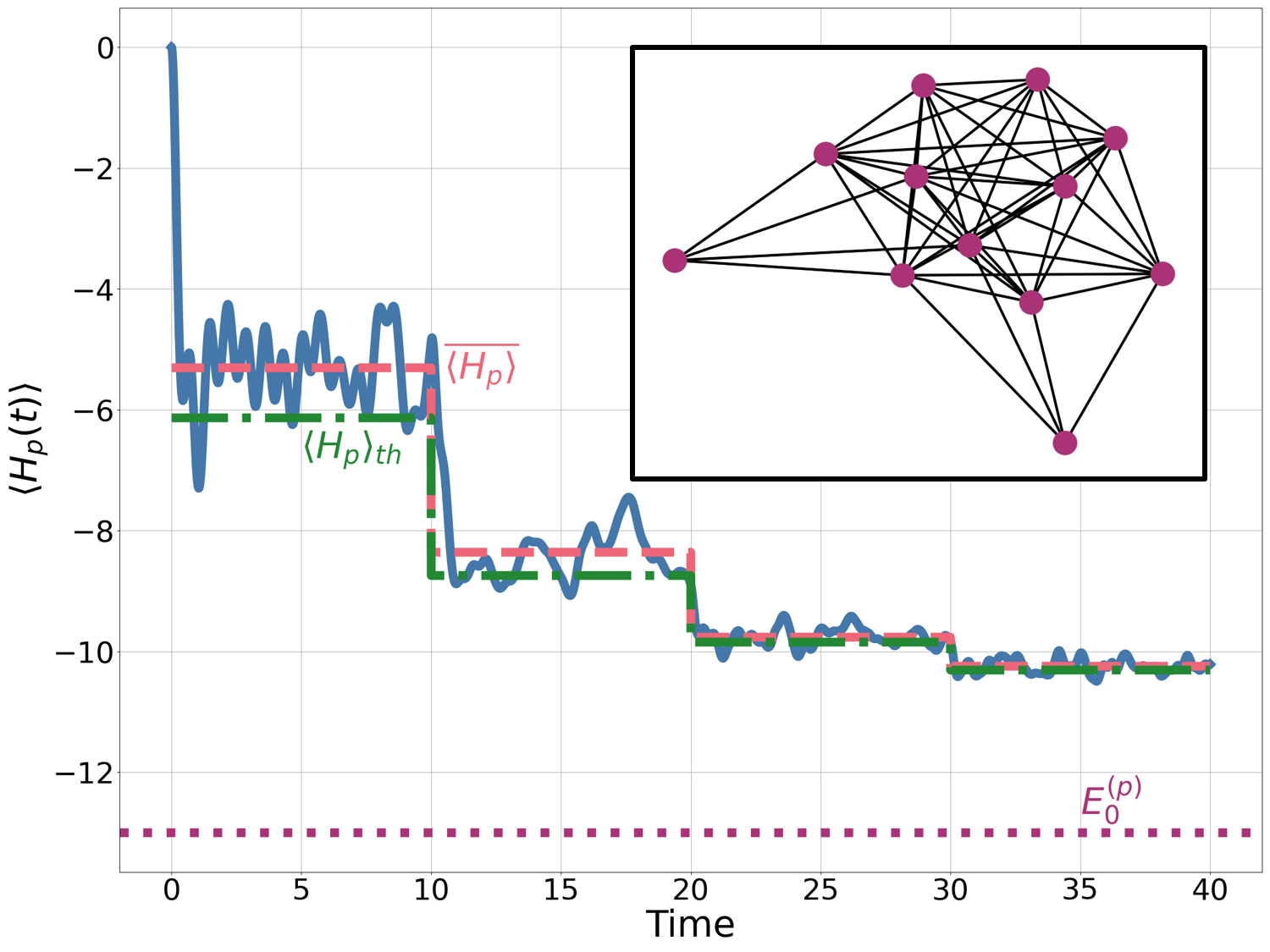}
    \caption{An example of an MSQW on a 12-qubit graph (shown in the inset of the figure). The schedule is shown in Fig.~\ref{fig:ex_sch}. The solid blue line shows the Schr\"odinger evolution. The dotted purple line shows $E_0^{(P)}$, the ground state energy of $H_p$. The dashed red line shows the time-averaged value of $\langle H_p \rangle$. The dash-dot green line shows prediction of $\langle H_p\rangle$ according to Assumption \ref{ass:eth}. }
    \label{fig:msqw_ex}
\end{figure}

To illustrate MSQWs, we consider a specific example. Consider the combinatorial optimisation problem MAX-CUT on binomial graphs (also known as Erd\H os-R\' enyi graphs). For a binomial graph $G=(V,E)$ consisting of $n$ nodes, each edge is selected with a given probability. Throughout this paper, this probability is taken to be 2/3. The optimisation problem is encoded as the Ising Hamiltonian
\begin{equation} 
    \label{eq:MCIsing}
    H_p^{(MC)}=\sum_{(i,j)\in E} Z_iZ_j.
\end{equation}
The driver Hamiltonian is taken to be the transverse-field:
\begin{equation}
    \label{eq:TF}
    H_{TF}=-\sum_{i} X_i.
\end{equation}
These choices of Hamiltonians are typical within the CTQO literature \cite{Cal19, Alb18,Far00, Kad98} and can be experimentally realised \cite{Kin22}. The initial state $\ket{+}$ is the ground state of $H_{TF}$. The inset of Fig.~\ref{fig:msqw_ex} shows the specific graph considered, and Fig.~\ref{fig:ex_sch} the associated schedule. Fig.~\ref{fig:msqw_ex} shows the time-dependent value of $\langle H_p(t)\rangle$ in blue. At each stage $\langle H_p (t)\rangle$ appears to fluctuate around some constant value. After each stage $\langle H_p(t)\rangle$ appears to be lower on average. The ground state energy of the problem Hamiltonian, denoted by $E_0^{(p)}$, is shown by the dotted purple line. The value of $\langle H_p (t)\rangle$ for the MSQW appears to be converging to some value above the ground state. The aim of this paper is to provide some physical reasoning into this behaviour. 

Moving away from monotonic schedules, reverse quantum annealing (RQA) is where the system starts in an eigenstate of $H_p$ and $H_d$ is turned on and then off. The idea is to use this cyclic process to explore the solution space locally \cite{Cha17}. More recently, it was proposed that the many-body localised phase transition in spin glasses could be used to cyclically cool the system \cite{Wan22}. This approach is essentially RQA where a bias Hamiltonian is added that has as its ground state the initial state. The bias considered was a sum over Pauli $Z$s. We discuss these approaches in Sec.~\ref{sec:cyc}.  RQA is one approach to warm-starting a quantum algorithm. Warm-starting refers to using prior information about the combinatorial optimisation problem to try to increase the performance of the quantum algorithm, typically by exploiting known reasonable solutions to the problem \cite{Egg21}. In this text, we refer to approaches such as conventional QA and MSQWs as examples of \textit{forward} approaches that interpolate monotonically between two distinct initial and final non-commuting Hamiltonians. We refer to approaches where the initial Hamiltonian is the same as the final Hamiltonian as \textit{cyclic}.

\section{Thermalisation, extractable work, and diagonal entropy}
\label{sec:ass}
Our discussion of the continuous-time quantum optimisation (CTQO) approaches is limited to isolated systems. The Hamiltonians consist of local terms, and we will restrict our focus to measuring $\langle H_p (t) \rangle$ as the metric of success, which ideally should be as small as possible. 

Consider the expectation value of some few-body observable $A$ in a closed system with initial state $\ket{\psi_i}$ and evolved under some time-independent Hamiltonian \cite{Dal16}. The eigenvectors and associated eigenvalues of the Hamiltonian are denoted by $\ket{E_k}$ and $E_k$ respectively. Let the overlap between the initial state ($\ket{\psi_i}$) and the $k^{\text{th}}$ eigenstate be written as $c_k$ (i.e. $c_k=\bra{E_k}\ket{\psi_i}$).  It follows that the expectation of $A$ is given by:
\begin{align}
    \langle A \rangle &=\sum_{m,n} e^{i (E_m-E_n) t} c_m^* c_n \bra{E_m} A \ket{E_n} \\
    &=\sum_{m} \abs{c_m}^2 \bra{E_m} A \ket{E_m} \nonumber\\
    \label{eq:TI_sum}
    &+ \sum_{m \neq n} e^{i (E_m-E_n) t} c_m^* c_n \bra{E_m} A \ket{E_n}\\
    &=\overline{\langle A \rangle}+\langle \Delta A \rangle,
\end{align}
here $\overline{\langle A \rangle}$ corresponds to the time independent sum in Eq.~\ref{eq:TI_sum} and $\langle \Delta A \rangle$ the time-dependent sum. As the system evolves, the difference in energy gaps between pairs of eigenstates will cause dispersion. As a consequence $\langle \Delta A \rangle\approx 0$ up to some negligible fluctuation for large systems. This means that the expectation of $A$ will approach some steady state after some time. This timescale will be problem specific; details on how it might be estimated can be found in \cite{Wil18, Oli18}. It has been numerically observed and justified that this timescale will not necessarily be exponential in the problem size \cite{Wil17, Wil18}. This leads to our first two assumptions:
\begin{assumption}[Stationary value]
\label{ass:ss}
Under evolution by a time-independent Hamiltonian, the expectation of a local observable $A$ can be replaced with its steady state value $\overline{\langle A \rangle}$ after some time $\tau_{d}$. This includes the expectation values of $H_p$ and $H_d$. The time $\tau_{d}$ is the timescale associated with dephasing in the energy basis.
\end{assumption}
\begin{assumption}[Diagonal in the energy eigenbasis]
\label{ass:diag}
Once the system approaches a steady state, we can approximate the state-vector by a density operator diagonal in the energy eigenbasis \cite{Dal16}.
\end{assumption}
For many systems, it is conjectured that $\overline{\langle A \rangle}$ will be well approximated by assuming a microcanonical distribution of energy eigenstates. This is typically referred to as the eigenstate thermalisation hypothesis (ETH) \cite{Dal16, Deu18}. This leads to our third assumption:
\begin{assumption}[ETH]
\label{ass:eth}
The steady state is locally indistinguishable from a Gibbs state. The temperature of $\rho_{Gibbs}$ is fixed according to the energy of the system. That is to say, the steady state value $\overline{\langle A \rangle}$ can be well approximated by using the Gibbs state, $\overline{\langle A \rangle}\approx \Tr\left( A \rho_{Gibbs}\right)$. The inverse temperature $\beta$ is fixed by replacing $A$ with the Hamiltonian. 
\end{assumption}
The ETH, as described above, is likely to be a better approximation for large problem sizes where subextensive corrections and fluctuations can be ignored \cite{Deu18}. The ETH has been explored in the context of CTQWs in \cite{Ban24}. To illustrate how these assumptions manifest in CTQO, specifically in MSQWs, we return to the MAX-CUT example considered in Sec.~\ref{sec:CTQO}. Fig.~\ref{fig:msqw_ex} shows the Schr\"odinger evolution of $\langle H_p (t) \rangle$. Each stage oscillates around some constant value after some short time at each stage, as expected from Assumption \ref{ass:ss}. The dashed red line shows $\overline{\langle H_p \rangle}$, which is calculated by assuming a density operator diagonal in the energy eigenbasis (Assumption \ref{ass:diag}). Finally, the dash-dot green line shows the prediction from the ETH (Assumption \ref{ass:eth}). As the evolution continues, the ETH becomes a better approximation to $\langle H_p \rangle$. 

The final assumption is:
\begin{assumption}[Planck's Principle]
\label{ass:work}
For any relevant cyclic process in an isolated system, no work can be extracted. That is for any cyclic process represented by a unitary $U$, with initial Hamiltonian $H$ and initial state $\rho$, the following holds true: 
\begin{equation}
    \label{eq:work}
    W=E_{\text{Initial}}-E_{\text{Final}}=Tr \left[ H\left(\rho-U\rho U^\dagger \right) \right]\leq0.
\end{equation}
\end{assumption}
This assumption, as we will show in the rest of this paper, has significant consequences for CTQO. To justify it, we show its consistency with Assumptions \ref{ass:eth} and \ref{ass:diag}. We also point the reader to \cite{Gol13} where  Goldstein et al.~showed that under certain assumptions, in the closed system setting, the extractable work is exponentially more likely to decrease than not. The assumptions include that knowledge of the initial state before the cyclic process is lost by time averaging under evolution by a Hamiltonian — which is the case in an MSQW. Assumption \ref{ass:work} is sometimes referred to as Kelvin's statement of the second law of thermodynamics \cite{Dal16}, or Planck's principle \cite{Gol13, Ito20}. See also \cite{Tak00,Ike15,Ito20, Kan19, Dal16} for further attempts to mathematically motivate the second law of thermodynamics in closed systems from the rules of quantum mechanics. 

\begin{figure}
    \centering
    \begin{tikzpicture}[domain=0:4]
        \draw[->] (-0.2,0) node[left] {$0$} -- (4.2,0) node[right] {$E$} ;
        \draw[->] (0,-0.2) -- (0,3.2) node[above] {$\Omega(E)$};
         \draw[domain=0:4, smooth, variable=\x, color=black, ultra thick] plot ({\x},{3*exp(-(\x-2)*(\x-2)});
         \draw[domain=0:4, smooth, variable=\x, color=myblue, ultra thick] plot ({\x},{2*exp(-(\x-0.5)*(\x-0.5)*100});
         \draw[domain=0:4, smooth, variable=\x, color=myred, ultra thick] plot ({\x},{1.5*exp(-(\x-1)*(\x-1)*20});
        \draw[-,color=myblue]  (0.5,-0.2) node[below] {$E_1$} -- (0.5,0);
        \draw[-, color=myred]  (1,-0.2) node[below] {$E_2$} -- (1,0);
    \end{tikzpicture}
    \caption{The solid black line sketches the typical density-of-states $\Omega(E)$, for a large non-integrable system \cite{Dal16}. The solid blue line shows a possible energy distribution for the initial state. The red line shows a possible resulting energy distribution as the result of a cyclic process. The energy distribution has been broadened and has moved towards the middle of the spectrum. }
    \label{fig:dos}
\end{figure}
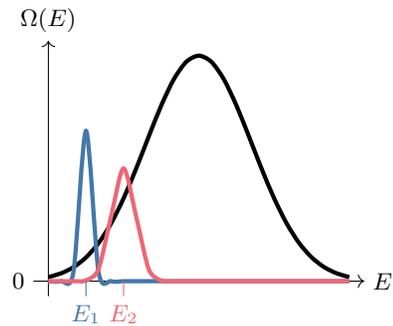

It is known that any Gibbs state, with a positive temperature, satisfies Eq.\ \ref{eq:work} \cite{Kou21}. A derivation of this can be found in Appendix \ref{app:gibbs_passive}. If we assume that the ETH (i.e.~Assumption $\ref{ass:eth}$) holds for the initial state, then it remains to determine if $U^\dagger H U$ is an observable that exhibits thermalisation \cite{Dal16,Deu18}. If $U^\dagger H U$, where $U$ corresponds to a cyclic process, does correspond to a thermalising observable, then Assumption \ref{ass:work} follows as a consequence of Assumption $\ref{ass:eth}$. 

A positive temperature Gibbs state is an example of a passive state. Ref.~\cite{Kou21} shows that, in the case of passive states, Planck’s principle applies exactly. A state is passive given an initial Hamiltonian if:
\begin{enumerate}
    \item The state is diagonal in the energy basis of the Hamiltonian.
    \item The populations of the state in the energy basis are non-increasing with energy.
\end{enumerate}
Beyond Gibbs states, other examples include ground states and uniform distributions with no low energy cut-off.

For further evidence of Planck's principle, away from the ETH, we might look to the diagonal entropy \cite{Pol11}. The diagonal entropy is defined as 
\begin{equation}
    S_d=-\sum_m \rho_{mm} \log \rho_{mm},
\end{equation}
where $\rho_{dd}$ are the diagonal elements of the density operator in the energy eigenbasis. If the initial state is diagonal in the energy eigenbasis (i.e, Assumption \ref{ass:diag}) it has been shown the diagonal entropy cannot decrease \cite{Pol11}. Coupled with a reasonable unimodal model of the density-of-states (especially for non-integrable systems) \cite{Dal16}, as sketched in Fig.~\ref{fig:dos}, this implies Assumption $\ref{ass:work}$. We expect this to hold for initial Hamiltonians like $H_p$ too (see for example \cite{Sch24}) for a wide range of problems. Essentially, under a cyclic process, the system is expected to heat up as there are more states towards the middle of the spectrum than the edges. Under continued periodic drive, this is sometimes referred to as Floquet heating \cite{Ho23,Dal16}. In order to apply Assumption \ref{ass:work}, it is required that the system is more likely to move up in energy than down in energy. This is unlikely to be the case given an initial state that is very high in energy (corresponding to a negative temperature Gibbs state). Typically, the system is expected to move towards the infinite temperature state as a result of a cyclic process. This places a restriction on the initial state, $\ket{\psi_i}$, for Assumption \ref{ass:work} to hold:
\begin{equation}
    \bra{\psi_i} H \ket{\psi_i} < \frac{1}{\mathcal{D}} \Tr H,
\end{equation}
where $\mathcal{D}$ is the dimension of the Hilbert space. Perhaps a more accurate statement of Assumption \ref{ass:work} is that a cyclic process will move the energy of the system towards its infinite temperature value \cite{Ho23,Dal16}. For the purpose of what is to follow, Assumption \ref{ass:work} as stated originally is sufficient. 

Although there are specific cases which violate each assumption, we expect them to hold true for a wide variety of problems, drivers and encodings, especially as the problem size is increased. The intuition in this paper relies on being able to approximate the steady-state behaviour of these isolated systems with a passive state. This might not be a valid assumption for all systems. For example, in many-body localised systems \cite{Hus13,Nan15, Alt18} and integrable systems \cite{Deu91, Rig09, Ess16, Bre20, Noh21}, the ETH is known not to hold. Determining if a specific observable of a specific system will correspond to a thermal state is a challenging problem \cite{Deu18,Gar18, Shi21}. It is typically assumed that the Hamiltonian is highly non-degenerate and that the observable associated with thermalisation is a few-body observable \cite{Sre94, Rig08}. However, thermalisation has been widely observed numerically \cite{Rig08,Bir10,San10,Ste13,Kim14,Ste14,Fra15,Kho15,Mon17,Yos18, Jan19, Ban24} and experimentally \cite{Tro12,Clo16,Kau16,Kuc18,Tan18}. 

Assumptions \ref{ass:diag}, \ref{ass:eth} and \ref{ass:work} all have the effect of imposing an arrow of time on the evolution. This comes from discarding  the coherences in the energy eigenbasis. In a closed system, this is a result of dispersion in the energy eigenstates. For an open system, this could come from being very weakly coupled to a bath.

The rest of this paper considers forward CTQO approaches in Sec.~\ref{sec:for} and cyclic CTQO approaches in Sec.~\ref{sec:cyc}. Each section considers passive states, where Planck's principle can be straightforwardly applied, to derive statements about the performance. For example, that $\langle H_p(t)\rangle$ can only decrease compared to its initial value subject to monotonically increasing schedules. Since, under the ETH (i.e. Assumption \ref{ass:eth}) observables such as $\langle H_p \rangle$ can be approximated by a Gibbs state -- an example of a passive state, this intuition is extended to the pure state setting. Therefore, in each section, we use the knowledge gained from studying passive states to provide an effective and intuitive explanation of CTQO dynamics.

\section{Forward approaches}
\label{sec:for}
In this section, we consider forward CTQO approaches with monotonically increasing schedules (categorised as forward approaches). In Sec.~\ref{sec:pass} we show that for passive states, $\langle H_p(t) \rangle$ is less than or equal to $\langle H_p (0) \rangle$, by using Planck's Principle. Motivated from the ETH (i.e.\ Assumption \ref{ass:eth}) we then use this to provide intuition for MSQWs in Sec.~\ref{sec:msqw}. In Sec.~\ref{sec:oqs} some intuition is provided for when the system is coupled to a bath.

\subsection{Passive states}
\label{sec:pass}

\begin{figure}
    \centering
    \begin{tikzpicture}[domain=0:4]
        \draw[->] (-0.2,0) node[left] {$0$} -- (4.2,0) node[right] {$t$} ;
        \draw[->] (0,-0.2) -- (0,3.2) node[above] {$\gamma(t)$};
         \draw[domain=0:4, smooth, variable=\x, color=myblue, ultra thick] plot ({\x},{2*exp(\x-4)+0.5}) node[above] {$\gamma(t)$};
         \draw[color=mygreen, ultra thick] plot coordinates{(0,{2*exp(0-4)+0.5}) (0.5,{2*exp(0-4)+0.5}) (0.5,{2*exp(1-4)+0.5}) (1,{2*exp(1-4)+0.5}) (1,{2*exp(1.5-4)+0.5}) (1.5,{2*exp(1.5-4)+0.5}) (1.5,{2*exp(2-4)+0.5}) (2,{2*exp(2-4)+0.5}) (2,{2*exp(2.7-4)+0.5}) (2.7,{2*exp(2.7-4)+0.5}) (2.7,{2*exp(3-4)+0.5}) (3,{2*exp(3-4)+0.5}) (3,{2*exp(3.5-4)+0.5}) (3.5,{2*exp(3.5-4)+0.5}) (3.5,{2*exp(4-4)+0.5}) (4,{2*exp(4-4)+0.5})};
         \draw[-,color=mygreen, thick, dash dot] plot coordinates{ (2.7,{2*exp(0-4)+0.5}) (-0.2,{2*exp(0-4)+0.5})} node[left] {$\gamma(t_0)$} ;
         \draw[-,color=mygreen, thick, dash dot] plot coordinates{(2.7,0) (2.7,{2*exp(2.7-4)+0.5}) (-0.2,{2*exp(2.7-4)+0.5})} node[left] {$\gamma(t_{m+1})$} ;
        \draw[-,color=mygreen]  (0,-0.2) node[below] {$t_0$} -- (0,0);
        \draw[-,color=mygreen]  (0.5,-0.2) node[below] {$t_1$} -- (0.5,0);
        \draw[-,color=mygreen]  (1,-0.2) node[below] {$t_2$} -- (1,0);
        \filldraw [mygreen] (1.35,-0.45) circle (0.5pt);
        \filldraw [mygreen] (1.5,-0.45) circle (0.5pt);
        \filldraw [mygreen] (1.65,-0.45) circle (0.5pt);
        \draw[-,color=mygreen]  (2.0,-0.2) node[below] {$t_m$} -- (2.0,0);
        \draw[-,color=mygreen]  (2.7,-0.2) node[below] {$t_{m+1}$} -- (2.7,0);
        \draw[-,color=mygreen]  (4,-0.2) node[below] {$t_f$} -- (4,0);
        \filldraw [mygreen] (3.2,-0.45) circle (0.5pt);
        \filldraw [mygreen] (3.35,-0.45) circle (0.5pt);
        \filldraw [mygreen] (3.5,-0.45) circle (0.5pt);
        \draw[color=myred, ultra thick, dashed] plot coordinates{(0,{2*exp(0-4)+0.5}) (0.5,{2*exp(0-4)+0.5}) (0.5,{2*exp(1-4)+0.5}) (1,{2*exp(1-4)+0.5}) (1,{2*exp(1.5-4)+0.5}) (1.5,{2*exp(1.5-4)+0.5}) (1.5,{2*exp(2-4)+0.5}) (2,{2*exp(2-4)+0.5}) (2,{2*exp(2.7-4)+0.5}) (2.7,{2*exp(2.7-4)+0.5}) (2.7,{2*exp(1-4)+0.5}) };
        
    \end{tikzpicture}
    \caption{The solid blue line sketches a possible $\gamma(t)$, which in this case is monotonically increasing. The green line shows the piece-wise constant function used to approximate $\gamma(t)$. The dashed red line shows the cyclic process used to demonstrate $\langle H_p(t)\rangle\leq \langle H_p(0) \rangle$.}
    \label{fig:cyc_p}
\end{figure}

As mentioned in Sec.~\ref{sec:ass}, Planck's principle holds for arbitrary cyclic unitaries and passive states. The aim of this section is to make clear a provable statement that follows from passivity. This section allows for the initial state to be a mixed state, but the system is otherwise isolated. The Hamiltonian under consideration is 
\begin{equation}
    H(t)=H_d+\gamma(t) H_p,
\end{equation}
with, $t$ varying from $t=t_0$ to $t=t_f$. The initial state is any passive state with Hamiltonian $H(t_0)$. The schedule $\gamma(t)$ is approximated with a step-wise constant function, shown in Fig.~\ref{fig:cyc_p}. The change in energy as a result of the piece-wise constant function is
\begin{equation}
    \langle H(t_f)\rangle =\sum_{k=0}^{f-1} \left[\gamma(t_{k+1})-\gamma(t_k)\right] \langle H_p(t_k) \rangle.
\end{equation}
At the end of the anneal we introduce a fictitious quench to return the Hamiltonian to $H(t_0)$. The extractable work from this process is 
\begin{multline}
    W=-\sum_{k=0}^{f-1} \left[\gamma(t_{k+1})-\gamma(t_k)\right] \langle H_p(t_k) \rangle\\+\left[\gamma(t_{f})-\gamma(t_0)\right]\langle H_p(t_f) \rangle.
\end{multline}
Since the initial state is passive $W \leq 0$, it follows that, 
\begin{equation}
   \langle H_p(t_f) \rangle\leq \sum_{k=0}^{f-1} \left[\frac{\gamma(t_{k+1})-\gamma(t_k)}{\gamma(t_{f})-\gamma(t_0)}\right] \langle H_p(t_k) \rangle.
\end{equation}
That is to say, $\langle H_p(t_f) \rangle$ is bounded by a weighted average over the previous stages. Consider now a monotonically increasing schedule for $\gamma(t)$, such that $\gamma_{k+1}>\gamma_k$. For $k=0$, and introducing a fictitious quench at $t_1$ to create a cyclic process we have
\begin{equation}
    W_{10}=\left[\gamma(t_1)-\gamma(t_0)\right] \left(\langle H_p(t_1)\rangle-\langle H_p(t_0)\rangle\right)\leq0,
\end{equation}
hence 
\begin{equation}
    \langle H_p(t_1)\rangle\leq \langle H_p(t_0)\rangle.
\end{equation}
Assuming that
\begin{equation}
    \label{eq:ind}
    \langle H_p(t_k)\rangle\leq \langle H_p(t_0)\rangle
\end{equation}
is true for all $k\leq m$, it remains to show that it is true for $k=m+1$. Again a fictitious quench is introduced at $t_{m+1}$ to create a cyclic process:
\begin{multline}
    W_{m+1,0}=-\sum_{k=0}^{m} \left[\gamma(t_{k+1})-\gamma(t_k)\right] \langle H_p(t_k) \rangle\\+\left[\gamma(t_{m+1})-\gamma(t_0)\right]\langle H_p(t_{m+1}) \rangle.
\end{multline}
Again, due to passivity, $W_{m+1,0}\leq0$. It follows that 
\begin{equation}
    \langle H_p(t_{m+1}) \rangle \leq \sum_{k=0}^{m} \left[\frac{\gamma(t_{k+1})-\gamma(t_k)}{\gamma(t_{m+1})-\gamma(t_0)}\right] \langle H_p(t_{k}) \rangle,
\end{equation}
from the assumption stated in Eq.~\ref{eq:ind}
\begin{equation}
    \langle H_p(t_{m+1}) \rangle \leq \langle H_p(t_{0}) \rangle \underbrace{\sum_{k=0}^{m} \left[\frac{\gamma(t_{k+1})-\gamma(t_k)}{\gamma(t_{m+1})-\gamma(t_0)}\right] }_{=1}.
\end{equation}
By induction
\begin{equation}
    \langle H_p(t_k) \rangle \leq \langle H_p(t_0) \rangle
\end{equation}
for all $k\geq0$ given a monotonically increasing $\gamma(t_k)$. Making the time-step arbitrarily small and returning to the continuous limit gives:
\begin{equation}
    \langle H_p(t) \rangle \leq \langle H_p(t_0) \rangle
\end{equation}
for all $t\geq t_0$ given a monotonically increasing $\gamma(t)$. This has previously been shown for the case where the initial state is the ground state of $H_d$ \cite{Cal21}. Here we have extended it to any passive state, including ground states and Gibbs states with Hamiltonian $H_d+\gamma(t_0) H_p$ that have a non-trivial value of $\langle H_p (t_0) \rangle$. This suggests that passive states might have a significant role to play in QA.

\subsection{Intuition for forward approaches}

\subsubsection{Multi-stage quantum walks}
\label{sec:msqw}

An MSQW is given by the following Hamiltonian:
\begin{equation}
    \label{eq:msqw}
    H_{MS}=H_d+\Gamma(t) H_p,
\end{equation}
the schedule $\Gamma(t)$ being chosen to be monotonically increasing. The schedule is piecewise constant and given by
\begin{equation}
    \Gamma(t)=\begin{cases*}
    \gamma_1 \text{ for } 0\leq t\leq t_1\\
    \gamma_2 \text{ for } t_1 < t \leq t_2\\
    \cdots\\
    \gamma_k \text{ for } t_{k-1} < t \leq t_k\\
    \cdots\\
    \gamma_l \text{ for } t_{l-1} < t.\\
    \end{cases*}
\end{equation}
where $\gamma_l>\dots>\gamma_2>\gamma_1>0$ and $t_{l-1}> \dots >t_2 > t_1$. The initial state of a MSQW is the ground-state of $H_d$, which is a passive state. Combined with the non-decreasing schedule, it follows that MSQWs satisfy $\langle H_p(t) \leq H_p(0)$.

We now return to the MSQW example in Fig.~\ref{fig:msqw_ex}. From the ETH we expect that at each stage the system can be well approximated by a Gibbs state, as shown by the green line in the figure. At the end of each stage, the system is subjected to an increasing schedule. Therefore, by approximating the system as a Gibbs state (a passive state) at the end of each stage, we conclude that $\langle H_p \rangle$ should decrease with each stage compared to the initial value at each stage. Planck's principle combined with the ETH correctly predicts the behaviour seen in Fig.~\ref{fig:msqw_ex}. This behaviour has also been observed in other numerical experiments \cite{Cal21, Ban24, Imp24}.

Given significant dynamics, we would under Planck's Principle expect $\langle H_p \rangle$ to decrease at each stage. But $\langle H_p \rangle$ is bounded, so cannot decrease forever. As $\Gamma(t)$ becomes very large, we might reasonably approximate $H_{MS}$ as $\Gamma(t) H_p$, no longer driving transitions between eigenstates of $H_p$. As a follow-up, we might ask if this process results in the ground state of $H_p$. From Assumption \ref{ass:diag}, $S_d$ increases with each stage. This means that $S_d=0$ (corresponding to an eigenstate of $H_p$ in the large $\Gamma(t)$ limit) becomes entropically forbidden if the system thermalises to any state with non-zero diagonal entropy.

From pure-state statistical physics, we have been able to reason about a typical system without resorting to numeric diagonalisation or recourse to adiabaticity. Assuming only that work cannot be extracted from a cyclic process, we have motivated MSQWs. As long as this always holds true, $\langle H_p \rangle$ can only decrease. At no point was the magnitude of the quenches specified, allowing for extension to QA-like schedules. This motivates monotonic schedules in QA. In Appendix \ref{sec:pstqa} the continuum limit of a MSQW is explored.

\subsubsection{Application of a bath}
\label{sec:oqs}
To explore some consequences of open quantum system effects, we add a time-independent bath with time-independent couplings to the quantum system. The resulting Hamiltonian is given by:
\begin{equation}
    H_{\text{total}}=H_S(t)+H_{SB}+H_B,
\end{equation}
where $H_B$ acts on the bath, $H_S(t)$ on the system and $H_{SB}$ is the system-bath interaction. For simplicity, we take all the operators to be trace-class and the joint system to be isolated.

Consider the case where $H_S(t)$ is time-independent and given by $H_S(t)=H_d+\gamma H_p$. After some time, under the ETH (Assumption \ref{ass:eth}), the total system will be locally indistinguishable from a Gibbs state, with Hamiltonian $H_{\text{total}}$. The system's state, given by tracing out the bath, need not be a Gibbs state at all.

The temperature is fixed by the initial conditions:
\begin{equation}
    \label{eq:beta_oqs}
    \langle H_{\text{total}}\rangle=-\frac{\partial \log \mathcal{Z}_{\text{total}}}{\partial \beta},
\end{equation}
where
\begin{equation}
    \mathcal{Z}_{\text{total}}=\Tr\left(e^{-\beta H_{\text{total}}}\right).
\end{equation}
As the system becomes small compared to the bath, the system can be neglected in Eq.~\ref{eq:beta_oqs}. At this point $\beta$ becomes a property of the bath and independent of $\gamma$.

The expectation value of $\langle H_p \rangle$ is given by
\begin{equation}
    \langle H_{\text{total}}\rangle=-\frac{1}{\beta}\frac{\partial \log \mathcal{Z}_{\text{total}}}{\partial \gamma}.
\end{equation}
On the assumption $\beta$ is fixed such that it is independent of $\gamma$, under the Peierls-Bogoliubov inequality the free-energy is concave \cite{Car10}, i.e.,
\begin{equation}
    -\frac{1}{\beta}\frac{\partial^2 \ln \mathcal{Z}_{\text{total}}}{\partial \gamma^2}\leq 0.
\end{equation}
It follows that:
\begin{equation}
    \frac{\partial \langle H_p \rangle }{\partial \gamma}\leq 0,
\end{equation}
so $\langle H_p \rangle$ is a monotonically decreasing function of $\gamma$ for a fixed temperature Gibbs state. It follows that the optimal choice of $\gamma$ is as large as possible. Beyond a certain point, the assumption that $\beta$ is independent of $\gamma$ breaks down. Note that this argument only made use of the fact that the joint system is locally indistinguishable from a Gibbs state, making no assumptions about the form of the terms beyond being Hermitian. In summary, optimising Gibbs states with fixed temperature is straightforward and is problem-instance independent. This simplicity is in stark contrast to the bath-free case \cite{Cal19,Ban24}.  Utilising a bath to solve optimisation problems has previously been explored by Imparato et al.~\cite{Imp24}. They used ancilla qubits as a bath to tune the effective temperature of the joint-system. The explicit measurement of temperature is however, not touched upon. 

Further to this, once the joint system has approached thermal equilibrium, if the system is then quenched such that $\gamma(t)$ is increased at $t=0$, it follows from the passivity of the Gibbs state that $\langle H_p(t)\rangle \leq \langle H_p(0)\rangle$. This provides insight into the suggestion CTQO has some inbuilt error resistance. The inbuilt resistance to errors in QA has been explored in \cite{sch24b, tri23, Chi01}.

\section{Cyclic approaches}
\label{sec:cyc}

In this section, we consider cyclic schedules, with Hamiltonians of the form
\begin{equation}
    \label{eq:rqa}
    H_{cyc}(t)=H_p+G(t) H_d,
\end{equation}
where $0\leq t\leq t_{cyc}$ and $G(t)$ corresponds to a cyclic process (i.e. $G(t=0)=G(t=t_{cyc})=0$). We refer to this process as Reverse Quantum Annealing (RQA). Sec.~\ref{sec:cyc_pass} explores RQA applied to passive states. Sec.~\ref{sec:notPP} explores the limitations of this approach when energy biases are introduced. Finally, Sec.~\ref{sec:warm_start} uses the intuition built up in Sec.~\ref{sec:cyc_pass} to explore warm-started continuous-time quantum walks (CTQWs).

\subsection{Passive states}
\label{sec:cyc_pass}
In this paper, we are considering RQA as a protocol that maps a classical probability density function of computational basis states to another probability density function, by an isolated quantum process. Under this definition, it follows that:
\begin{enumerate}
    \item The entropy of the resulting distribution is greater than or equal to the entropy of the original distribution \cite{Pol11}. The cyclic process causes the distribution to become more spread.
    \item If the initial state is passive, for example a thermal state of $H_p$, then the mean of the resulting distribution can only increase. That is to say, if the initial state is passive, then there exists no unitary cyclic process that will reduce $\langle H_p \rangle$ averaged over the distribution. This is a direct consequence of Planck's principle.
\end{enumerate}

Now we will consider the ground-state probability. In order to do this, we will outline our model of RQA in more detail. The initial state is the ensemble 
\begin{equation}
    \rho_0= \sum_{s} p(s) \ket{s}\bra{s},
\end{equation}
where $\ket{s}$ is an eigenstate of $H_p$ with eigenvalue $s$. Denoting the unitary associated with one RQA cycle to be $U_{\text{cyc}}$, then the transition probability between states $\ket{s}$ and $\ket{j}$ is given by
\begin{equation}
    p(j|s)=\abs{\bra{j} U_{\text{cyc}} \ket{s}}^2.
\end{equation}
Note that the elements $p(j|s)$ constitute a doubly-stochastic matrix, $\mathbf{P}$ \cite{Dal16}. From Birkhoff's theorem \cite{Dal16,Mar10},  $\mathbf{P}$ can be written as a convex sum of permutation matrices $\Pi_\alpha$, such that the resulting state $\rho_1$ can be written as:
\begin{align}
    \rho_1=&\mathbf{P}\rho_0\\
    \rho_1=&\sum_\alpha q_\alpha \Pi_{\alpha}\rho_0
\end{align}
where $0\leq q_{\alpha} \leq 1$ and $\sum_{\alpha}q_\alpha=1$. Here $\rho_i$, $i=0,1$ can be viewed as the diagonal part of the density operator in the computational basis. Interpreting $q_\alpha$ as a probability, we can then view RQA as applying a permutation $\Pi_{\alpha}$ with probability $q_{\alpha}$. The distribution of $q_{\alpha}$ is determined by the schedule of the RQA protocol. This is likely to consist of some local and global moves, with locality being driver dependent. 

The doubly-stochastic evolution places restrictions on the resulting distribution $\rho_1$. For instance, if $\rho_0$ is passive, then $\mathbf{P}$ can only reduce the average quality of solution. We can also look at the change in ground-state probability $\Delta p(0)$ with initial ground-state probability $p(0)$:
\begin{align}
    \Delta p(0)=& \sum_m p(0|m)p(m)-p(m|0)p(0)\\
    \Delta p(0)=& \sum_m p(0|m) \left(p(m)-p(0) \right)\\
\end{align}
where we have used $p(0|m)=p(m|0)$. Bounding $\Delta p(0)$ gives:
\begin{multline}
    \min_m \left(p(m)-p(0)\right) \sum_m p(0|m) \leq \Delta p(0) \leq\\ \max_m \left(p(m)-p(0)\right) \sum_m p(0|m).\\
\end{multline}
Since $\mathbf{P}$ is doubly-stochastic,
\begin{equation}
    \min_m \left(p(m)-p(0)\right) \leq \Delta p(0) \leq \max_m \left(p(m)-p(0)\right). \\
\end{equation}
This means if the ground state is already the most-likely state (even if it is exponentially small) $\mathbf{P}$ can only reduce the probability of finding the ground state. If the ground state is the least likely state, then RQA can only improve the ground-state probability.  More generally, the ground-state probability is bounded by $\max_m p(m)$, which if exponentially small restricts finding the ground state to being exponentially unlikely. If $p(m)$ is sampled from a system with an extensive amount of entropy, then this is likely the case. For passive states the ground state is already the most likely state, hence the ground state probability cannot be increased for these states by RQA. 

\subsection{Biasing means that the system is no longer isolated}
\label{sec:notPP}

In this section, we discuss the limitations of the above analysis when a bias is applied. Consider biasing the RQA Hamiltonian, such that the modified Hamiltonian is given by 
\begin{equation}
    \label{eq:BQA}
    H_{cyc}(t)=H_p+H_b+G(t) H_d,
\end{equation}
where $H_b$ is the biasing Hamiltonian, which is diagonal in the computational basis. To distinguish RQA from RQA with a bias, we refer to this as biased quantum annealing (BQA). Consider the following BQA process, where the initial-state
\begin{equation}
    \rho_0=\sum_z p_z \ket{z} \bra{z}
\end{equation}
is diagonal in the computational basis. For each run, the state $\ket{z}$ is fed into the quantum annealer with probability $p_z$. The initial Hamiltonian is $H_p$. The system interpolates between $H_p$ and $H_b$, where $H_b$ has been chosen such that $\ket{z}$ is the ground-state of $H_b$. The system is then adiabatically evolved to the ground-state of $H_d$ (which by assumption commutes with neither $H_p$ nor $H_b$). From here, the system is adiabatically evolved to the ground-state of $H_p$. The result of this cyclic process is to take $\rho_0$ and map it to the ground-state of the problem Hamiltonian. Hence, we have constructed a cyclic process that leads to extractable work and a decrease in entropy, violating the second law of thermodynamics. This is true even for passive states. This process is typically referred to as adiabatic reverse quantum annealing \cite{Ohk18,Yam19}. 

The apparent violation arises because the evolution is not unitary (hence not isolated), evidenced by the change in von Neumann entropy. The choice of cyclic process is predicated on the initial state loaded into the quantum annealer. Therefore, Planck's principle cannot be applied. This is explored further in Appendix \ref{app:bqa}. This has also been exploited in works such as \cite{Wan22} and \cite{Zha24}.

\subsection{Warm starting continuous-time quantum walks}
\label{sec:warm_start}

Under Assumption \ref{ass:work}, the total energy of the system has to rise as a result of a cyclic process. This rules out cooling $H_p$ cyclically to find good quality solutions given a Hamiltonian of the form shown in Eq.~\ref{eq:rqa}. This is at odds with approaches such as RQA. But an average increase in $\langle H_p \rangle$ does not rule out finding states with a lower value of $\langle H_p \rangle$ - this is simply not the average case. Since in RQA the system starts in an energy eigenstate, the uncertainty can only increase as a result of the cyclic process. Provided the shift in average energy is less than the increased uncertainty in energy, RQA has a chance of doing better than its initial guess. 

In this section, we numerically investigate warm-started CTQWs. This is an example of a cyclic process, as an example of RQA. As discussed in Sec.~\ref{sec:CTQO} warm-starting refers to using previously obtained information to boost the performance of the algorithm. Here we consider the case of a CTQW, where instead of starting in an eigenstate of $H_d$ we start in an eigenstate of $H_p$. In order to apply Assumption \ref{ass:work}, we require that the initial state $\ket{z^*}$ is better than random guessing (see Sec.~\ref{sec:ass} for further discussion):
\begin{equation}
    \label{eq:init_cond}
    \bra{z^*}H_p\ket{z^*}<\Tr'H_p.
\end{equation}
We consider the Hamiltonian:
\begin{equation}
    \label{eq:ws_ham}
    H_{WS}=G(t) H_d+H_p,
\end{equation}
where
\begin{equation}
    \label{eq:ws_cyc}
    G(t)=\begin{cases*}
        0 \text{  for } t\leq 0 \\
        g \text{  for } 0<t\leq t_1 \\
        0 \text{  for } t_1\leq t.
    \end{cases*}
\end{equation}
Note that the time-dependent schedule is now appended to $H_d$, not $H_p$. The period between $0<t\leq t_1$ corresponds to a CTQW.

\begin{figure}
    \centering
    \includegraphics[width=0.48 \textwidth]{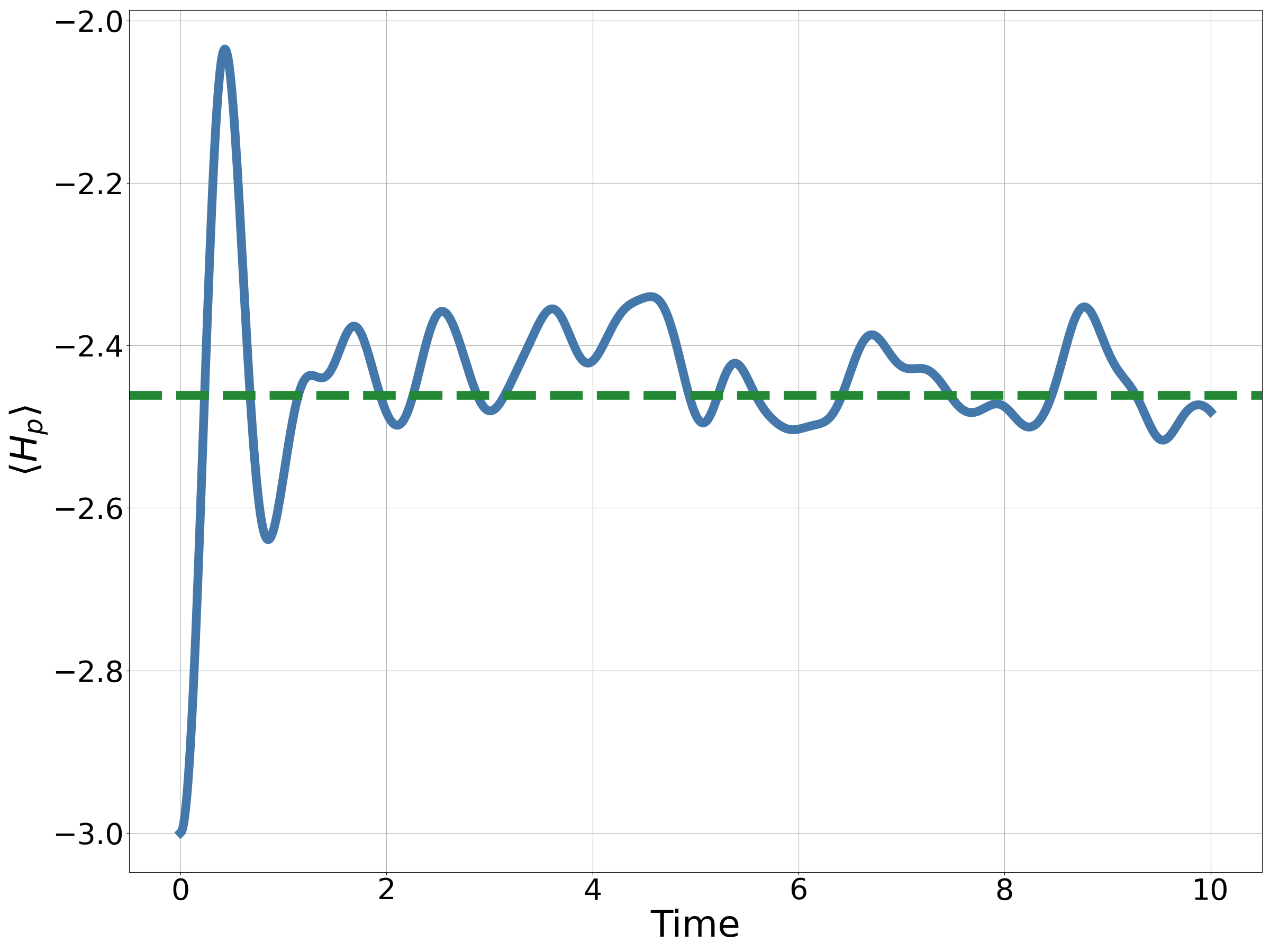}
     \caption{The time-evolution of $\langle H_p \rangle$ for a 12 qubit MAX-CUT instance. The dashed green line shows the infinite time average, $\overline{\langle H_p \rangle}$. }
    \label{fig:ws_mc_ex_t}
\end{figure}

\begin{figure}
    \centering
    \includegraphics[width=0.48 \textwidth]{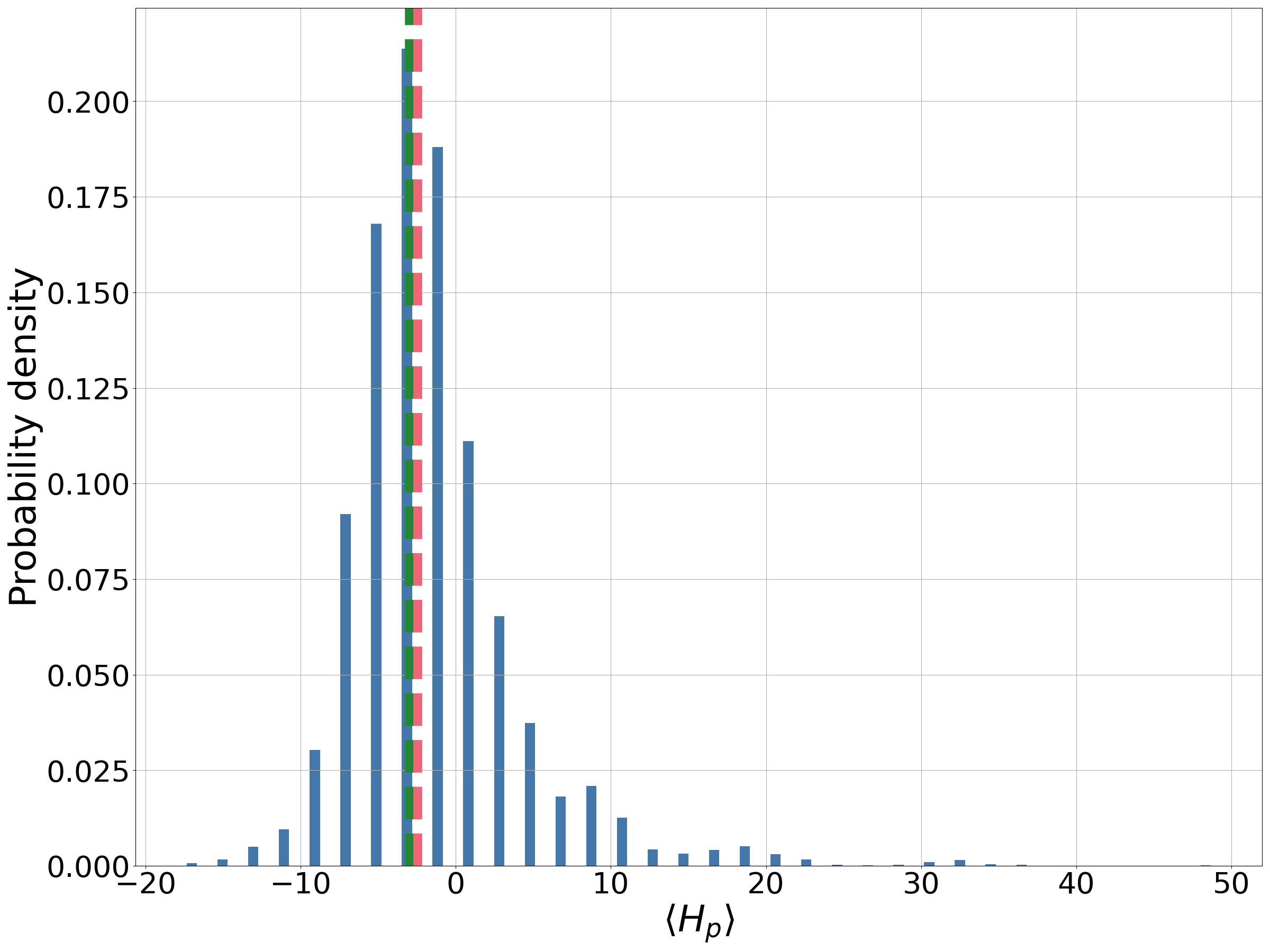}
    \caption{The time-averaged distribution of $\langle H_p \rangle$ for the problem instance shown in Fig.~\ref{fig:ws_mc_ex_t}. The dashed green line shows the original value of $\langle H_p \rangle$. The dashed red line shows the average of the distribution.}
    \label{fig:ws_mc_ex_dist}
\end{figure}

\begin{figure}
    \centering
        \includegraphics[width=0.48\textwidth]{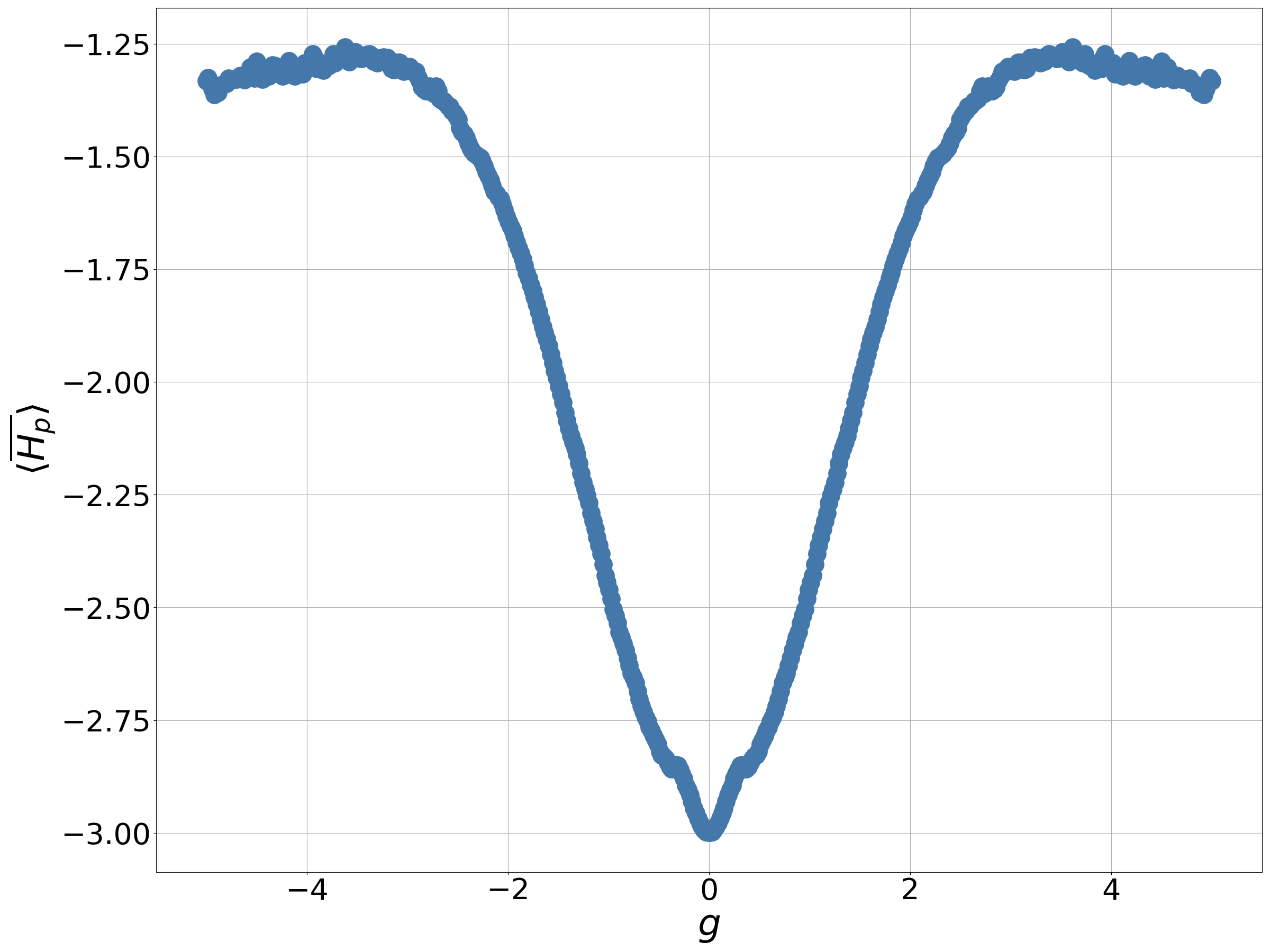}
        \caption{The time-averaged value of $\langle H_p \rangle$ as $g$ is varied for the MAX-CUT instance considered in Fig.~\ref{fig:ws_mc_ex_t}.}
        \label{fig:ws_mc_g}
\end{figure}

\begin{figure}
    \centering
    \includegraphics[width=0.48\textwidth]{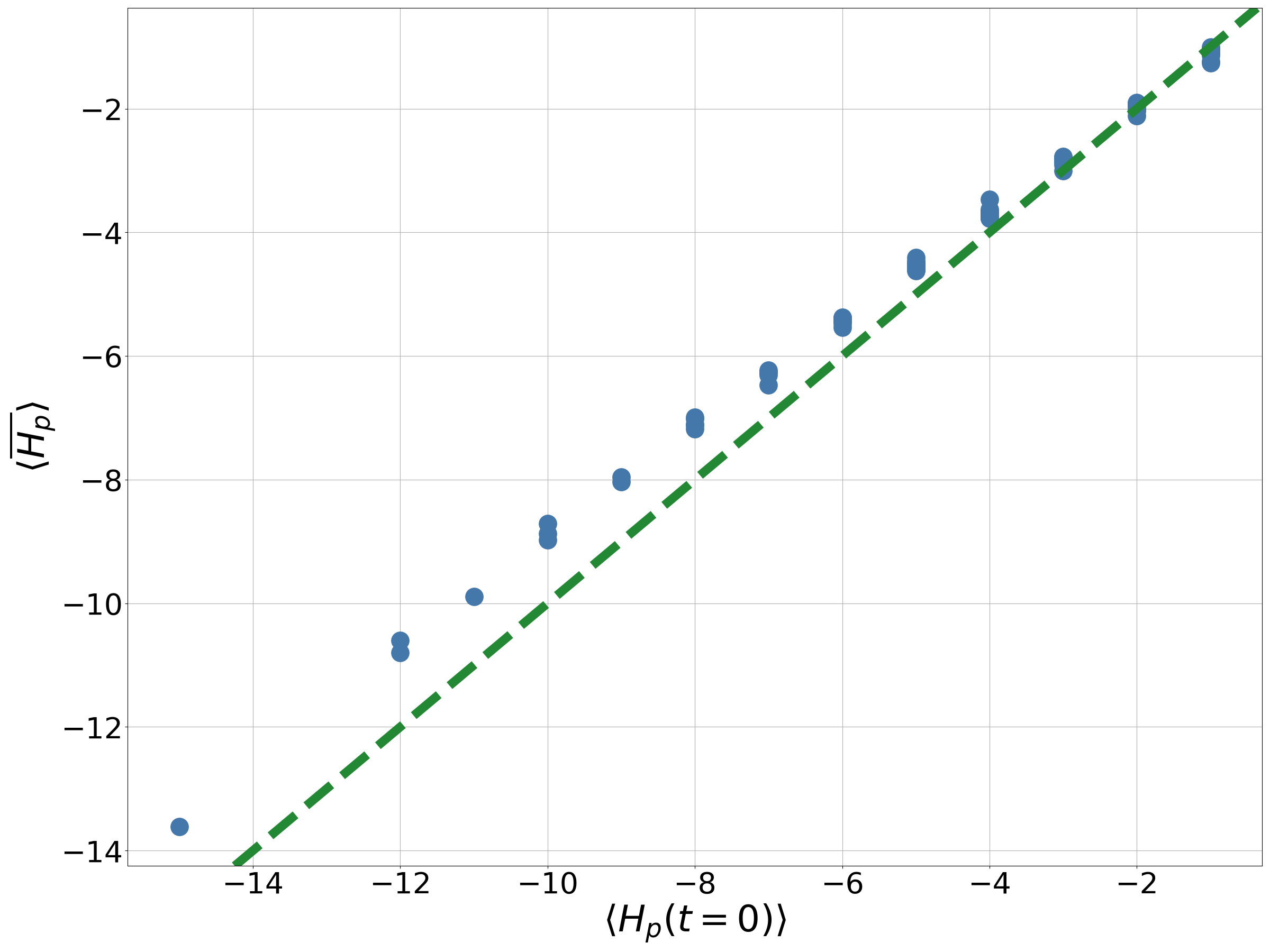}
    \caption{The initial value of $\langle H_p \rangle$ for a warm-started CTQW compared to the infinite time averaged value. For each instance $g=0.5$. The blue circles show 100 MAX-CUT instances.The number of qubits is 12 in all cases. The initial state is randomly selected given that it satisfies Eq.~\ref{eq:init_cond}.}
    \label{fig:ws_num}
\end{figure}

In order to provide numerical evidence of the challenges of warm-starting CTQWs, we consider MAX-CUT on binomial graphs with each edge is selected with probability 2/3. The Ising Hamiltonian for MAX-CUT is shown in Fig.~\ref{eq:MCIsing}. We take $H_d$ to be the transverse-field driver (shown in Eq.~\ref{eq:TF}). The problems in this section consist of 12-qubits. Fig.~\ref{fig:ws_mc_ex_t} shows a MAX-CUT instance with $g=1$, where $g$ is the coefficient in front of $H_d$. The evolution after some time approaches a steady state, as expected for a time-independent Hamiltonian. The dashed green line shows the infinite time average of $\langle H_p \rangle$, denoted by $\overline{\langle H_p \rangle}$. As predicted by Assumption \ref{ass:work}, $\langle H_p \rangle$ is always greater than its initial value.

As mentioned, since the initial state is an energy eigenstate, the uncertainty can only increase, giving the warm-stared CTQW a chance to find better solutions. To see this for the warm-started CTQWs, we look at the probability distribution of observing a given value of $\langle H_p \rangle$ for the infinite time averaged density operator. We consider the same example as Fig.~\ref{fig:ws_mc_ex_t}. The distribution of $\langle H_p \rangle$ for the MAX-CUT instance can be found in Fig.~\ref{fig:ws_mc_ex_dist}. Despite the average value of $\langle H_p \rangle$ having increased, there is significant overlap with states with a lower value of $\langle H_p \rangle$ than the initial state.    

To demonstrate that the average increase in $\langle H_p \rangle$ is not specific to the choice of $g$, Fig.~\ref{fig:ws_mc_g} shows how $\overline{\langle H_p \rangle}$ varies with $g$. The problem instance is the same as Fig.~\ref{fig:ws_mc_ex_t}. It demonstrates clear heating, as for any value of $g$ the value of $\overline{\langle H_p \rangle}$ increases or stays the same compared to its initial value. The initial value of $\overline{\langle H_p \rangle}$ corresponds to $g=0$. So for the cyclic process set out in Sec.~\ref{sec:warm_start} there is an increase in energy, this corresponds to heating.
    
Fig.~\ref{fig:ws_num} shows how $\overline{\langle H_p \rangle}$ compares with the initial value of $\langle H_p(t=0) \rangle$ for 100 instances of the MAX-CUT problem. In each instance $g=0.5$. The dashed green line marks no change in $\langle H_p \rangle$.  For a few initial states, there is a minor cooling effect from the CTQW. This is perhaps reflective of the problem having more structure than the bell shape sketched in Fig.~\ref{fig:dos}.

\begin{figure}
    \centering
    \includegraphics[width=0.48\textwidth]{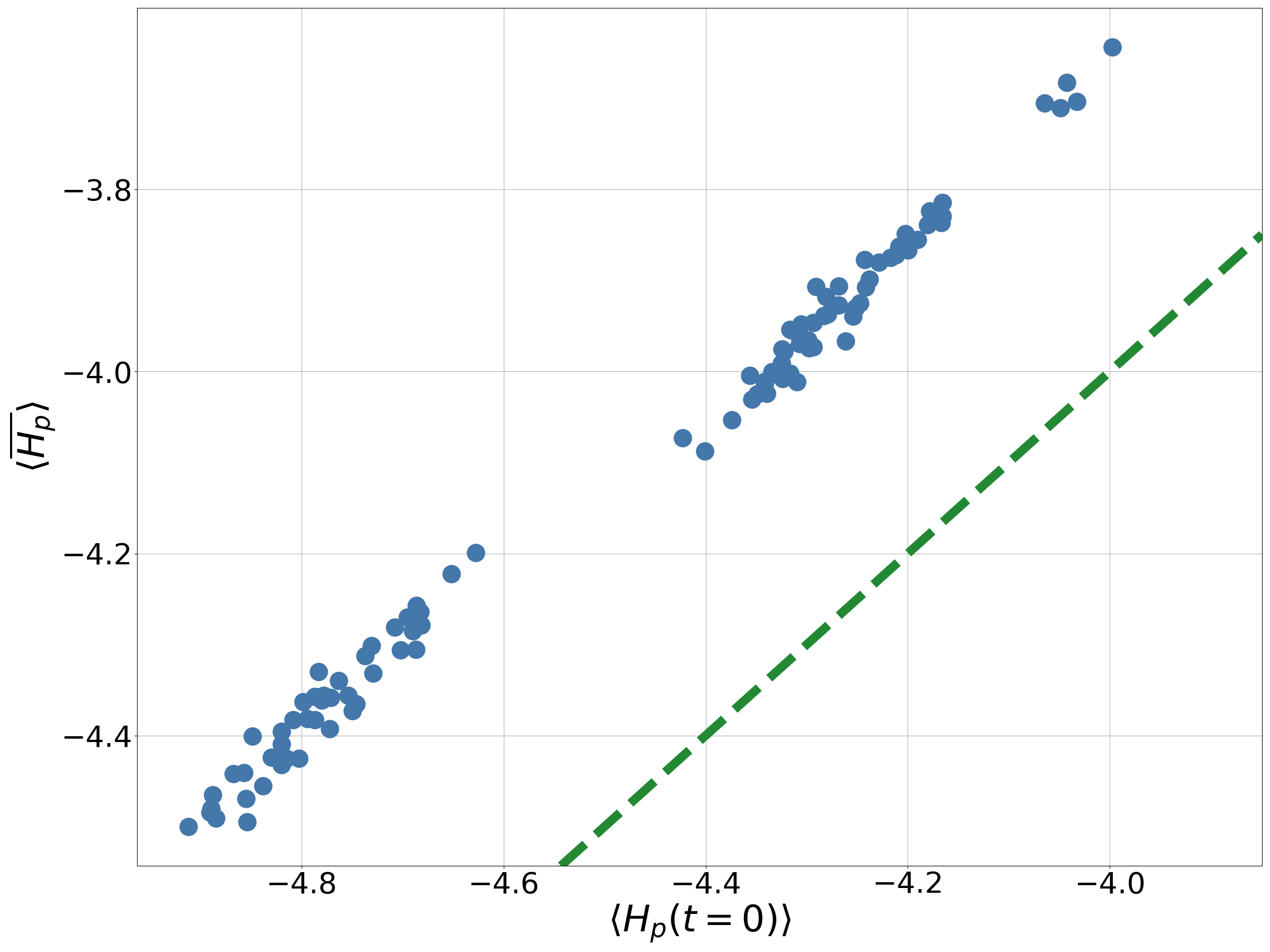}
    \caption{The initial value of $\langle H_p \rangle$ for a warm-started CTQW compared to the infinite time averaged value. For each instance $g=0.5$. The blue circles show 100 MAX-CUT instances. The number of qubits is 12 in all cases. The initial state is given by Eq.~\ref{eq:ens_rho0}.}
    \label{fig:ws_num_ens}
\end{figure}

In each instance so far we have shown what has happened when the system is initialised randomly in an eigenstate of $H_p$, that satisfies the condition $\bra{s} H_p \ket{s}<0$ where $\ket{s}$ is the initial-state. For a few instances in Fig.~\ref{fig:ws_num} we have seen $\overline{\langle H_p \rangle}$ decrease — does this imply cooling and contradict the work in the previous section? Firstly, we have taken Planck's principle to be a physically motivated principle and broadly true. Secondly, this cooling effect disappears once the system is averaged over the full ensemble, taking into account all possible starting states. The full ensemble can be taken into account by using the density operator:
\begin{equation}
    \label{eq:ens_rho0}
    \rho_0=\frac{1}{\mathcal{N}} \sum_{s:\bra{s} H_p \ket{s}<\Tr' H_p} \ket{s}\bra{s},
\end{equation}
where $\ket{s}$ is an eigenstate of $H_p$ with eigenvalue $s$. The normalisation $\mathcal{N}$ of the initial state is
\begin{equation}
    \mathcal{N}=\sum_{s:\bra{s} H_p \ket{s}<0} 1.
\end{equation}
Repeating Fig.~\ref{fig:ws_num} with Eq.~\ref{eq:ens_rho0} gives Fig.~\ref{fig:ws_num_ens}. Note that $\rho_0$ is a passive state. From Fig.~\ref{fig:ws_num_ens} it is clear that by averaging over the ensemble given by Eq.~\ref{eq:ens_rho0} there is no exception. 

In summary, Planck's principle remains a good predictor of the behaviour of cyclic approaches. Further numerical studies can be found in Appendix \ref{app:cyc_num}.

\section{Conclusion}

Planck's principle is a physically reasonable assumption in a broad range of circumstances. In this work, we have used it to motivate continuous-time quantum optimisation (CTQO) without appealing to the adiabatic theorem. Pure state statistical physics provides a novel way of investigating CTQO without appealing to the minutiae of the energy spectrum that would typically be inaccessible for large quantum systems. Statistical physics also provides new ways of discussing and reasoning about CTQO, by invoking (for example) temperature. 

Planck's principle motivates forward quantum annealing (QA) with monotonic schedules, but raises questions over cyclic processes (such as reverse quantum annealing). We demonstrated that $\langle H_p \rangle$ can only improve under monotonic quenching. Planck's principle was also able to capture the behaviour of both monotonic schedules and cyclic schedules, despite their obvious differences. In this work we have demonstrated that Planck's principle can provide an intuitive explanation for CTQO, reflected in the numerical experiments considered.

With improved knowledge of how systems behave away from the adiabatic limit, it may be possible to design better heuristic continuous-time quantum algorithms for optimisation. Studying CTQO is to understand the limitations of Planck's principle, or equivalently Kelvin's statement of the second law of thermodynamics, within the quantum regime. 

\section*{Acknowledgements}
We gratefully acknowledge Henry Chew, Natasha Feinstein, and Leon Guerrero for inspiring discussion and helpful comments. The authors acknowledge the use of the UCL Myriad High Performance Computing Facility (Myriad@UCL), and associated support services, in the completion of this work. This work was supported by the Engineering and Physical Sciences Research Council through the Centre for Doctoral Training in Delivering Quantum Technologies [grant number EP/S021582/1]. 

\bibliography{references.bib}

\begin{thebibliography}{73}%
\makeatletter
\providecommand \@ifxundefined [1]{%
 \@ifx{#1\undefined}
}%
\providecommand \@ifnum [1]{%
 \ifnum #1\expandafter \@firstoftwo
 \else \expandafter \@secondoftwo
 \fi
}%
\providecommand \@ifx [1]{%
 \ifx #1\expandafter \@firstoftwo
 \else \expandafter \@secondoftwo
 \fi
}%
\providecommand \natexlab [1]{#1}%
\providecommand \enquote  [1]{``#1''}%
\providecommand \bibnamefont  [1]{#1}%
\providecommand \bibfnamefont [1]{#1}%
\providecommand \citenamefont [1]{#1}%
\providecommand \href@noop [0]{\@secondoftwo}%
\providecommand \href [0]{\begingroup \@sanitize@url \@href}%
\providecommand \@href[1]{\@@startlink{#1}\@@href}%
\providecommand \@@href[1]{\endgroup#1\@@endlink}%
\providecommand \@sanitize@url [0]{\catcode `\\12\catcode `\$12\catcode `\&12\catcode `\#12\catcode `\^12\catcode `\_12\catcode `\%12\relax}%
\providecommand \@@startlink[1]{}%
\providecommand \@@endlink[0]{}%
\providecommand \url  [0]{\begingroup\@sanitize@url \@url }%
\providecommand \@url [1]{\endgroup\@href {#1}{\urlprefix }}%
\providecommand \urlprefix  [0]{URL }%
\providecommand \Eprint [0]{\href }%
\providecommand \doibase [0]{https://doi.org/}%
\providecommand \selectlanguage [0]{\@gobble}%
\providecommand \bibinfo  [0]{\@secondoftwo}%
\providecommand \bibfield  [0]{\@secondoftwo}%
\providecommand \translation [1]{[#1]}%
\providecommand \BibitemOpen [0]{}%
\providecommand \bibitemStop [0]{}%
\providecommand \bibitemNoStop [0]{.\EOS\space}%
\providecommand \EOS [0]{\spacefactor3000\relax}%
\providecommand \BibitemShut  [1]{\csname bibitem#1\endcsname}%
\let\auto@bib@innerbib\@empty
\bibitem [{\citenamefont {Farhi}\ \emph {et~al.}(2000)\citenamefont {Farhi}, \citenamefont {Goldstone}, \citenamefont {Gutmann},\ and\ \citenamefont {Sipser}}]{Far00}%
  \BibitemOpen
  \bibfield  {author} {\bibinfo {author} {\bibfnamefont {E.}~\bibnamefont {Farhi}}, \bibinfo {author} {\bibfnamefont {J.}~\bibnamefont {Goldstone}}, \bibinfo {author} {\bibfnamefont {S.}~\bibnamefont {Gutmann}},\ and\ \bibinfo {author} {\bibfnamefont {M.}~\bibnamefont {Sipser}},\ }\href@noop {} {\bibinfo {title} {Quantum computation by adiabatic evolution}} (\bibinfo {year} {2000}),\ \Eprint {https://arxiv.org/abs/quant-ph/0001106} {arXiv:quant-ph/0001106 [quant-ph]} \BibitemShut {NoStop}%
\bibitem [{\citenamefont {Albash}\ and\ \citenamefont {Lidar}(2018)}]{Alb18}%
  \BibitemOpen
  \bibfield  {author} {\bibinfo {author} {\bibfnamefont {T.}~\bibnamefont {Albash}}\ and\ \bibinfo {author} {\bibfnamefont {D.~A.}\ \bibnamefont {Lidar}},\ }\bibfield  {title} {\bibinfo {title} {Adiabatic quantum computation},\ }\href {https://doi.org/10.1103/RevModPhys.90.015002} {\bibfield  {journal} {\bibinfo  {journal} {Rev. Mod. Phys.}\ }\textbf {\bibinfo {volume} {90}},\ \bibinfo {pages} {015002} (\bibinfo {year} {2018})}\BibitemShut {NoStop}%
\bibitem [{\citenamefont {Kadowaki}\ and\ \citenamefont {Nishimori}(1998)}]{Kad98}%
  \BibitemOpen
  \bibfield  {author} {\bibinfo {author} {\bibfnamefont {T.}~\bibnamefont {Kadowaki}}\ and\ \bibinfo {author} {\bibfnamefont {H.}~\bibnamefont {Nishimori}},\ }\bibfield  {title} {\bibinfo {title} {Quantum annealing in the transverse ising model},\ }\href {https://doi.org/10.1103/PhysRevE.58.5355} {\bibfield  {journal} {\bibinfo  {journal} {Phys. Rev. E}\ }\textbf {\bibinfo {volume} {58}},\ \bibinfo {pages} {5355} (\bibinfo {year} {1998})}\BibitemShut {NoStop}%
\bibitem [{\citenamefont {Hauke}\ \emph {et~al.}(2020)\citenamefont {Hauke}, \citenamefont {Katzgraber}, \citenamefont {Lechner}, \citenamefont {Nishimori},\ and\ \citenamefont {Oliver}}]{Hau20}%
  \BibitemOpen
  \bibfield  {author} {\bibinfo {author} {\bibfnamefont {P.}~\bibnamefont {Hauke}}, \bibinfo {author} {\bibfnamefont {H.~G.}\ \bibnamefont {Katzgraber}}, \bibinfo {author} {\bibfnamefont {W.}~\bibnamefont {Lechner}}, \bibinfo {author} {\bibfnamefont {H.}~\bibnamefont {Nishimori}},\ and\ \bibinfo {author} {\bibfnamefont {W.~D.}\ \bibnamefont {Oliver}},\ }\bibfield  {title} {\bibinfo {title} {Perspectives of quantum annealing: methods and implementations},\ }\href {https://doi.org/10.1088/1361-6633/ab85b8} {\bibfield  {journal} {\bibinfo  {journal} {Reports on Progress in Physics}\ }\textbf {\bibinfo {volume} {83}},\ \bibinfo {pages} {054401} (\bibinfo {year} {2020})}\BibitemShut {NoStop}%
\bibitem [{\citenamefont {Crosson}\ and\ \citenamefont {Lidar}(2021)}]{Cro21}%
  \BibitemOpen
  \bibfield  {author} {\bibinfo {author} {\bibfnamefont {E.~J.}\ \bibnamefont {Crosson}}\ and\ \bibinfo {author} {\bibfnamefont {D.~A.}\ \bibnamefont {Lidar}},\ }\bibfield  {title} {\bibinfo {title} {Prospects for quantum enhancement with diabatic quantum annealing},\ }\href {https://doi.org/10.1038/s42254-021-00313-6} {\bibfield  {journal} {\bibinfo  {journal} {Nature Reviews Physics}\ }\textbf {\bibinfo {volume} {3}},\ \bibinfo {pages} {466} (\bibinfo {year} {2021})}\BibitemShut {NoStop}%
\bibitem [{\citenamefont {Callison}\ \emph {et~al.}(2019)\citenamefont {Callison}, \citenamefont {Chancellor}, \citenamefont {Mintert},\ and\ \citenamefont {Kendon}}]{Cal19}%
  \BibitemOpen
  \bibfield  {author} {\bibinfo {author} {\bibfnamefont {A.}~\bibnamefont {Callison}}, \bibinfo {author} {\bibfnamefont {N.}~\bibnamefont {Chancellor}}, \bibinfo {author} {\bibfnamefont {F.}~\bibnamefont {Mintert}},\ and\ \bibinfo {author} {\bibfnamefont {V.}~\bibnamefont {Kendon}},\ }\bibfield  {title} {\bibinfo {title} {Finding spin glass ground states using quantum walks},\ }\href {https://doi.org/10.1088/1367-2630/ab5ca2} {\bibfield  {journal} {\bibinfo  {journal} {New Journal of Physics}\ }\textbf {\bibinfo {volume} {21}},\ \bibinfo {pages} {123022} (\bibinfo {year} {2019})}\BibitemShut {NoStop}%
\bibitem [{\citenamefont {Callison}\ \emph {et~al.}(2021)\citenamefont {Callison}, \citenamefont {Festenstein}, \citenamefont {Chen}, \citenamefont {Nita}, \citenamefont {Kendon},\ and\ \citenamefont {Chancellor}}]{Cal21}%
  \BibitemOpen
  \bibfield  {author} {\bibinfo {author} {\bibfnamefont {A.}~\bibnamefont {Callison}}, \bibinfo {author} {\bibfnamefont {M.}~\bibnamefont {Festenstein}}, \bibinfo {author} {\bibfnamefont {J.}~\bibnamefont {Chen}}, \bibinfo {author} {\bibfnamefont {L.}~\bibnamefont {Nita}}, \bibinfo {author} {\bibfnamefont {V.}~\bibnamefont {Kendon}},\ and\ \bibinfo {author} {\bibfnamefont {N.}~\bibnamefont {Chancellor}},\ }\bibfield  {title} {\bibinfo {title} {Energetic perspective on rapid quenches in quantum annealing},\ }\href {https://doi.org/10.1103/PRXQuantum.2.010338} {\bibfield  {journal} {\bibinfo  {journal} {PRX Quantum}\ }\textbf {\bibinfo {volume} {2}},\ \bibinfo {pages} {010338} (\bibinfo {year} {2021})}\BibitemShut {NoStop}%
\bibitem [{\citenamefont {Chancellor}(2017)}]{Cha17}%
  \BibitemOpen
  \bibfield  {author} {\bibinfo {author} {\bibfnamefont {N.}~\bibnamefont {Chancellor}},\ }\bibfield  {title} {\bibinfo {title} {Modernizing quantum annealing using local searches},\ }\href {https://doi.org/10.1088/1367-2630/aa59c4} {\bibfield  {journal} {\bibinfo  {journal} {New Journal of Physics}\ }\textbf {\bibinfo {volume} {19}},\ \bibinfo {pages} {023024} (\bibinfo {year} {2017})}\BibitemShut {NoStop}%
\bibitem [{\citenamefont {Wang}\ \emph {et~al.}(2022)\citenamefont {Wang}, \citenamefont {Yeh},\ and\ \citenamefont {Kamenev}}]{Wan22}%
  \BibitemOpen
  \bibfield  {author} {\bibinfo {author} {\bibfnamefont {H.}~\bibnamefont {Wang}}, \bibinfo {author} {\bibfnamefont {H.-C.}\ \bibnamefont {Yeh}},\ and\ \bibinfo {author} {\bibfnamefont {A.}~\bibnamefont {Kamenev}},\ }\bibfield  {title} {\bibinfo {title} {Many-body localization enables iterative quantum optimization},\ }\href {https://doi.org/10.1038/s41467-022-33179-y} {\bibfield  {journal} {\bibinfo  {journal} {Nature Communications}\ }\textbf {\bibinfo {volume} {13}},\ \bibinfo {pages} {5503} (\bibinfo {year} {2022})}\BibitemShut {NoStop}%
\bibitem [{\citenamefont {Zhang}\ \emph {et~al.}(2024)\citenamefont {Zhang}, \citenamefont {Boothby},\ and\ \citenamefont {Kamenev}}]{Zha24}%
  \BibitemOpen
  \bibfield  {author} {\bibinfo {author} {\bibfnamefont {H.}~\bibnamefont {Zhang}}, \bibinfo {author} {\bibfnamefont {K.}~\bibnamefont {Boothby}},\ and\ \bibinfo {author} {\bibfnamefont {A.}~\bibnamefont {Kamenev}},\ }\href@noop {} {\bibinfo {title} {Cyclic quantum annealing: Searching for deep low-energy states in 5000-qubit spin glass}} (\bibinfo {year} {2024}),\ \Eprint {https://arxiv.org/abs/2403.01034} {arXiv:2403.01034 [cond-mat.dis-nn]} \BibitemShut {NoStop}%
\bibitem [{\citenamefont {Blundell}\ and\ \citenamefont {Blundell}(2009)}]{Blu09}%
  \BibitemOpen
  \bibfield  {author} {\bibinfo {author} {\bibfnamefont {S.~J.}\ \bibnamefont {Blundell}}\ and\ \bibinfo {author} {\bibfnamefont {K.~M.}\ \bibnamefont {Blundell}},\ }\href {https://doi.org/10.1093/acprof:oso/9780199562091.001.0001} {\emph {\bibinfo {title} {{Concepts in Thermal Physics}}}}\ (\bibinfo  {publisher} {Oxford University Press},\ \bibinfo {year} {2009})\BibitemShut {NoStop}%
\bibitem [{\citenamefont {Gogolin}(2010)}]{Gog10}%
  \BibitemOpen
  \bibfield  {author} {\bibinfo {author} {\bibfnamefont {C.}~\bibnamefont {Gogolin}},\ }\href@noop {} {\bibinfo {title} {Pure state quantum statistical mechanics}} (\bibinfo {year} {2010}),\ \Eprint {https://arxiv.org/abs/1003.5058} {arXiv:1003.5058 [quant-ph]} \BibitemShut {NoStop}%
\bibitem [{\citenamefont {D’Alessio}\ \emph {et~al.}(2016)\citenamefont {D’Alessio}, \citenamefont {Kafri}, \citenamefont {Polkovnikov},\ and\ \citenamefont {Rigol}}]{Dal16}%
  \BibitemOpen
  \bibfield  {author} {\bibinfo {author} {\bibfnamefont {L.}~\bibnamefont {D’Alessio}}, \bibinfo {author} {\bibfnamefont {Y.}~\bibnamefont {Kafri}}, \bibinfo {author} {\bibfnamefont {A.}~\bibnamefont {Polkovnikov}},\ and\ \bibinfo {author} {\bibfnamefont {M.}~\bibnamefont {Rigol}},\ }\bibfield  {title} {\bibinfo {title} {From quantum chaos and eigenstate thermalization to statistical mechanics and thermodynamics},\ }\href {https://doi.org/10.1080/00018732.2016.1198134} {\bibfield  {journal} {\bibinfo  {journal} {Advances in Physics}\ }\textbf {\bibinfo {volume} {65}},\ \bibinfo {pages} {239–362} (\bibinfo {year} {2016})}\BibitemShut {NoStop}%
\bibitem [{\citenamefont {Koukoulekidis}\ \emph {et~al.}(2021)\citenamefont {Koukoulekidis}, \citenamefont {Alexander}, \citenamefont {Hebdige},\ and\ \citenamefont {Jennings}}]{Kou21}%
  \BibitemOpen
  \bibfield  {author} {\bibinfo {author} {\bibfnamefont {N.}~\bibnamefont {Koukoulekidis}}, \bibinfo {author} {\bibfnamefont {R.}~\bibnamefont {Alexander}}, \bibinfo {author} {\bibfnamefont {T.}~\bibnamefont {Hebdige}},\ and\ \bibinfo {author} {\bibfnamefont {D.}~\bibnamefont {Jennings}},\ }\bibfield  {title} {\bibinfo {title} {The geometry of passivity for quantum systems and a novel elementary derivation of the {G}ibbs state},\ }\href {https://doi.org/10.22331/q-2021-03-15-411} {\bibfield  {journal} {\bibinfo  {journal} {{Quantum}}\ }\textbf {\bibinfo {volume} {5}},\ \bibinfo {pages} {411} (\bibinfo {year} {2021})}\BibitemShut {NoStop}%
\bibitem [{\citenamefont {Johansson}\ \emph {et~al.}(2012)\citenamefont {Johansson}, \citenamefont {Nation},\ and\ \citenamefont {Nori}}]{Joh12}%
  \BibitemOpen
  \bibfield  {author} {\bibinfo {author} {\bibfnamefont {J.}~\bibnamefont {Johansson}}, \bibinfo {author} {\bibfnamefont {P.}~\bibnamefont {Nation}},\ and\ \bibinfo {author} {\bibfnamefont {F.}~\bibnamefont {Nori}},\ }\bibfield  {title} {\bibinfo {title} {Qutip: An open-source python framework for the dynamics of open quantum systems},\ }\href {https://doi.org/https://doi.org/10.1016/j.cpc.2012.02.021} {\bibfield  {journal} {\bibinfo  {journal} {Computer Physics Communications}\ }\textbf {\bibinfo {volume} {183}},\ \bibinfo {pages} {1760} (\bibinfo {year} {2012})}\BibitemShut {NoStop}%
\bibitem [{\citenamefont {Johansson}\ \emph {et~al.}(2013)\citenamefont {Johansson}, \citenamefont {Nation},\ and\ \citenamefont {Nori}}]{Joh13}%
  \BibitemOpen
  \bibfield  {author} {\bibinfo {author} {\bibfnamefont {J.}~\bibnamefont {Johansson}}, \bibinfo {author} {\bibfnamefont {P.}~\bibnamefont {Nation}},\ and\ \bibinfo {author} {\bibfnamefont {F.}~\bibnamefont {Nori}},\ }\bibfield  {title} {\bibinfo {title} {Qutip 2: A python framework for the dynamics of open quantum systems},\ }\href {https://doi.org/https://doi.org/10.1016/j.cpc.2012.11.019} {\bibfield  {journal} {\bibinfo  {journal} {Computer Physics Communications}\ }\textbf {\bibinfo {volume} {184}},\ \bibinfo {pages} {1234} (\bibinfo {year} {2013})}\BibitemShut {NoStop}%
\bibitem [{\citenamefont {Roland}\ and\ \citenamefont {Cerf}(2002)}]{Rol02}%
  \BibitemOpen
  \bibfield  {author} {\bibinfo {author} {\bibfnamefont {J.}~\bibnamefont {Roland}}\ and\ \bibinfo {author} {\bibfnamefont {N.~J.}\ \bibnamefont {Cerf}},\ }\bibfield  {title} {\bibinfo {title} {Quantum search by local adiabatic evolution},\ }\href {https://doi.org/10.1103/PhysRevA.65.042308} {\bibfield  {journal} {\bibinfo  {journal} {Phys. Rev. A}\ }\textbf {\bibinfo {volume} {65}},\ \bibinfo {pages} {042308} (\bibinfo {year} {2002})}\BibitemShut {NoStop}%
\bibitem [{\citenamefont {Hastings}(2021)}]{Has21}%
  \BibitemOpen
  \bibfield  {author} {\bibinfo {author} {\bibfnamefont {M.~B.}\ \bibnamefont {Hastings}},\ }\bibfield  {title} {\bibinfo {title} {The power of adiabatic quantum computation with no sign problem},\ }\href {https://doi.org/10.22331/q-2021-12-06-597} {\bibfield  {journal} {\bibinfo  {journal} {Quantum}\ }\textbf {\bibinfo {volume} {5}},\ \bibinfo {pages} {597} (\bibinfo {year} {2021})}\BibitemShut {NoStop}%
\bibitem [{\citenamefont {Aharonov}\ \emph {et~al.}(2005)\citenamefont {Aharonov}, \citenamefont {van Dam}, \citenamefont {Kempe}, \citenamefont {Landau}, \citenamefont {Lloyd},\ and\ \citenamefont {Regev}}]{aha05}%
  \BibitemOpen
  \bibfield  {author} {\bibinfo {author} {\bibfnamefont {D.}~\bibnamefont {Aharonov}}, \bibinfo {author} {\bibfnamefont {W.}~\bibnamefont {van Dam}}, \bibinfo {author} {\bibfnamefont {J.}~\bibnamefont {Kempe}}, \bibinfo {author} {\bibfnamefont {Z.}~\bibnamefont {Landau}}, \bibinfo {author} {\bibfnamefont {S.}~\bibnamefont {Lloyd}},\ and\ \bibinfo {author} {\bibfnamefont {O.}~\bibnamefont {Regev}},\ }\href@noop {} {\bibinfo {title} {Adiabatic quantum computation is equivalent to standard quantum computation}} (\bibinfo {year} {2005}),\ \Eprint {https://arxiv.org/abs/quant-ph/0405098} {arXiv:quant-ph/0405098 [quant-ph]} \BibitemShut {NoStop}%
\bibitem [{\citenamefont {Banks}\ \emph {et~al.}(2024)\citenamefont {Banks}, \citenamefont {Haque}, \citenamefont {Nazef}, \citenamefont {Fethallah}, \citenamefont {Ruqaya}, \citenamefont {Ahsan}, \citenamefont {Vora}, \citenamefont {Tahir}, \citenamefont {Ahmad}, \citenamefont {Hewins}, \citenamefont {Shah}, \citenamefont {Baranwal}, \citenamefont {Arora}, \citenamefont {Asad}, \citenamefont {Khan}, \citenamefont {Hasan}, \citenamefont {Azad}, \citenamefont {Fedaiee}, \citenamefont {Majeed}, \citenamefont {Bhuyan}, \citenamefont {Tarannum}, \citenamefont {Ali}, \citenamefont {Browne},\ and\ \citenamefont {Warburton}}]{Ban24}%
  \BibitemOpen
  \bibfield  {author} {\bibinfo {author} {\bibfnamefont {R.~J.}\ \bibnamefont {Banks}}, \bibinfo {author} {\bibfnamefont {E.}~\bibnamefont {Haque}}, \bibinfo {author} {\bibfnamefont {F.}~\bibnamefont {Nazef}}, \bibinfo {author} {\bibfnamefont {F.}~\bibnamefont {Fethallah}}, \bibinfo {author} {\bibfnamefont {F.}~\bibnamefont {Ruqaya}}, \bibinfo {author} {\bibfnamefont {H.}~\bibnamefont {Ahsan}}, \bibinfo {author} {\bibfnamefont {H.}~\bibnamefont {Vora}}, \bibinfo {author} {\bibfnamefont {H.}~\bibnamefont {Tahir}}, \bibinfo {author} {\bibfnamefont {I.}~\bibnamefont {Ahmad}}, \bibinfo {author} {\bibfnamefont {I.}~\bibnamefont {Hewins}}, \bibinfo {author} {\bibfnamefont {I.}~\bibnamefont {Shah}}, \bibinfo {author} {\bibfnamefont {K.}~\bibnamefont {Baranwal}}, \bibinfo {author} {\bibfnamefont {M.}~\bibnamefont {Arora}}, \bibinfo {author} {\bibfnamefont {M.}~\bibnamefont {Asad}}, \bibinfo {author} {\bibfnamefont {M.}~\bibnamefont {Khan}}, \bibinfo {author} {\bibfnamefont {N.}~\bibnamefont {Hasan}}, \bibinfo
  {author} {\bibfnamefont {N.}~\bibnamefont {Azad}}, \bibinfo {author} {\bibfnamefont {S.}~\bibnamefont {Fedaiee}}, \bibinfo {author} {\bibfnamefont {S.}~\bibnamefont {Majeed}}, \bibinfo {author} {\bibfnamefont {S.}~\bibnamefont {Bhuyan}}, \bibinfo {author} {\bibfnamefont {T.}~\bibnamefont {Tarannum}}, \bibinfo {author} {\bibfnamefont {Y.}~\bibnamefont {Ali}}, \bibinfo {author} {\bibfnamefont {D.~E.}\ \bibnamefont {Browne}},\ and\ \bibinfo {author} {\bibfnamefont {P.~A.}\ \bibnamefont {Warburton}},\ }\bibfield  {title} {\bibinfo {title} {Continuous-time quantum walks for max-cut are hot},\ }\href {https://doi.org/10.22331/q-2024-02-13-1254} {\bibfield  {journal} {\bibinfo  {journal} {Quantum}\ }\textbf {\bibinfo {volume} {8}},\ \bibinfo {pages} {1254} (\bibinfo {year} {2024})}\BibitemShut {NoStop}%
\bibitem [{\citenamefont {King}\ \emph {et~al.}(2022)\citenamefont {King}, \citenamefont {Suzuki}, \citenamefont {Raymond}, \citenamefont {Zucca}, \citenamefont {Lanting}, \citenamefont {Altomare}, \citenamefont {Berkley}, \citenamefont {Ejtemaee}, \citenamefont {Hoskinson}, \citenamefont {Huang}, \citenamefont {Ladizinsky}, \citenamefont {MacDonald}, \citenamefont {Marsden}, \citenamefont {Oh}, \citenamefont {Poulin-Lamarre}, \citenamefont {Reis}, \citenamefont {Rich}, \citenamefont {Sato}, \citenamefont {Whittaker}, \citenamefont {Yao}, \citenamefont {Harris}, \citenamefont {Lidar}, \citenamefont {Nishimori},\ and\ \citenamefont {Amin}}]{Kin22}%
  \BibitemOpen
  \bibfield  {author} {\bibinfo {author} {\bibfnamefont {A.~D.}\ \bibnamefont {King}}, \bibinfo {author} {\bibfnamefont {S.}~\bibnamefont {Suzuki}}, \bibinfo {author} {\bibfnamefont {J.}~\bibnamefont {Raymond}}, \bibinfo {author} {\bibfnamefont {A.}~\bibnamefont {Zucca}}, \bibinfo {author} {\bibfnamefont {T.}~\bibnamefont {Lanting}}, \bibinfo {author} {\bibfnamefont {F.}~\bibnamefont {Altomare}}, \bibinfo {author} {\bibfnamefont {A.~J.}\ \bibnamefont {Berkley}}, \bibinfo {author} {\bibfnamefont {S.}~\bibnamefont {Ejtemaee}}, \bibinfo {author} {\bibfnamefont {E.}~\bibnamefont {Hoskinson}}, \bibinfo {author} {\bibfnamefont {S.}~\bibnamefont {Huang}}, \bibinfo {author} {\bibfnamefont {E.}~\bibnamefont {Ladizinsky}}, \bibinfo {author} {\bibfnamefont {A.~J.~R.}\ \bibnamefont {MacDonald}}, \bibinfo {author} {\bibfnamefont {G.}~\bibnamefont {Marsden}}, \bibinfo {author} {\bibfnamefont {T.}~\bibnamefont {Oh}}, \bibinfo {author} {\bibfnamefont {G.}~\bibnamefont {Poulin-Lamarre}}, \bibinfo {author} {\bibfnamefont
  {M.}~\bibnamefont {Reis}}, \bibinfo {author} {\bibfnamefont {C.}~\bibnamefont {Rich}}, \bibinfo {author} {\bibfnamefont {Y.}~\bibnamefont {Sato}}, \bibinfo {author} {\bibfnamefont {J.~D.}\ \bibnamefont {Whittaker}}, \bibinfo {author} {\bibfnamefont {J.}~\bibnamefont {Yao}}, \bibinfo {author} {\bibfnamefont {R.}~\bibnamefont {Harris}}, \bibinfo {author} {\bibfnamefont {D.~A.}\ \bibnamefont {Lidar}}, \bibinfo {author} {\bibfnamefont {H.}~\bibnamefont {Nishimori}},\ and\ \bibinfo {author} {\bibfnamefont {M.~H.}\ \bibnamefont {Amin}},\ }\bibfield  {title} {\bibinfo {title} {Coherent quantum annealing in a programmable 2,000{\thinspace}qubit ising chain},\ }\href {https://doi.org/10.1038/s41567-022-01741-6} {\bibfield  {journal} {\bibinfo  {journal} {Nature Physics}\ }\textbf {\bibinfo {volume} {18}},\ \bibinfo {pages} {1324} (\bibinfo {year} {2022})}\BibitemShut {NoStop}%
\bibitem [{\citenamefont {Egger}\ \emph {et~al.}(2021)\citenamefont {Egger}, \citenamefont {Mare{\v{c}}ek},\ and\ \citenamefont {Woerner}}]{Egg21}%
  \BibitemOpen
  \bibfield  {author} {\bibinfo {author} {\bibfnamefont {D.~J.}\ \bibnamefont {Egger}}, \bibinfo {author} {\bibfnamefont {J.}~\bibnamefont {Mare{\v{c}}ek}},\ and\ \bibinfo {author} {\bibfnamefont {S.}~\bibnamefont {Woerner}},\ }\bibfield  {title} {\bibinfo {title} {Warm-starting quantum optimization},\ }\href {https://doi.org/10.22331/q-2021-06-17-479} {\bibfield  {journal} {\bibinfo  {journal} {{Quantum}}\ }\textbf {\bibinfo {volume} {5}},\ \bibinfo {pages} {479} (\bibinfo {year} {2021})}\BibitemShut {NoStop}%
\bibitem [{\citenamefont {Wilming}\ \emph {et~al.}(2018)\citenamefont {Wilming}, \citenamefont {de~Oliveira}, \citenamefont {Short},\ and\ \citenamefont {Eisert}}]{Wil18}%
  \BibitemOpen
  \bibfield  {author} {\bibinfo {author} {\bibfnamefont {H.}~\bibnamefont {Wilming}}, \bibinfo {author} {\bibfnamefont {T.~R.}\ \bibnamefont {de~Oliveira}}, \bibinfo {author} {\bibfnamefont {A.~J.}\ \bibnamefont {Short}},\ and\ \bibinfo {author} {\bibfnamefont {J.}~\bibnamefont {Eisert}},\ }\bibinfo {title} {Equilibration times in closed quantum many-body systems},\ in\ \href {https://doi.org/10.1007/978-3-319-99046-0_18} {\emph {\bibinfo {booktitle} {Thermodynamics in the Quantum Regime: Fundamental Aspects and New Directions}}},\ \bibinfo {editor} {edited by\ \bibinfo {editor} {\bibfnamefont {F.}~\bibnamefont {Binder}}, \bibinfo {editor} {\bibfnamefont {L.~A.}\ \bibnamefont {Correa}}, \bibinfo {editor} {\bibfnamefont {C.}~\bibnamefont {Gogolin}}, \bibinfo {editor} {\bibfnamefont {J.}~\bibnamefont {Anders}},\ and\ \bibinfo {editor} {\bibfnamefont {G.}~\bibnamefont {Adesso}}}\ (\bibinfo  {publisher} {Springer International Publishing},\ \bibinfo {address} {Cham},\ \bibinfo {year} {2018})\ pp.\ \bibinfo
  {pages} {435--455}\BibitemShut {NoStop}%
\bibitem [{\citenamefont {de~Oliveira}\ \emph {et~al.}(2018)\citenamefont {de~Oliveira}, \citenamefont {Charalambous}, \citenamefont {Jonathan}, \citenamefont {Lewenstein},\ and\ \citenamefont {Riera}}]{Oli18}%
  \BibitemOpen
  \bibfield  {author} {\bibinfo {author} {\bibfnamefont {T.~R.}\ \bibnamefont {de~Oliveira}}, \bibinfo {author} {\bibfnamefont {C.}~\bibnamefont {Charalambous}}, \bibinfo {author} {\bibfnamefont {D.}~\bibnamefont {Jonathan}}, \bibinfo {author} {\bibfnamefont {M.}~\bibnamefont {Lewenstein}},\ and\ \bibinfo {author} {\bibfnamefont {A.}~\bibnamefont {Riera}},\ }\bibfield  {title} {\bibinfo {title} {Equilibration time scales in closed many-body quantum systems},\ }\href {https://doi.org/10.1088/1367-2630/aab03b} {\bibfield  {journal} {\bibinfo  {journal} {New Journal of Physics}\ }\textbf {\bibinfo {volume} {20}},\ \bibinfo {pages} {033032} (\bibinfo {year} {2018})}\BibitemShut {NoStop}%
\bibitem [{\citenamefont {Wilming}\ \emph {et~al.}(2017)\citenamefont {Wilming}, \citenamefont {Goihl}, \citenamefont {Krumnow},\ and\ \citenamefont {Eisert}}]{Wil17}%
  \BibitemOpen
  \bibfield  {author} {\bibinfo {author} {\bibfnamefont {H.}~\bibnamefont {Wilming}}, \bibinfo {author} {\bibfnamefont {M.}~\bibnamefont {Goihl}}, \bibinfo {author} {\bibfnamefont {C.}~\bibnamefont {Krumnow}},\ and\ \bibinfo {author} {\bibfnamefont {J.}~\bibnamefont {Eisert}},\ }\href@noop {} {\bibinfo {title} {Towards local equilibration in closed interacting quantum many-body systems}} (\bibinfo {year} {2017}),\ \Eprint {https://arxiv.org/abs/1704.06291} {arXiv:1704.06291 [quant-ph]} \BibitemShut {NoStop}%
\bibitem [{\citenamefont {Deutsch}(2018)}]{Deu18}%
  \BibitemOpen
  \bibfield  {author} {\bibinfo {author} {\bibfnamefont {J.~M.}\ \bibnamefont {Deutsch}},\ }\bibfield  {title} {\bibinfo {title} {Eigenstate thermalization hypothesis},\ }\href {https://doi.org/10.1088/1361-6633/aac9f1} {\bibfield  {journal} {\bibinfo  {journal} {Reports on Progress in Physics}\ }\textbf {\bibinfo {volume} {81}},\ \bibinfo {pages} {082001} (\bibinfo {year} {2018})}\BibitemShut {NoStop}%
\bibitem [{\citenamefont {Goldstein}\ \emph {et~al.}(2013)\citenamefont {Goldstein}, \citenamefont {Hara},\ and\ \citenamefont {Tasaki}}]{Gol13}%
  \BibitemOpen
  \bibfield  {author} {\bibinfo {author} {\bibfnamefont {S.}~\bibnamefont {Goldstein}}, \bibinfo {author} {\bibfnamefont {T.}~\bibnamefont {Hara}},\ and\ \bibinfo {author} {\bibfnamefont {H.}~\bibnamefont {Tasaki}},\ }\href@noop {} {\bibinfo {title} {The second law of thermodynamics for pure quantum states}} (\bibinfo {year} {2013}),\ \Eprint {https://arxiv.org/abs/1303.6393} {arXiv:1303.6393 [cond-mat.stat-mech]} \BibitemShut {NoStop}%
\bibitem [{\citenamefont {Itoi}\ and\ \citenamefont {Amano}(2020)}]{Ito20}%
  \BibitemOpen
  \bibfield  {author} {\bibinfo {author} {\bibfnamefont {C.}~\bibnamefont {Itoi}}\ and\ \bibinfo {author} {\bibfnamefont {M.}~\bibnamefont {Amano}},\ }\bibfield  {title} {\bibinfo {title} {The second law of thermodynamics from concavity of energy eigenvalues},\ }\href {https://doi.org/10.7566/jpsj.89.104001} {\bibfield  {journal} {\bibinfo  {journal} {Journal of the Physical Society of Japan}\ }\textbf {\bibinfo {volume} {89}},\ \bibinfo {pages} {104001} (\bibinfo {year} {2020})}\BibitemShut {NoStop}%
\bibitem [{\citenamefont {Tasaki}(2000)}]{Tak00}%
  \BibitemOpen
  \bibfield  {author} {\bibinfo {author} {\bibfnamefont {H.}~\bibnamefont {Tasaki}},\ }\href@noop {} {\bibinfo {title} {The second law of thermodynamics as a theorem in quantum mechanics}} (\bibinfo {year} {2000}),\ \Eprint {https://arxiv.org/abs/cond-mat/0011321} {arXiv:cond-mat/0011321 [cond-mat.stat-mech]} \BibitemShut {NoStop}%
\bibitem [{\citenamefont {Ikeda}\ \emph {et~al.}(2015)\citenamefont {Ikeda}, \citenamefont {Sakumichi}, \citenamefont {Polkovnikov},\ and\ \citenamefont {Ueda}}]{Ike15}%
  \BibitemOpen
  \bibfield  {author} {\bibinfo {author} {\bibfnamefont {T.~N.}\ \bibnamefont {Ikeda}}, \bibinfo {author} {\bibfnamefont {N.}~\bibnamefont {Sakumichi}}, \bibinfo {author} {\bibfnamefont {A.}~\bibnamefont {Polkovnikov}},\ and\ \bibinfo {author} {\bibfnamefont {M.}~\bibnamefont {Ueda}},\ }\bibfield  {title} {\bibinfo {title} {The second law of thermodynamics under unitary evolution and external operations},\ }\href {https://doi.org/10.1016/j.aop.2015.01.003} {\bibfield  {journal} {\bibinfo  {journal} {Annals of Physics}\ }\textbf {\bibinfo {volume} {354}},\ \bibinfo {pages} {338–352} (\bibinfo {year} {2015})}\BibitemShut {NoStop}%
\bibitem [{\citenamefont {Kaneko}\ \emph {et~al.}(2019)\citenamefont {Kaneko}, \citenamefont {Iyoda},\ and\ \citenamefont {Sagawa}}]{Kan19}%
  \BibitemOpen
  \bibfield  {author} {\bibinfo {author} {\bibfnamefont {K.}~\bibnamefont {Kaneko}}, \bibinfo {author} {\bibfnamefont {E.}~\bibnamefont {Iyoda}},\ and\ \bibinfo {author} {\bibfnamefont {T.}~\bibnamefont {Sagawa}},\ }\bibfield  {title} {\bibinfo {title} {Work extraction from a single energy eigenstate},\ }\href {https://doi.org/10.1103/PhysRevE.99.032128} {\bibfield  {journal} {\bibinfo  {journal} {Phys. Rev. E}\ }\textbf {\bibinfo {volume} {99}},\ \bibinfo {pages} {032128} (\bibinfo {year} {2019})}\BibitemShut {NoStop}%
\bibitem [{\citenamefont {Polkovnikov}(2011)}]{Pol11}%
  \BibitemOpen
  \bibfield  {author} {\bibinfo {author} {\bibfnamefont {A.}~\bibnamefont {Polkovnikov}},\ }\bibfield  {title} {\bibinfo {title} {Microscopic diagonal entropy and its connection to basic thermodynamic relations},\ }\href {https://doi.org/https://doi.org/10.1016/j.aop.2010.08.004} {\bibfield  {journal} {\bibinfo  {journal} {Annals of Physics}\ }\textbf {\bibinfo {volume} {326}},\ \bibinfo {pages} {486} (\bibinfo {year} {2011})}\BibitemShut {NoStop}%
\bibitem [{\citenamefont {Schulz}\ \emph {et~al.}(2024)\citenamefont {Schulz}, \citenamefont {Willsch},\ and\ \citenamefont {Michielsen}}]{Sch24}%
  \BibitemOpen
  \bibfield  {author} {\bibinfo {author} {\bibfnamefont {S.}~\bibnamefont {Schulz}}, \bibinfo {author} {\bibfnamefont {D.}~\bibnamefont {Willsch}},\ and\ \bibinfo {author} {\bibfnamefont {K.}~\bibnamefont {Michielsen}},\ }\bibfield  {title} {\bibinfo {title} {Guided quantum walk},\ }\bibfield  {journal} {\bibinfo  {journal} {Physical Review Research}\ }\textbf {\bibinfo {volume} {6}},\ \href {https://doi.org/10.1103/physrevresearch.6.013312} {10.1103/physrevresearch.6.013312} (\bibinfo {year} {2024})\BibitemShut {NoStop}%
\bibitem [{\citenamefont {Ho}\ \emph {et~al.}(2023)\citenamefont {Ho}, \citenamefont {Mori}, \citenamefont {Abanin},\ and\ \citenamefont {Dalla~Torre}}]{Ho23}%
  \BibitemOpen
  \bibfield  {author} {\bibinfo {author} {\bibfnamefont {W.~W.}\ \bibnamefont {Ho}}, \bibinfo {author} {\bibfnamefont {T.}~\bibnamefont {Mori}}, \bibinfo {author} {\bibfnamefont {D.~A.}\ \bibnamefont {Abanin}},\ and\ \bibinfo {author} {\bibfnamefont {E.~G.}\ \bibnamefont {Dalla~Torre}},\ }\bibfield  {title} {\bibinfo {title} {Quantum and classical floquet prethermalization},\ }\href {https://doi.org/10.1016/j.aop.2023.169297} {\bibfield  {journal} {\bibinfo  {journal} {Annals of Physics}\ }\textbf {\bibinfo {volume} {454}},\ \bibinfo {pages} {169297} (\bibinfo {year} {2023})}\BibitemShut {NoStop}%
\bibitem [{\citenamefont {Huse}\ \emph {et~al.}(2013)\citenamefont {Huse}, \citenamefont {Nandkishore}, \citenamefont {Oganesyan}, \citenamefont {Pal},\ and\ \citenamefont {Sondhi}}]{Hus13}%
  \BibitemOpen
  \bibfield  {author} {\bibinfo {author} {\bibfnamefont {D.~A.}\ \bibnamefont {Huse}}, \bibinfo {author} {\bibfnamefont {R.}~\bibnamefont {Nandkishore}}, \bibinfo {author} {\bibfnamefont {V.}~\bibnamefont {Oganesyan}}, \bibinfo {author} {\bibfnamefont {A.}~\bibnamefont {Pal}},\ and\ \bibinfo {author} {\bibfnamefont {S.~L.}\ \bibnamefont {Sondhi}},\ }\bibfield  {title} {\bibinfo {title} {Localization-protected quantum order},\ }\href {https://doi.org/10.1103/PhysRevB.88.014206} {\bibfield  {journal} {\bibinfo  {journal} {Phys. Rev. B}\ }\textbf {\bibinfo {volume} {88}},\ \bibinfo {pages} {014206} (\bibinfo {year} {2013})}\BibitemShut {NoStop}%
\bibitem [{\citenamefont {Nandkishore}\ and\ \citenamefont {Huse}(2015)}]{Nan15}%
  \BibitemOpen
  \bibfield  {author} {\bibinfo {author} {\bibfnamefont {R.}~\bibnamefont {Nandkishore}}\ and\ \bibinfo {author} {\bibfnamefont {D.~A.}\ \bibnamefont {Huse}},\ }\bibfield  {title} {\bibinfo {title} {Many-body localization and thermalization in quantum statistical mechanics},\ }\href {https://doi.org/10.1146/annurev-conmatphys-031214-014726} {\bibfield  {journal} {\bibinfo  {journal} {Annual Review of Condensed Matter Physics}\ }\textbf {\bibinfo {volume} {6}},\ \bibinfo {pages} {15} (\bibinfo {year} {2015})},\ \Eprint {https://arxiv.org/abs/https://doi.org/10.1146/annurev-conmatphys-031214-014726} {https://doi.org/10.1146/annurev-conmatphys-031214-014726} \BibitemShut {NoStop}%
\bibitem [{\citenamefont {Altman}(2018)}]{Alt18}%
  \BibitemOpen
  \bibfield  {author} {\bibinfo {author} {\bibfnamefont {E.}~\bibnamefont {Altman}},\ }\bibfield  {title} {\bibinfo {title} {Many-body localization and quantum thermalization},\ }\href {https://doi.org/10.1038/s41567-018-0305-7} {\bibfield  {journal} {\bibinfo  {journal} {Nature Physics}\ }\textbf {\bibinfo {volume} {14}},\ \bibinfo {pages} {979} (\bibinfo {year} {2018})}\BibitemShut {NoStop}%
\bibitem [{\citenamefont {Deutsch}(1991)}]{Deu91}%
  \BibitemOpen
  \bibfield  {author} {\bibinfo {author} {\bibfnamefont {J.~M.}\ \bibnamefont {Deutsch}},\ }\bibfield  {title} {\bibinfo {title} {Quantum statistical mechanics in a closed system},\ }\href {https://doi.org/10.1103/PhysRevA.43.2046} {\bibfield  {journal} {\bibinfo  {journal} {Phys. Rev. A}\ }\textbf {\bibinfo {volume} {43}},\ \bibinfo {pages} {2046} (\bibinfo {year} {1991})}\BibitemShut {NoStop}%
\bibitem [{\citenamefont {Rigol}(2009)}]{Rig09}%
  \BibitemOpen
  \bibfield  {author} {\bibinfo {author} {\bibfnamefont {M.}~\bibnamefont {Rigol}},\ }\bibfield  {title} {\bibinfo {title} {Breakdown of thermalization in finite one-dimensional systems},\ }\href {https://doi.org/10.1103/PhysRevLett.103.100403} {\bibfield  {journal} {\bibinfo  {journal} {Phys. Rev. Lett.}\ }\textbf {\bibinfo {volume} {103}},\ \bibinfo {pages} {100403} (\bibinfo {year} {2009})}\BibitemShut {NoStop}%
\bibitem [{\citenamefont {Essler}\ and\ \citenamefont {Fagotti}(2016)}]{Ess16}%
  \BibitemOpen
  \bibfield  {author} {\bibinfo {author} {\bibfnamefont {F.~H.~L.}\ \bibnamefont {Essler}}\ and\ \bibinfo {author} {\bibfnamefont {M.}~\bibnamefont {Fagotti}},\ }\bibfield  {title} {\bibinfo {title} {Quench dynamics and relaxation in isolated integrable quantum spin chains},\ }\href {https://doi.org/10.1088/1742-5468/2016/06/064002} {\bibfield  {journal} {\bibinfo  {journal} {Journal of Statistical Mechanics: Theory and Experiment}\ }\textbf {\bibinfo {volume} {2016}},\ \bibinfo {pages} {064002} (\bibinfo {year} {2016})}\BibitemShut {NoStop}%
\bibitem [{\citenamefont {Brenes}\ \emph {et~al.}(2020)\citenamefont {Brenes}, \citenamefont {LeBlond}, \citenamefont {Goold},\ and\ \citenamefont {Rigol}}]{Bre20}%
  \BibitemOpen
  \bibfield  {author} {\bibinfo {author} {\bibfnamefont {M.}~\bibnamefont {Brenes}}, \bibinfo {author} {\bibfnamefont {T.}~\bibnamefont {LeBlond}}, \bibinfo {author} {\bibfnamefont {J.}~\bibnamefont {Goold}},\ and\ \bibinfo {author} {\bibfnamefont {M.}~\bibnamefont {Rigol}},\ }\bibfield  {title} {\bibinfo {title} {Eigenstate thermalization in a locally perturbed integrable system},\ }\href {https://doi.org/10.1103/PhysRevLett.125.070605} {\bibfield  {journal} {\bibinfo  {journal} {Phys. Rev. Lett.}\ }\textbf {\bibinfo {volume} {125}},\ \bibinfo {pages} {070605} (\bibinfo {year} {2020})}\BibitemShut {NoStop}%
\bibitem [{\citenamefont {Noh}(2021)}]{Noh21}%
  \BibitemOpen
  \bibfield  {author} {\bibinfo {author} {\bibfnamefont {J.~D.}\ \bibnamefont {Noh}},\ }\bibfield  {title} {\bibinfo {title} {Eigenstate thermalization hypothesis and eigenstate-to-eigenstate fluctuations},\ }\href {https://doi.org/10.1103/PhysRevE.103.012129} {\bibfield  {journal} {\bibinfo  {journal} {Phys. Rev. E}\ }\textbf {\bibinfo {volume} {103}},\ \bibinfo {pages} {012129} (\bibinfo {year} {2021})}\BibitemShut {NoStop}%
\bibitem [{\citenamefont {Garrison}\ and\ \citenamefont {Grover}(2018)}]{Gar18}%
  \BibitemOpen
  \bibfield  {author} {\bibinfo {author} {\bibfnamefont {J.~R.}\ \bibnamefont {Garrison}}\ and\ \bibinfo {author} {\bibfnamefont {T.}~\bibnamefont {Grover}},\ }\bibfield  {title} {\bibinfo {title} {Does a single eigenstate encode the full hamiltonian?},\ }\bibfield  {journal} {\bibinfo  {journal} {Physical Review X}\ }\textbf {\bibinfo {volume} {8}},\ \href {https://doi.org/10.1103/physrevx.8.021026} {10.1103/physrevx.8.021026} (\bibinfo {year} {2018})\BibitemShut {NoStop}%
\bibitem [{\citenamefont {Shiraishi}\ and\ \citenamefont {Matsumoto}(2021)}]{Shi21}%
  \BibitemOpen
  \bibfield  {author} {\bibinfo {author} {\bibfnamefont {N.}~\bibnamefont {Shiraishi}}\ and\ \bibinfo {author} {\bibfnamefont {K.}~\bibnamefont {Matsumoto}},\ }\bibfield  {title} {\bibinfo {title} {Undecidability in quantum thermalization},\ }\href {https://doi.org/10.1038/s41467-021-25053-0} {\bibfield  {journal} {\bibinfo  {journal} {Nature Communications}\ }\textbf {\bibinfo {volume} {12}},\ \bibinfo {pages} {5084} (\bibinfo {year} {2021})}\BibitemShut {NoStop}%
\bibitem [{\citenamefont {Srednicki}(1994)}]{Sre94}%
  \BibitemOpen
  \bibfield  {author} {\bibinfo {author} {\bibfnamefont {M.}~\bibnamefont {Srednicki}},\ }\bibfield  {title} {\bibinfo {title} {Chaos and quantum thermalization},\ }\href {https://doi.org/10.1103/PhysRevE.50.888} {\bibfield  {journal} {\bibinfo  {journal} {Phys. Rev. E}\ }\textbf {\bibinfo {volume} {50}},\ \bibinfo {pages} {888} (\bibinfo {year} {1994})}\BibitemShut {NoStop}%
\bibitem [{\citenamefont {Rigol}\ \emph {et~al.}(2008)\citenamefont {Rigol}, \citenamefont {Dunjko},\ and\ \citenamefont {Olshanii}}]{Rig08}%
  \BibitemOpen
  \bibfield  {author} {\bibinfo {author} {\bibfnamefont {M.}~\bibnamefont {Rigol}}, \bibinfo {author} {\bibfnamefont {V.}~\bibnamefont {Dunjko}},\ and\ \bibinfo {author} {\bibfnamefont {M.}~\bibnamefont {Olshanii}},\ }\bibfield  {title} {\bibinfo {title} {Thermalization and its mechanism for generic isolated quantum systems},\ }\href {https://doi.org/10.1038/nature06838} {\bibfield  {journal} {\bibinfo  {journal} {Nature}\ }\textbf {\bibinfo {volume} {452}},\ \bibinfo {pages} {854} (\bibinfo {year} {2008})}\BibitemShut {NoStop}%
\bibitem [{\citenamefont {Biroli}\ \emph {et~al.}(2010)\citenamefont {Biroli}, \citenamefont {Kollath},\ and\ \citenamefont {L\"auchli}}]{Bir10}%
  \BibitemOpen
  \bibfield  {author} {\bibinfo {author} {\bibfnamefont {G.}~\bibnamefont {Biroli}}, \bibinfo {author} {\bibfnamefont {C.}~\bibnamefont {Kollath}},\ and\ \bibinfo {author} {\bibfnamefont {A.~M.}\ \bibnamefont {L\"auchli}},\ }\bibfield  {title} {\bibinfo {title} {Effect of rare fluctuations on the thermalization of isolated quantum systems},\ }\href {https://doi.org/10.1103/PhysRevLett.105.250401} {\bibfield  {journal} {\bibinfo  {journal} {Phys. Rev. Lett.}\ }\textbf {\bibinfo {volume} {105}},\ \bibinfo {pages} {250401} (\bibinfo {year} {2010})}\BibitemShut {NoStop}%
\bibitem [{\citenamefont {Santos}\ and\ \citenamefont {Rigol}(2010)}]{San10}%
  \BibitemOpen
  \bibfield  {author} {\bibinfo {author} {\bibfnamefont {L.~F.}\ \bibnamefont {Santos}}\ and\ \bibinfo {author} {\bibfnamefont {M.}~\bibnamefont {Rigol}},\ }\bibfield  {title} {\bibinfo {title} {Onset of quantum chaos in one-dimensional bosonic and fermionic systems and its relation to thermalization},\ }\href {https://doi.org/10.1103/PhysRevE.81.036206} {\bibfield  {journal} {\bibinfo  {journal} {Phys. Rev. E}\ }\textbf {\bibinfo {volume} {81}},\ \bibinfo {pages} {036206} (\bibinfo {year} {2010})}\BibitemShut {NoStop}%
\bibitem [{\citenamefont {Steinigeweg}\ \emph {et~al.}(2013)\citenamefont {Steinigeweg}, \citenamefont {Herbrych},\ and\ \citenamefont {Prelov{\ifmmode \check{s}\else \v{s}\fi{}}ek}}]{Ste13}%
  \BibitemOpen
  \bibfield  {author} {\bibinfo {author} {\bibfnamefont {R.}~\bibnamefont {Steinigeweg}}, \bibinfo {author} {\bibfnamefont {J.}~\bibnamefont {Herbrych}},\ and\ \bibinfo {author} {\bibfnamefont {P.}~\bibnamefont {Prelov{\ifmmode \check{s}\else \v{s}\fi{}}ek}},\ }\bibfield  {title} {\bibinfo {title} {Eigenstate thermalization within isolated spin-chain systems},\ }\href {https://doi.org/10.1103/PhysRevE.87.012118} {\bibfield  {journal} {\bibinfo  {journal} {Phys. Rev. E}\ }\textbf {\bibinfo {volume} {87}},\ \bibinfo {pages} {012118} (\bibinfo {year} {2013})}\BibitemShut {NoStop}%
\bibitem [{\citenamefont {Kim}\ \emph {et~al.}(2014)\citenamefont {Kim}, \citenamefont {Ikeda},\ and\ \citenamefont {Huse}}]{Kim14}%
  \BibitemOpen
  \bibfield  {author} {\bibinfo {author} {\bibfnamefont {H.}~\bibnamefont {Kim}}, \bibinfo {author} {\bibfnamefont {T.~N.}\ \bibnamefont {Ikeda}},\ and\ \bibinfo {author} {\bibfnamefont {D.~A.}\ \bibnamefont {Huse}},\ }\bibfield  {title} {\bibinfo {title} {Testing whether all eigenstates obey the eigenstate thermalization hypothesis},\ }\href {https://doi.org/10.1103/PhysRevE.90.052105} {\bibfield  {journal} {\bibinfo  {journal} {Phys. Rev. E}\ }\textbf {\bibinfo {volume} {90}},\ \bibinfo {pages} {052105} (\bibinfo {year} {2014})}\BibitemShut {NoStop}%
\bibitem [{\citenamefont {Steinigeweg}\ \emph {et~al.}(2014)\citenamefont {Steinigeweg}, \citenamefont {Khodja}, \citenamefont {Niemeyer}, \citenamefont {Gogolin},\ and\ \citenamefont {Gemmer}}]{Ste14}%
  \BibitemOpen
  \bibfield  {author} {\bibinfo {author} {\bibfnamefont {R.}~\bibnamefont {Steinigeweg}}, \bibinfo {author} {\bibfnamefont {A.}~\bibnamefont {Khodja}}, \bibinfo {author} {\bibfnamefont {H.}~\bibnamefont {Niemeyer}}, \bibinfo {author} {\bibfnamefont {C.}~\bibnamefont {Gogolin}},\ and\ \bibinfo {author} {\bibfnamefont {J.}~\bibnamefont {Gemmer}},\ }\bibfield  {title} {\bibinfo {title} {Pushing the limits of the eigenstate thermalization hypothesis towards mesoscopic quantum systems},\ }\href {https://doi.org/10.1103/PhysRevLett.112.130403} {\bibfield  {journal} {\bibinfo  {journal} {Phys. Rev. Lett.}\ }\textbf {\bibinfo {volume} {112}},\ \bibinfo {pages} {130403} (\bibinfo {year} {2014})}\BibitemShut {NoStop}%
\bibitem [{\citenamefont {Fratus}\ and\ \citenamefont {Srednicki}(2015)}]{Fra15}%
  \BibitemOpen
  \bibfield  {author} {\bibinfo {author} {\bibfnamefont {K.~R.}\ \bibnamefont {Fratus}}\ and\ \bibinfo {author} {\bibfnamefont {M.}~\bibnamefont {Srednicki}},\ }\bibfield  {title} {\bibinfo {title} {Eigenstate thermalization in systems with spontaneously broken symmetry},\ }\href {https://doi.org/10.1103/PhysRevE.92.040103} {\bibfield  {journal} {\bibinfo  {journal} {Phys. Rev. E}\ }\textbf {\bibinfo {volume} {92}},\ \bibinfo {pages} {040103} (\bibinfo {year} {2015})}\BibitemShut {NoStop}%
\bibitem [{\citenamefont {Khodja}\ \emph {et~al.}(2015)\citenamefont {Khodja}, \citenamefont {Steinigeweg},\ and\ \citenamefont {Gemmer}}]{Kho15}%
  \BibitemOpen
  \bibfield  {author} {\bibinfo {author} {\bibfnamefont {A.}~\bibnamefont {Khodja}}, \bibinfo {author} {\bibfnamefont {R.}~\bibnamefont {Steinigeweg}},\ and\ \bibinfo {author} {\bibfnamefont {J.}~\bibnamefont {Gemmer}},\ }\bibfield  {title} {\bibinfo {title} {Relevance of the eigenstate thermalization hypothesis for thermal relaxation},\ }\href {https://doi.org/10.1103/PhysRevE.91.012120} {\bibfield  {journal} {\bibinfo  {journal} {Phys. Rev. E}\ }\textbf {\bibinfo {volume} {91}},\ \bibinfo {pages} {012120} (\bibinfo {year} {2015})}\BibitemShut {NoStop}%
\bibitem [{\citenamefont {Mondaini}\ and\ \citenamefont {Rigol}(2017)}]{Mon17}%
  \BibitemOpen
  \bibfield  {author} {\bibinfo {author} {\bibfnamefont {R.}~\bibnamefont {Mondaini}}\ and\ \bibinfo {author} {\bibfnamefont {M.}~\bibnamefont {Rigol}},\ }\bibfield  {title} {\bibinfo {title} {Eigenstate thermalization in the two-dimensional transverse field ising model. ii. off-diagonal matrix elements of observables},\ }\href {https://doi.org/10.1103/PhysRevE.96.012157} {\bibfield  {journal} {\bibinfo  {journal} {Phys. Rev. E}\ }\textbf {\bibinfo {volume} {96}},\ \bibinfo {pages} {012157} (\bibinfo {year} {2017})}\BibitemShut {NoStop}%
\bibitem [{\citenamefont {Yoshizawa}\ \emph {et~al.}(2018)\citenamefont {Yoshizawa}, \citenamefont {Iyoda},\ and\ \citenamefont {Sagawa}}]{Yos18}%
  \BibitemOpen
  \bibfield  {author} {\bibinfo {author} {\bibfnamefont {T.}~\bibnamefont {Yoshizawa}}, \bibinfo {author} {\bibfnamefont {E.}~\bibnamefont {Iyoda}},\ and\ \bibinfo {author} {\bibfnamefont {T.}~\bibnamefont {Sagawa}},\ }\bibfield  {title} {\bibinfo {title} {Numerical large deviation analysis of the eigenstate thermalization hypothesis},\ }\href {https://doi.org/10.1103/PhysRevLett.120.200604} {\bibfield  {journal} {\bibinfo  {journal} {Phys. Rev. Lett.}\ }\textbf {\bibinfo {volume} {120}},\ \bibinfo {pages} {200604} (\bibinfo {year} {2018})}\BibitemShut {NoStop}%
\bibitem [{\citenamefont {Jansen}\ \emph {et~al.}(2019)\citenamefont {Jansen}, \citenamefont {Stolpp}, \citenamefont {Vidmar},\ and\ \citenamefont {Heidrich-Meisner}}]{Jan19}%
  \BibitemOpen
  \bibfield  {author} {\bibinfo {author} {\bibfnamefont {D.}~\bibnamefont {Jansen}}, \bibinfo {author} {\bibfnamefont {J.}~\bibnamefont {Stolpp}}, \bibinfo {author} {\bibfnamefont {L.}~\bibnamefont {Vidmar}},\ and\ \bibinfo {author} {\bibfnamefont {F.}~\bibnamefont {Heidrich-Meisner}},\ }\bibfield  {title} {\bibinfo {title} {Eigenstate thermalization and quantum chaos in the holstein polaron model},\ }\href {https://doi.org/10.1103/PhysRevB.99.155130} {\bibfield  {journal} {\bibinfo  {journal} {Phys. Rev. B}\ }\textbf {\bibinfo {volume} {99}},\ \bibinfo {pages} {155130} (\bibinfo {year} {2019})}\BibitemShut {NoStop}%
\bibitem [{\citenamefont {Trotzky}\ \emph {et~al.}(2012)\citenamefont {Trotzky}, \citenamefont {Chen}, \citenamefont {Flesch}, \citenamefont {McCulloch}, \citenamefont {Schollwöck}, \citenamefont {Eisert},\ and\ \citenamefont {Bloch}}]{Tro12}%
  \BibitemOpen
  \bibfield  {author} {\bibinfo {author} {\bibfnamefont {S.}~\bibnamefont {Trotzky}}, \bibinfo {author} {\bibfnamefont {Y.-A.}\ \bibnamefont {Chen}}, \bibinfo {author} {\bibfnamefont {A.}~\bibnamefont {Flesch}}, \bibinfo {author} {\bibfnamefont {I.~P.}\ \bibnamefont {McCulloch}}, \bibinfo {author} {\bibfnamefont {U.}~\bibnamefont {Schollwöck}}, \bibinfo {author} {\bibfnamefont {J.}~\bibnamefont {Eisert}},\ and\ \bibinfo {author} {\bibfnamefont {I.}~\bibnamefont {Bloch}},\ }\bibfield  {title} {\bibinfo {title} {Probing the relaxation towards equilibrium in an isolated strongly correlated one-dimensional bose~gas},\ }\href {https://doi.org/10.1038/nphys2232} {\bibfield  {journal} {\bibinfo  {journal} {Nature Physics}\ }\textbf {\bibinfo {volume} {8}},\ \bibinfo {pages} {325} (\bibinfo {year} {2012})}\BibitemShut {NoStop}%
\bibitem [{\citenamefont {Clos}\ \emph {et~al.}(2016)\citenamefont {Clos}, \citenamefont {Porras}, \citenamefont {Warring},\ and\ \citenamefont {Schaetz}}]{Clo16}%
  \BibitemOpen
  \bibfield  {author} {\bibinfo {author} {\bibfnamefont {G.}~\bibnamefont {Clos}}, \bibinfo {author} {\bibfnamefont {D.}~\bibnamefont {Porras}}, \bibinfo {author} {\bibfnamefont {U.}~\bibnamefont {Warring}},\ and\ \bibinfo {author} {\bibfnamefont {T.}~\bibnamefont {Schaetz}},\ }\bibfield  {title} {\bibinfo {title} {Time-resolved observation of thermalization in an isolated quantum system},\ }\href {https://doi.org/10.1103/PhysRevLett.117.170401} {\bibfield  {journal} {\bibinfo  {journal} {Phys. Rev. Lett.}\ }\textbf {\bibinfo {volume} {117}},\ \bibinfo {pages} {170401} (\bibinfo {year} {2016})}\BibitemShut {NoStop}%
\bibitem [{\citenamefont {Kaufman}\ \emph {et~al.}(2016)\citenamefont {Kaufman}, \citenamefont {Tai}, \citenamefont {Lukin}, \citenamefont {Rispoli}, \citenamefont {Schittko}, \citenamefont {Preiss},\ and\ \citenamefont {Greiner}}]{Kau16}%
  \BibitemOpen
  \bibfield  {author} {\bibinfo {author} {\bibfnamefont {A.~M.}\ \bibnamefont {Kaufman}}, \bibinfo {author} {\bibfnamefont {M.~E.}\ \bibnamefont {Tai}}, \bibinfo {author} {\bibfnamefont {A.}~\bibnamefont {Lukin}}, \bibinfo {author} {\bibfnamefont {M.}~\bibnamefont {Rispoli}}, \bibinfo {author} {\bibfnamefont {R.}~\bibnamefont {Schittko}}, \bibinfo {author} {\bibfnamefont {P.~M.}\ \bibnamefont {Preiss}},\ and\ \bibinfo {author} {\bibfnamefont {M.}~\bibnamefont {Greiner}},\ }\bibfield  {title} {\bibinfo {title} {Quantum thermalization through entanglement in an isolated many-body system},\ }\href {https://doi.org/10.1126/science.aaf6725} {\bibfield  {journal} {\bibinfo  {journal} {Science}\ }\textbf {\bibinfo {volume} {353}},\ \bibinfo {pages} {794} (\bibinfo {year} {2016})}\BibitemShut {NoStop}%
\bibitem [{\citenamefont {Kucsko}\ \emph {et~al.}(2018)\citenamefont {Kucsko}, \citenamefont {Choi}, \citenamefont {Choi}, \citenamefont {Maurer}, \citenamefont {Zhou}, \citenamefont {Landig}, \citenamefont {Sumiya}, \citenamefont {Onoda}, \citenamefont {Isoya}, \citenamefont {Jelezko}, \citenamefont {Demler}, \citenamefont {Yao},\ and\ \citenamefont {Lukin}}]{Kuc18}%
  \BibitemOpen
  \bibfield  {author} {\bibinfo {author} {\bibfnamefont {G.}~\bibnamefont {Kucsko}}, \bibinfo {author} {\bibfnamefont {S.}~\bibnamefont {Choi}}, \bibinfo {author} {\bibfnamefont {J.}~\bibnamefont {Choi}}, \bibinfo {author} {\bibfnamefont {P.~C.}\ \bibnamefont {Maurer}}, \bibinfo {author} {\bibfnamefont {H.}~\bibnamefont {Zhou}}, \bibinfo {author} {\bibfnamefont {R.}~\bibnamefont {Landig}}, \bibinfo {author} {\bibfnamefont {H.}~\bibnamefont {Sumiya}}, \bibinfo {author} {\bibfnamefont {S.}~\bibnamefont {Onoda}}, \bibinfo {author} {\bibfnamefont {J.}~\bibnamefont {Isoya}}, \bibinfo {author} {\bibfnamefont {F.}~\bibnamefont {Jelezko}}, \bibinfo {author} {\bibfnamefont {E.}~\bibnamefont {Demler}}, \bibinfo {author} {\bibfnamefont {N.~Y.}\ \bibnamefont {Yao}},\ and\ \bibinfo {author} {\bibfnamefont {M.~D.}\ \bibnamefont {Lukin}},\ }\bibfield  {title} {\bibinfo {title} {Critical thermalization of a disordered dipolar spin system in diamond},\ }\href {https://doi.org/10.1103/PhysRevLett.121.023601} {\bibfield
  {journal} {\bibinfo  {journal} {Phys. Rev. Lett.}\ }\textbf {\bibinfo {volume} {121}},\ \bibinfo {pages} {023601} (\bibinfo {year} {2018})}\BibitemShut {NoStop}%
\bibitem [{\citenamefont {Tang}\ \emph {et~al.}(2018)\citenamefont {Tang}, \citenamefont {Kao}, \citenamefont {Li}, \citenamefont {Seo}, \citenamefont {Mallayya}, \citenamefont {Rigol}, \citenamefont {Gopalakrishnan},\ and\ \citenamefont {Lev}}]{Tan18}%
  \BibitemOpen
  \bibfield  {author} {\bibinfo {author} {\bibfnamefont {Y.}~\bibnamefont {Tang}}, \bibinfo {author} {\bibfnamefont {W.}~\bibnamefont {Kao}}, \bibinfo {author} {\bibfnamefont {K.-Y.}\ \bibnamefont {Li}}, \bibinfo {author} {\bibfnamefont {S.}~\bibnamefont {Seo}}, \bibinfo {author} {\bibfnamefont {K.}~\bibnamefont {Mallayya}}, \bibinfo {author} {\bibfnamefont {M.}~\bibnamefont {Rigol}}, \bibinfo {author} {\bibfnamefont {S.}~\bibnamefont {Gopalakrishnan}},\ and\ \bibinfo {author} {\bibfnamefont {B.~L.}\ \bibnamefont {Lev}},\ }\bibfield  {title} {\bibinfo {title} {Thermalization near integrability in a dipolar quantum newton's cradle},\ }\href {https://doi.org/10.1103/PhysRevX.8.021030} {\bibfield  {journal} {\bibinfo  {journal} {Phys. Rev. X}\ }\textbf {\bibinfo {volume} {8}},\ \bibinfo {pages} {021030} (\bibinfo {year} {2018})}\BibitemShut {NoStop}%
\bibitem [{\citenamefont {Imparato}\ \emph {et~al.}(2024)\citenamefont {Imparato}, \citenamefont {Chancellor},\ and\ \citenamefont {De~Chiara}}]{Imp24}%
  \BibitemOpen
  \bibfield  {author} {\bibinfo {author} {\bibfnamefont {A.}~\bibnamefont {Imparato}}, \bibinfo {author} {\bibfnamefont {N.}~\bibnamefont {Chancellor}},\ and\ \bibinfo {author} {\bibfnamefont {G.}~\bibnamefont {De~Chiara}},\ }\bibfield  {title} {\bibinfo {title} {A thermodynamic approach to optimization in complex quantum systems},\ }\href {https://doi.org/10.1088/2058-9565/ad26b3} {\bibfield  {journal} {\bibinfo  {journal} {Quantum Science and Technology}\ }\textbf {\bibinfo {volume} {9}},\ \bibinfo {pages} {025011} (\bibinfo {year} {2024})}\BibitemShut {NoStop}%
\bibitem [{\citenamefont {Carlen}(2010)}]{Car10}%
  \BibitemOpen
  \bibfield  {author} {\bibinfo {author} {\bibfnamefont {E.}~\bibnamefont {Carlen}},\ }\bibfield  {title} {\bibinfo {title} {Trace inequalities and quantum entropy: an introductory course},\ }\href@noop {} {\bibfield  {journal} {\bibinfo  {journal} {Entropy and the Quantum}\ }\textbf {\bibinfo {volume} {529}},\ \bibinfo {pages} {73} (\bibinfo {year} {2010})}\BibitemShut {NoStop}%
\bibitem [{\citenamefont {Schiffer}\ \emph {et~al.}(2024)\citenamefont {Schiffer}, \citenamefont {Rubio}, \citenamefont {Trivedi},\ and\ \citenamefont {Cirac}}]{sch24b}%
  \BibitemOpen
  \bibfield  {author} {\bibinfo {author} {\bibfnamefont {B.~F.}\ \bibnamefont {Schiffer}}, \bibinfo {author} {\bibfnamefont {A.~F.}\ \bibnamefont {Rubio}}, \bibinfo {author} {\bibfnamefont {R.}~\bibnamefont {Trivedi}},\ and\ \bibinfo {author} {\bibfnamefont {J.~I.}\ \bibnamefont {Cirac}},\ }\href@noop {} {\bibinfo {title} {The quantum adiabatic algorithm suppresses the proliferation of errors}} (\bibinfo {year} {2024}),\ \Eprint {https://arxiv.org/abs/2404.15397} {arXiv:2404.15397 [quant-ph]} \BibitemShut {NoStop}%
\bibitem [{\citenamefont {Trivedi}\ \emph {et~al.}(2023)\citenamefont {Trivedi}, \citenamefont {Rubio},\ and\ \citenamefont {Cirac}}]{tri23}%
  \BibitemOpen
  \bibfield  {author} {\bibinfo {author} {\bibfnamefont {R.}~\bibnamefont {Trivedi}}, \bibinfo {author} {\bibfnamefont {A.~F.}\ \bibnamefont {Rubio}},\ and\ \bibinfo {author} {\bibfnamefont {J.~I.}\ \bibnamefont {Cirac}},\ }\href@noop {} {\bibinfo {title} {Quantum advantage and stability to errors in analogue quantum simulators}} (\bibinfo {year} {2023}),\ \Eprint {https://arxiv.org/abs/2212.04924} {arXiv:2212.04924 [quant-ph]} \BibitemShut {NoStop}%
\bibitem [{\citenamefont {Childs}\ \emph {et~al.}(2001)\citenamefont {Childs}, \citenamefont {Farhi},\ and\ \citenamefont {Preskill}}]{Chi01}%
  \BibitemOpen
  \bibfield  {author} {\bibinfo {author} {\bibfnamefont {A.~M.}\ \bibnamefont {Childs}}, \bibinfo {author} {\bibfnamefont {E.}~\bibnamefont {Farhi}},\ and\ \bibinfo {author} {\bibfnamefont {J.}~\bibnamefont {Preskill}},\ }\bibfield  {title} {\bibinfo {title} {Robustness of adiabatic quantum computation},\ }\bibfield  {journal} {\bibinfo  {journal} {Physical Review A}\ }\textbf {\bibinfo {volume} {65}},\ \href {https://doi.org/10.1103/physreva.65.012322} {10.1103/physreva.65.012322} (\bibinfo {year} {2001})\BibitemShut {NoStop}%
\bibitem [{\citenamefont {Marshall}\ \emph {et~al.}(2010)\citenamefont {Marshall}, \citenamefont {Olkin},\ and\ \citenamefont {C.~Arnold}}]{Mar10}%
  \BibitemOpen
  \bibfield  {author} {\bibinfo {author} {\bibfnamefont {A.~W.}\ \bibnamefont {Marshall}}, \bibinfo {author} {\bibfnamefont {I.}~\bibnamefont {Olkin}},\ and\ \bibinfo {author} {\bibfnamefont {B.}~\bibnamefont {C.~Arnold}},\ }\href {https://doi.org/https://doi.org/10.1007/978-0-387-68276-1} {\emph {\bibinfo {title} {Inequalities: Theory of Majorization and Its Applications}}}\ (\bibinfo  {publisher} {Springer New York, NY},\ \bibinfo {year} {2010})\BibitemShut {NoStop}%
\bibitem [{\citenamefont {Ohkuwa}\ \emph {et~al.}(2018)\citenamefont {Ohkuwa}, \citenamefont {Nishimori},\ and\ \citenamefont {Lidar}}]{Ohk18}%
  \BibitemOpen
  \bibfield  {author} {\bibinfo {author} {\bibfnamefont {M.}~\bibnamefont {Ohkuwa}}, \bibinfo {author} {\bibfnamefont {H.}~\bibnamefont {Nishimori}},\ and\ \bibinfo {author} {\bibfnamefont {D.~A.}\ \bibnamefont {Lidar}},\ }\bibfield  {title} {\bibinfo {title} {Reverse annealing for the fully connected $p$-spin model},\ }\href {https://doi.org/10.1103/PhysRevA.98.022314} {\bibfield  {journal} {\bibinfo  {journal} {Phys. Rev. A}\ }\textbf {\bibinfo {volume} {98}},\ \bibinfo {pages} {022314} (\bibinfo {year} {2018})}\BibitemShut {NoStop}%
\bibitem [{\citenamefont {Yamashiro}\ \emph {et~al.}(2019)\citenamefont {Yamashiro}, \citenamefont {Ohkuwa}, \citenamefont {Nishimori},\ and\ \citenamefont {Lidar}}]{Yam19}%
  \BibitemOpen
  \bibfield  {author} {\bibinfo {author} {\bibfnamefont {Y.}~\bibnamefont {Yamashiro}}, \bibinfo {author} {\bibfnamefont {M.}~\bibnamefont {Ohkuwa}}, \bibinfo {author} {\bibfnamefont {H.}~\bibnamefont {Nishimori}},\ and\ \bibinfo {author} {\bibfnamefont {D.~A.}\ \bibnamefont {Lidar}},\ }\bibfield  {title} {\bibinfo {title} {Dynamics of reverse annealing for the fully connected $p$-spin model},\ }\href {https://doi.org/10.1103/PhysRevA.100.052321} {\bibfield  {journal} {\bibinfo  {journal} {Phys. Rev. A}\ }\textbf {\bibinfo {volume} {100}},\ \bibinfo {pages} {052321} (\bibinfo {year} {2019})}\BibitemShut {NoStop}%
\bibitem [{\citenamefont {Prosen}(1999)}]{Pro99}%
  \BibitemOpen
  \bibfield  {author} {\bibinfo {author} {\bibfnamefont {T.}~\bibnamefont {Prosen}},\ }\bibfield  {title} {\bibinfo {title} {Ergodic properties of a generic nonintegrable quantum many-body system in the thermodynamic limit},\ }\href {https://doi.org/10.1103/PhysRevE.60.3949} {\bibfield  {journal} {\bibinfo  {journal} {Phys. Rev. E}\ }\textbf {\bibinfo {volume} {60}},\ \bibinfo {pages} {3949} (\bibinfo {year} {1999})}\BibitemShut {NoStop}%
\bibitem [{\citenamefont {Beugeling}\ \emph {et~al.}(2015)\citenamefont {Beugeling}, \citenamefont {Moessner},\ and\ \citenamefont {Haque}}]{Beu15}%
  \BibitemOpen
  \bibfield  {author} {\bibinfo {author} {\bibfnamefont {W.}~\bibnamefont {Beugeling}}, \bibinfo {author} {\bibfnamefont {R.}~\bibnamefont {Moessner}},\ and\ \bibinfo {author} {\bibfnamefont {M.}~\bibnamefont {Haque}},\ }\bibfield  {title} {\bibinfo {title} {Off-diagonal matrix elements of local operators in many-body quantum systems},\ }\href {https://doi.org/10.1103/PhysRevE.91.012144} {\bibfield  {journal} {\bibinfo  {journal} {Phys. Rev. E}\ }\textbf {\bibinfo {volume} {91}},\ \bibinfo {pages} {012144} (\bibinfo {year} {2015})}\BibitemShut {NoStop}%
\bibitem [{\citenamefont {Hastings}(1970)}]{Has70}%
  \BibitemOpen
  \bibfield  {author} {\bibinfo {author} {\bibfnamefont {W.~K.}\ \bibnamefont {Hastings}},\ }\bibfield  {title} {\bibinfo {title} {{Monte Carlo sampling methods using Markov chains and their applications}},\ }\href {https://doi.org/10.1093/biomet/57.1.97} {\bibfield  {journal} {\bibinfo  {journal} {Biometrika}\ }\textbf {\bibinfo {volume} {57}},\ \bibinfo {pages} {97} (\bibinfo {year} {1970})},\ \Eprint {https://arxiv.org/abs/https://academic.oup.com/biomet/article-pdf/57/1/97/23940249/57-1-97.pdf} {https://academic.oup.com/biomet/article-pdf/57/1/97/23940249/57-1-97.pdf} \BibitemShut {NoStop}%
\bibitem [{\citenamefont {Chancellor}(2023)}]{Cha23}%
  \BibitemOpen
  \bibfield  {author} {\bibinfo {author} {\bibfnamefont {N.}~\bibnamefont {Chancellor}},\ }\bibfield  {title} {\bibinfo {title} {Modernizing quantum annealing ii: genetic algorithms with the inference primitive formalism},\ }\href {https://doi.org/10.1007/s11047-022-09905-2} {\bibfield  {journal} {\bibinfo  {journal} {Natural Computing}\ }\textbf {\bibinfo {volume} {22}},\ \bibinfo {pages} {737} (\bibinfo {year} {2023})}\BibitemShut {NoStop}%
\end{thebibliography}%

\clearpage
\appendix
\section{A simple derivation of the passivity of Gibbs states}
\label{app:gibbs_passive}
In this section, we provide a very simple derivation of the passivity of Gibbs states. The initial state $\rho_0$ is a Gibbs state at inverse temperature $\beta$ with Hamiltonian $H$. The state after unitary evolution is denoted by $\rho(t)$. The Gibbs state minimises the free energy: 
\begin{equation}
    \Tr{H \rho_0}-\frac{1}{\beta} S(\rho_0)\leq \Tr{H \rho(t)}-\frac{1}{\beta} S(\rho(t))
\end{equation}
where $S(\cdot)$ is the von Neumann entropy. Rearranging the above gives:
\begin{equation}
     S(\rho(t))- S(\rho_0) \leq \beta \left(\Tr{H \rho(t)}-\Tr{H \rho_0}\right).
\end{equation}
Since the von Neumann entropy is conserved under unitary evolution,
\begin{equation}
     0 \leq \beta \left(\Tr{H \rho(t)}-\Tr{H \rho_0}\right),
\end{equation}
provided the temperature is positive
\begin{equation}
     \Tr{H \rho_0} \leq \Tr{H \rho(t)}.
\end{equation}
Hence for any unitary (including cyclic unitaries) the energy can only increase for a Gibbs state; meaning Gibbs states satisfy Planck's principle.

\section{Pure State Thermal Quantum Annealing}
\label{sec:pstqa}
In this section, we consider the continuum limit of an MSQW, such that the rate at which $\Gamma(t)$ changes is sufficiently slow compared to the dephasing timescale of the system. By applying the ETH (Assumption \ref{ass:eth}), we then model the system as a Gibbs state at all times. We also allow the coefficient in front of the driver Hamiltonian to change too, under the same assumptions. The Hamiltonian is given by:
\begin{equation}
    \label{eq:hamtqa}
    H_{TQA}=A(t) H_d +B(t) H_p,
\end{equation}
where $A(t)$ and $B(t)$ are changed slowly compared to the dephasing timescale. Intuition into this timescale can be found in \cite{Wil17,Oli18, Wil18}. We refer to this limit as pure state thermal quantum annealing (PSTQA). Since the system is isolated, we can write:
\begin{align}
   \frac{\dd \langle H_{TQA}(t) \rangle }{\dd t}&=\langle \frac{\partial H_{TQA}}{\partial t} \rangle\\
   \label{eq:dHdqa}
   &=\dot{A}(t)\langle H_{d} \rangle+\dot{B}(t)\langle H_{p} \rangle,
\end{align}
where the dot denotes a time derivative. Denoting the time-dependent partition function of the system as $\mathcal{Z}(t)$, we have:
\begin{align}
    \label{eq:hdqa_th}
    \langle H_{TQA}(t) \rangle=&-\frac{\partial \ln \mathcal{Z}(t)}{\partial \beta}\\
    \label{eq:hd_th}
    \langle H_{d}(t) \rangle=&-\frac{1}{\beta(t)} \frac{\partial \ln \mathcal{Z}(t)}{\partial A}\\
    \label{eq:hp_th}
    \langle H_{p}(t) \rangle=&-\frac{1}{\beta(t)} \frac{\partial \ln \mathcal{Z}(t)}{\partial B},
\end{align}
where $\beta$ is the inverse temperature. The rest of this section investigates the consequences of these equations.

\subsection{Pure state thermal quantum annealing is path-independent}

Returning to Eq.\ \ref{eq:dHdqa}, the equation can be rewritten into normalised time with the substitution $s=t/t_f$,
\begin{equation}
    \label{eq:dhdqa_s}
    \frac{\dd \langle H_{TQA}(s) \rangle }{\dd s}=\langle H_{d} \rangle\frac{\dd A(s)}{\dd s}+\langle H_{p} \rangle \frac{\dd B(s)}{\dd s}.
\end{equation}
The above equation depends only on the normalised time and the expectation values depend only on $A,B$ and $\beta$ and not their derivatives. This implies $\langle H_p \rangle$ in the thermal model is independent of scaling by $t_f$, although obviously too small a $t_f$ will cause the thermal predictions to break down. Further rewriting Eq.\ \ref{eq:dhdqa_s} gives:
\begin{equation}
    \label{eq:dhdqa_path}
    \langle H_{TQA}(1) \rangle =\int_{A(0),B(0)}^{A(1),B(1)}
    \begin{pmatrix}
        \langle H_{d} \rangle\\
        \langle H_{p} \rangle
    \end{pmatrix}
    \cdot
    \begin{pmatrix}
         \dd A \\
         \dd B 
    \end{pmatrix},
\end{equation}
hence we can view Eq.\ \ref{eq:dhdqa_s} as a path integral under the ``force''
\begin{equation}
    \vec{D}=
    -\begin{pmatrix}
        \langle H_{d} \rangle\\
        \langle H_{p} \rangle
    \end{pmatrix}=-\vec{\nabla} F,
\end{equation}
where $F$ is the Helmholtz free energy, $F=-\ln \left(\mathcal{Z}\right)/\beta $ and $\vec{\nabla}=(\partial_A, \partial_B)^T$. Since each of $\langle H_{d} \rangle$ and $\langle H_{p} \rangle$ depend only on $A$ and $B$ and not their derivatives and the force can be written as the gradient of a scalar field, $\langle H_{TQA} \rangle$ is path independent. It follows that $\langle H_p \rangle$ will also be path-independent.  
\begin{figure*}
    \centering
    \begin{subfigure}[l]{0.48\textwidth}
        \centering
        \includegraphics[width=\textwidth]{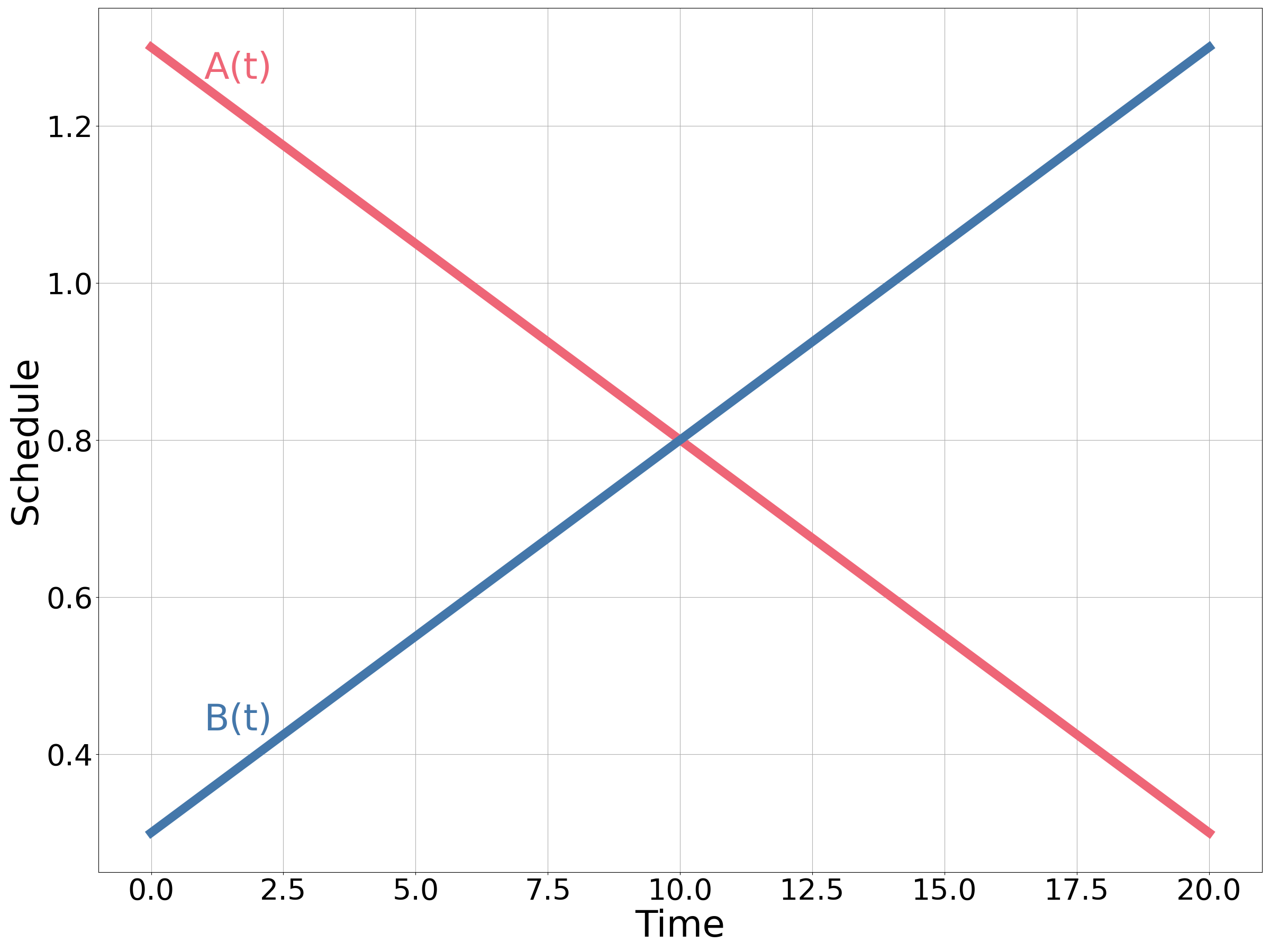}
        \caption{Schedule}
        \label{fig:mc_sched}
    \end{subfigure}
    \hfill
    \begin{subfigure}[l]{0.48\textwidth}
        \centering
        \includegraphics[width=\textwidth]{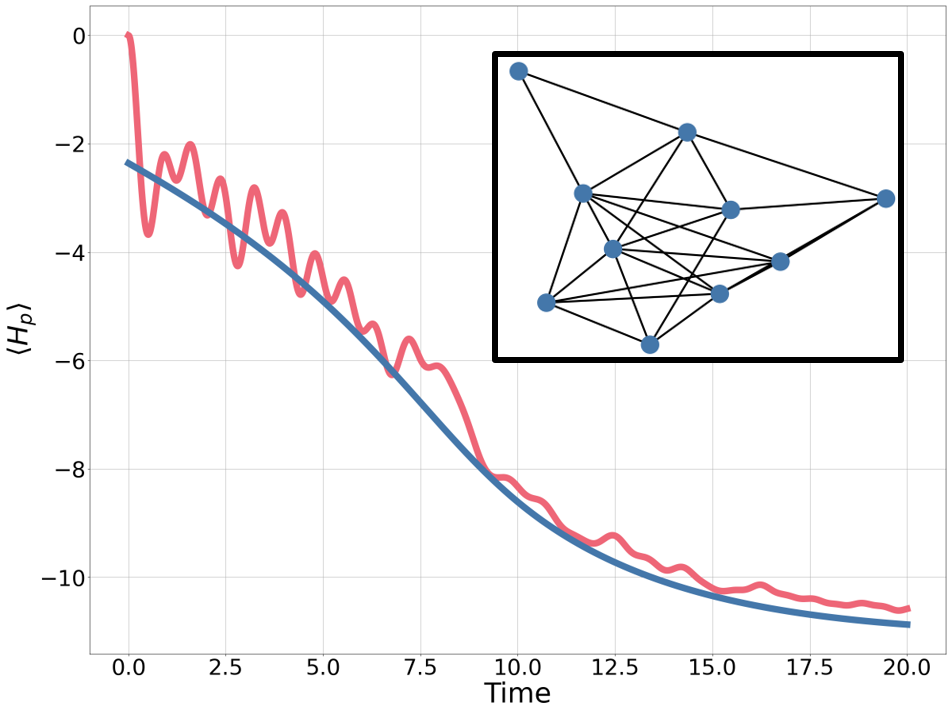}
        \caption{A comparison of $\langle H_p(t)\rangle$ calculated from the Schr\"odinger equation (red) and the PSTQA equations (blue).}
        \label{fig:mc_pstqa_inst}
    \end{subfigure}
    \\
    \begin{subfigure}[l]{0.48\textwidth}
        \centering
        \includegraphics[width=\textwidth]{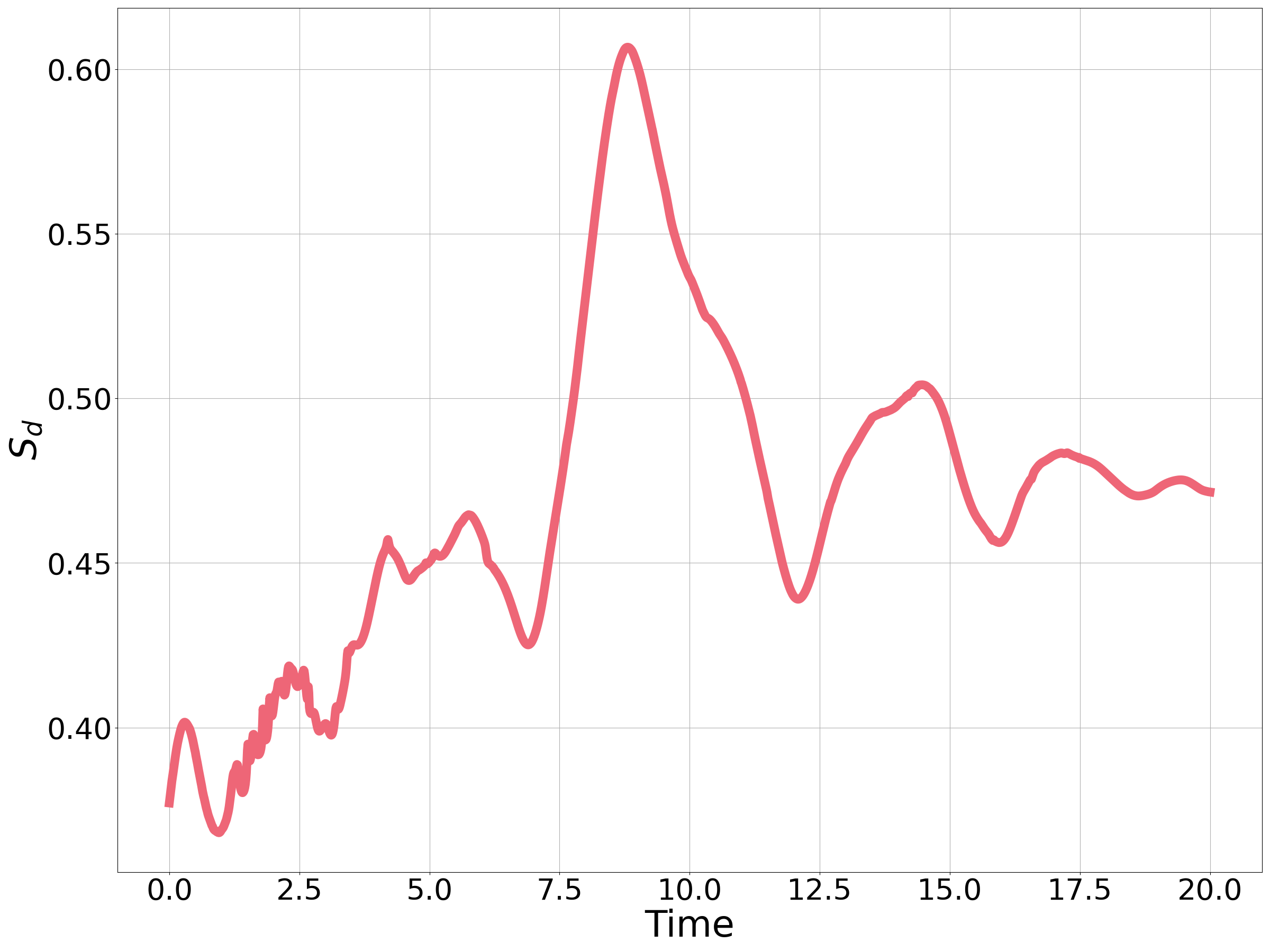}
        \caption{The diagonal entropy calculated from the Schr\"odinger equation.}
        \label{fig:mc_pstqa_sd}
    \end{subfigure}
    \hfill
    \begin{subfigure}[l]{0.48\textwidth}
        \centering
        \includegraphics[width=\textwidth]{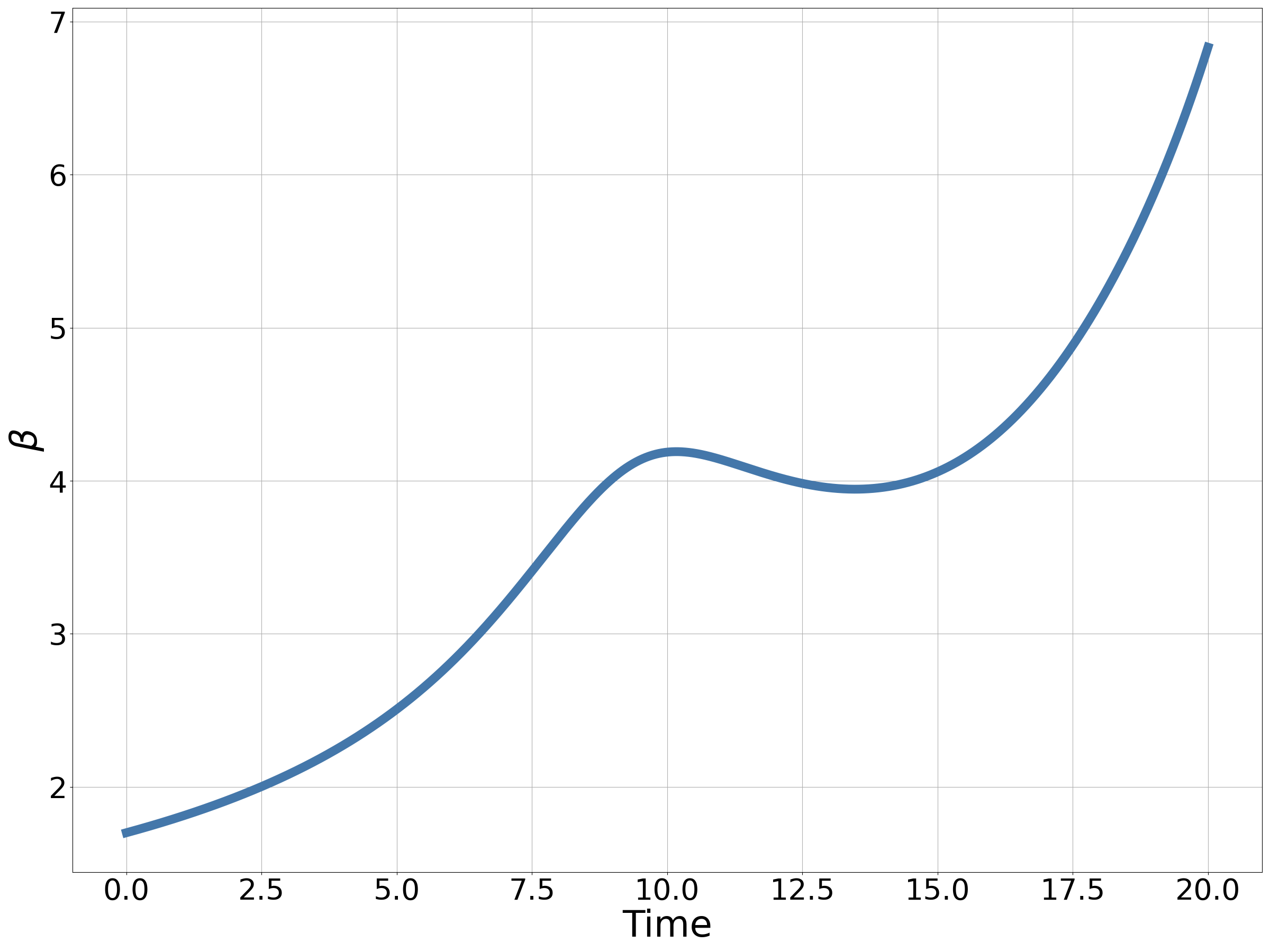}
        \caption{The inverse temperature calculated from the PSTQA equations.}
        \label{fig:mc_pstqa_beta}
    \end{subfigure}
    \caption{Comparing the PSTQA equations to the Schr\"odinger equation for a 10 qubit MAX-CUT graph, shown in the inset of Fig.~\ref{fig:mc_pstqa_inst}.}
    \label{fig:enter-label}
\end{figure*}

\subsection{Diagonal entropy is conserved}

The diagonal entropy, assuming a Gibbs state, is given by
\begin{equation}
    S_d(\beta, E, A, B)=\ln \mathcal{Z}(\beta,A,B)+\beta E,
\end{equation}
where $E=\langle H_{TQA}(t)\rangle$. Taking the time derivative of $S_d$ gives
\begin{multline}
    \frac{\dd S_d}{\dd t}=\frac{\partial\ln \mathcal{Z}}{\partial A} \frac{\dd A}{\dd t}+\frac{\partial\ln \mathcal{Z}}{\partial B} \frac{\dd B}{\dd t}\\+\left(\frac{\partial\ln \mathcal{Z}}{\partial \beta} +E\right)\frac{\dd \beta}{\dd t}+\beta \frac{\dd E}{\dd t},
\end{multline}
substituting in Eq.~\ref{eq:hdqa_th} gives:
\begin{equation}
    \frac{\dd S_d}{\dd t}=\frac{\partial\ln \mathcal{Z}}{\partial A} \frac{\dd A}{\dd t}+\frac{\partial\ln \mathcal{Z}}{\partial B} \frac{\dd B}{\dd t}+\beta \frac{\dd E}{\dd t},
\end{equation}
Using Eqs.~\ref{eq:hd_th}, \ref{eq:hp_th} and Eq.~\ref{eq:dHdqa} we conclude
\begin{equation}
    \frac{\dd S_d}{\dd t}=0.
\end{equation}

In conclusion, PSTQA is the adiabatic limit of an MSQW, with Assumption $\ref{ass:eth}$. The standard quantum adiabatic theorem would imply that the relevant timescale would be associated with the minimum spectral gap, hence a timescale that grows exponentially with the number of qubits. But to reach PSTQA the relevant timescale invoked was the dephasing timescale. To make Assumption \ref{ass:eth} we assumed that non-extensive corrections to the observables could be neglected. Since $S_d$ is an extensive quantity, it is perhaps more reasonable to say that PSTQA is adiabatic up to non-extensive corrections.

\subsection{Numerical evidence of PSTQA}

To illustrate PSTQA we numerically consider an example. The example is a single instance of MAX-CUT on a randomly generated graph with 10 qubits, with $H_p=H_{MC}$ (Eq.~\ref{eq:MCIsing}) and $H_d=H_{TF}$ (Eq.~\ref{eq:TF}). Each edge in the graph has been selected with probability 2/3. The MAX-CUT graph is shown in  the inset of Fig.~\ref{fig:mc_pstqa_inst}. The schedule is shown in Fig.~\ref{fig:mc_sched}, where  $B(0)\neq0$ and $A(t_f)\neq0$ breaking integrability and the conventional assumption in AQO.

Fig.~\ref{fig:mc_pstqa_inst} shows $\langle H_p(t)\rangle$ for the MAX-CUT example. The red line shows the Schr\"odinger equation. The blue line shows the prediction from directly solving the PSTQA equations (i.e. Eqs.~\ref{eq:dHdqa}-\ref{eq:hp_th}). There is very good agreement throughout the evolution. Fig.~\ref{fig:mc_pstqa_sd} shows the diagonal entropy for the same evolution, according to the Schr\"odinger evolution. The change is relatively small, however non-zero. This is evidence of diabatic transitions between energy eigenstates. Indeed, at the end of the evolution, there is a net increase in diagonal entropy. The inverse temperature calculated from Eq.~\ref{eq:hdqa_th} is shown in Fig.~\ref{fig:mc_pstqa_beta}. Generally, the temperature is seen to be decreasing, except at around $t\approx 11$ where the system heats. 

Changing both $H_d$ and $H_p$ creates effects not previously observed in MSQWs \cite{Ban24}, such as cooling as the system progresses. 

\begin{figure}
    \centering
    \begin{subfigure}[l]{0.48\textwidth}
        \centering
        \includegraphics[width=\textwidth]{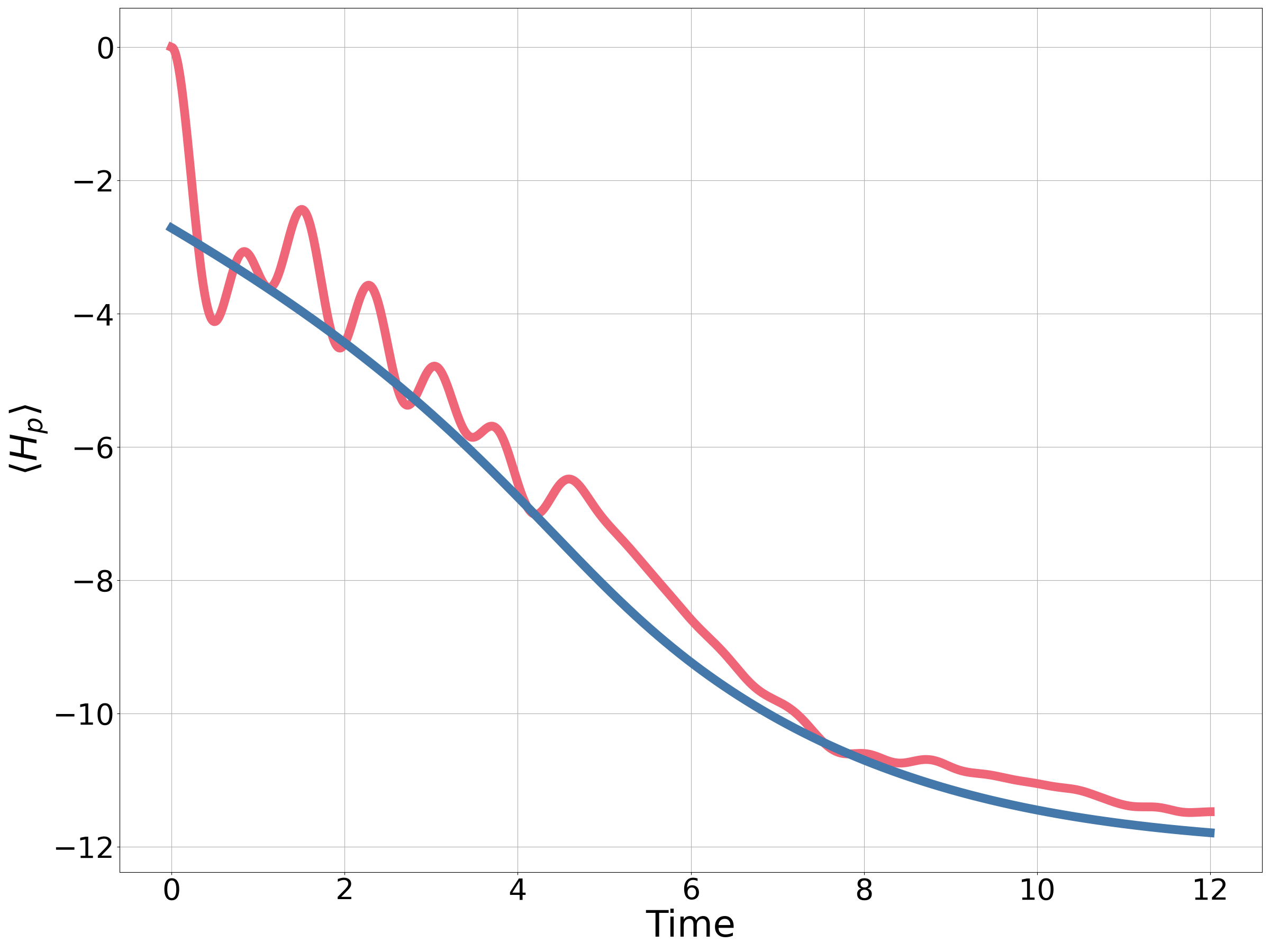}
        \caption{The time evolution of $\langle H_p (t)\rangle$. The red line shows the Schr\"odinger evolution. The blue line shows the solution of the PSTQA equations for this instance.}
        \label{fig:hpt12}
    \end{subfigure}
    \vfill
    \begin{subfigure}[l]{0.48\textwidth}
        \centering
        \includegraphics[width=\textwidth]{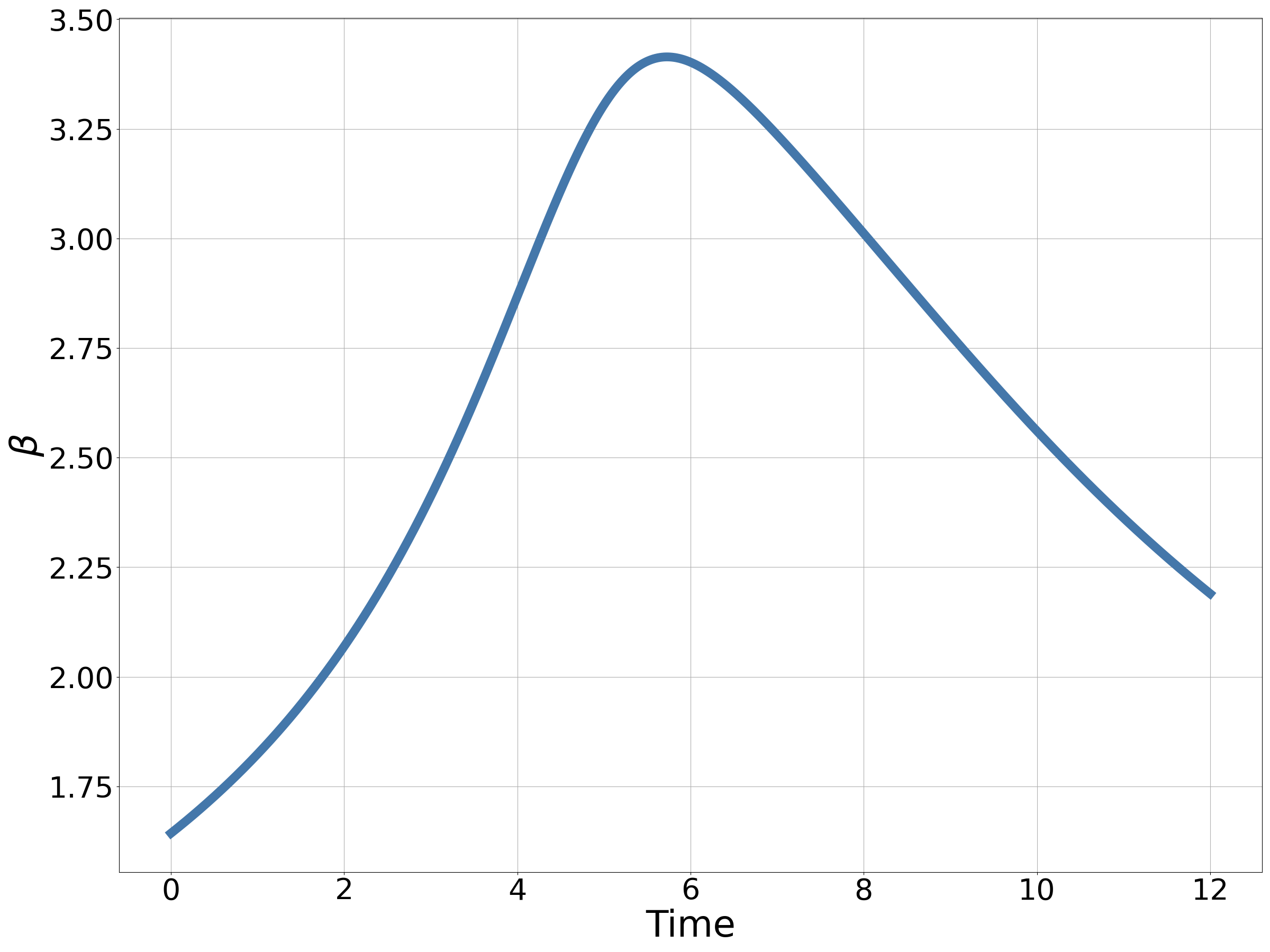}
        \caption{The inverse temperature from the PSTQA equations. }
        \label{fig:betat12}
    \end{subfigure}
    \vfill
    \begin{subfigure}[l]{0.48\textwidth}
        \centering
        \includegraphics[width=\textwidth]{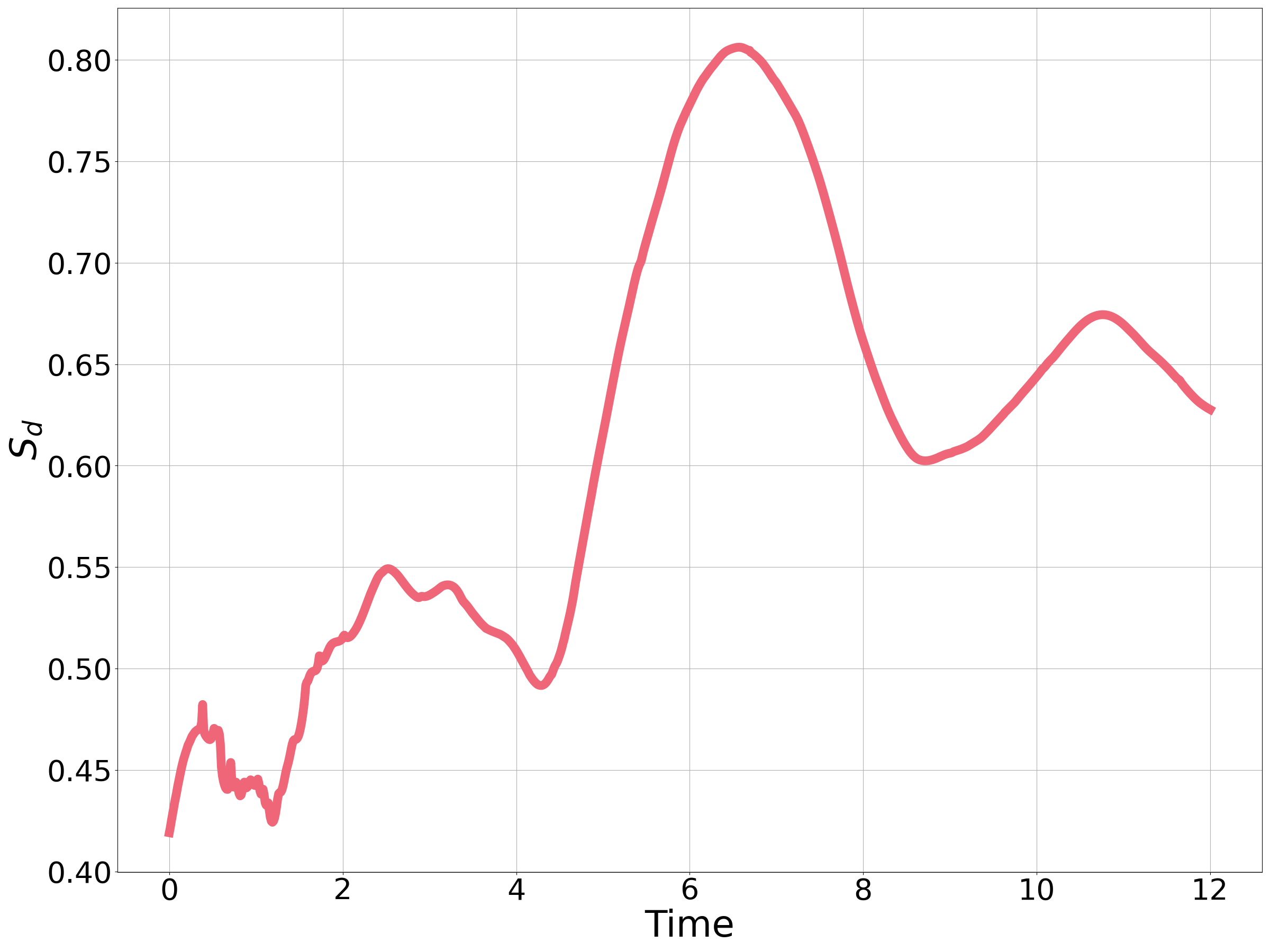}
        \caption{The diagonal entropy from the Schr\"odinger equation.}
        \label{fig:sdt12}
    \end{subfigure}
    \caption{A second 10-qubit MAX-CUT example. The schedule is a linear ramp, with a minimum value of 0.3 and maximum value 1.3.}
    \label{fig:enter-label}
\end{figure}

\begin{figure}
    \centering
    \includegraphics[width=0.48\textwidth]{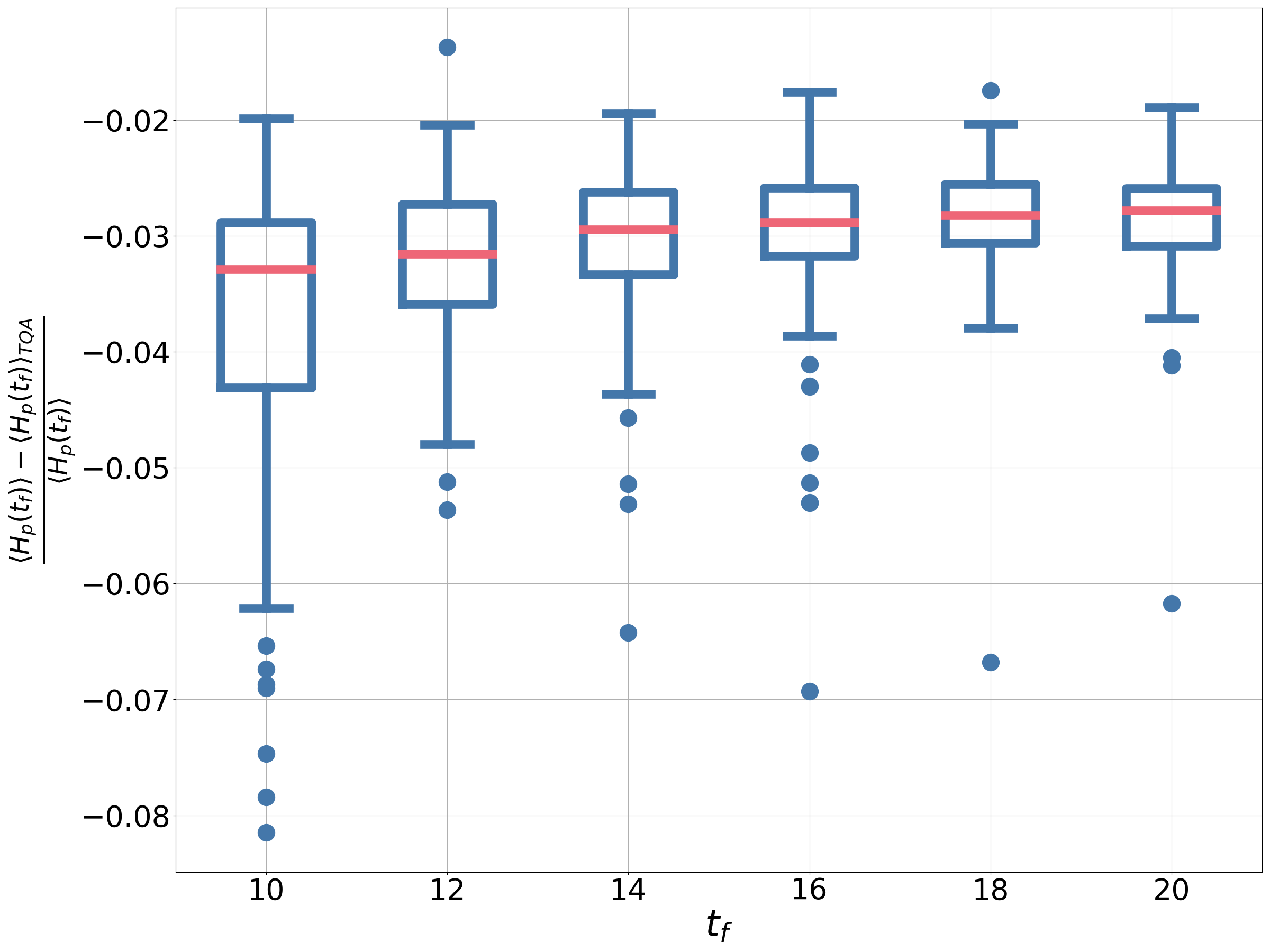}
    \caption{A box-plot showing $(\langle H_p(t_f) \rangle-\langle H_p(t_f) \rangle_{TQA})/\langle H_p(t_f) \rangle$ for MAX-CUT on 10 qubit binomial graphs. The final value of the Schr\"odinger evolution is denoted by $\langle H_p(t_f) \rangle$. The value predicted by the PSTQA equations is denoted by $\langle H_p(t_f) \rangle_{TQA}$. In each case, a linear schedule is used with $A(0)=B(t_f)=1.3$ and $A(t_f)=B(0)=0.3$. At each value of $t_f$, 90 instances are considered.}
    \label{fig:PSTQAManyHp}
\end{figure}

\begin{figure}
    \centering
    \includegraphics[width=0.48\textwidth]{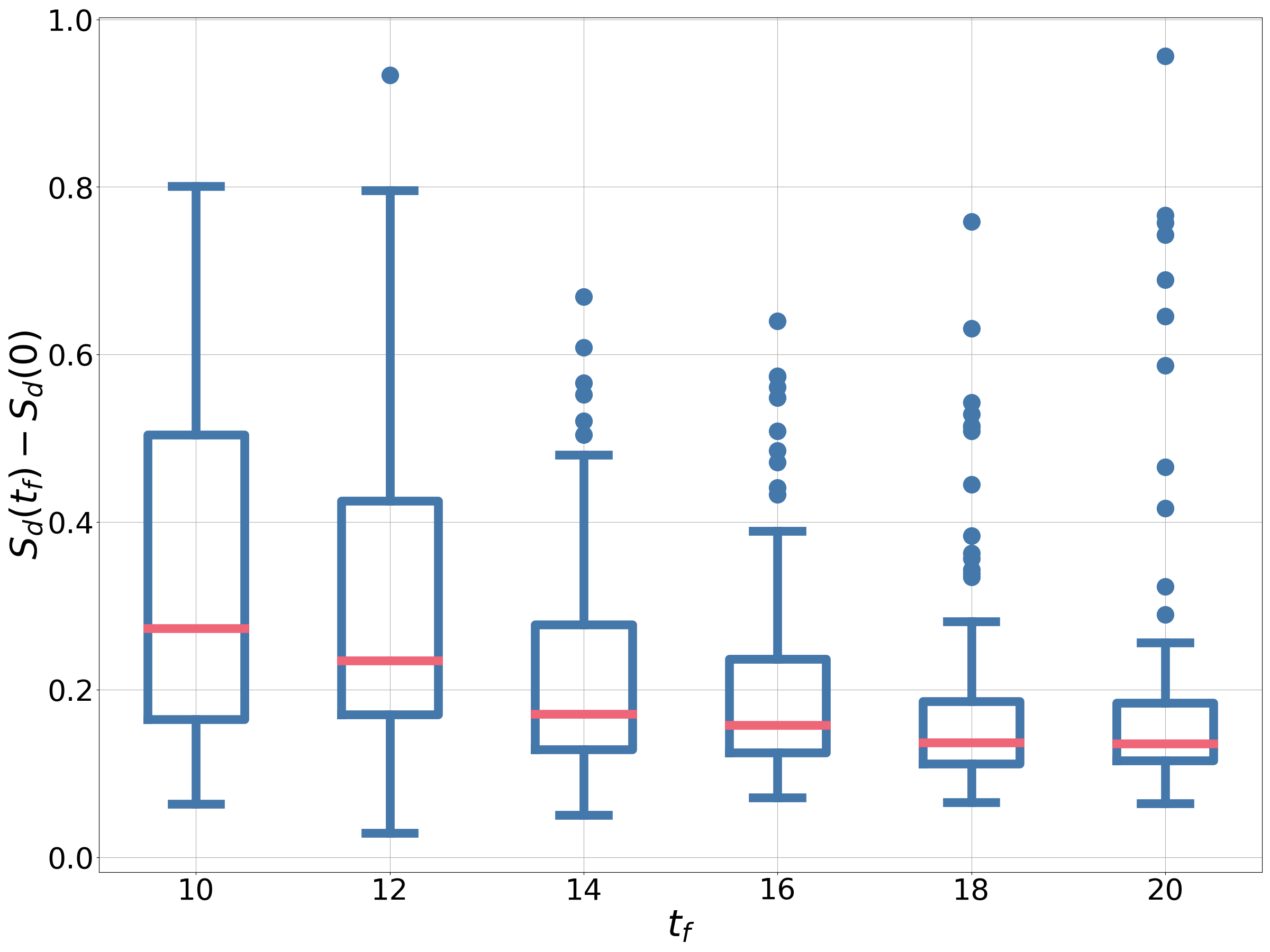}
    \caption{A box plot showing the change in diagonal entropy for MAX-CUT on 10 qubit binomial graphs for the instances shown in Fig.~\ref{fig:PSTQAManyHp}.}
    \label{fig:PSTQAManySd}
\end{figure}

In Fig.~\ref{fig:hpt12}, we consider other 10-qubit MAX-CUT instances with different annealing times. The schedule in each case is linear with $A(0)=B(t_f)=1.3$ and $A(t_f)=B(0)=0.3$. First, we consider a single 10-qubit example with $t_f=12$. Focusing on $\langle H_p(t)\rangle$, shown in Fig.~\ref{fig:hpt12}, the red line shows the Schr\"odinger equation and the blue line the PSTQA equations. There is a gap between the two curves at the beginning of the evolution, as the coherences in the Schr\"odinger evolution hide the thermal state. At the end of the evolution, the PSTQA equations overestimate the performance of the evolution. The  inverse temperature (shown in Fig.~\ref{fig:betat12}) tells a different story to the MAX-CUT example in Sec.~\ref{sec:pstqa}. In this example, the temperature cools for approximately half the interval before heating up again. The diagonal entropy for this example is shown in Fig.~\ref{fig:sdt12}. The evolution is not strictly adiabatic, evidenced by the varying diagonal entropy. 

Fig.~\ref{fig:PSTQAManyHp} shows the error between the error between $\langle H_p (t_f) \rangle$ according to the Schr\"odinger equation and the prediction from the PSTQA equations ($\langle H_p (t_f) \rangle_{TQA}$) for 90 instances as $t_f$ is changed. Each graph is a 10-qubit binomial graph. The error is typically within a few percent and decreasing on average as the run-time increases. Note that in all instances, the PSTQA equations predict a lower value of $\langle H_p (t_f) \rangle$ than the true value. This is reflected in Fig.~\ref{fig:PSTQAManySd} where there is a net increase in diagonal entropy for each instance which is not captured by the PSTQA equations. Despite this, we still see relatively good agreement.

\begin{figure}
    \centering
    \begin{subfigure}[l]{0.48\textwidth}
        \centering
        \includegraphics[width=\textwidth]{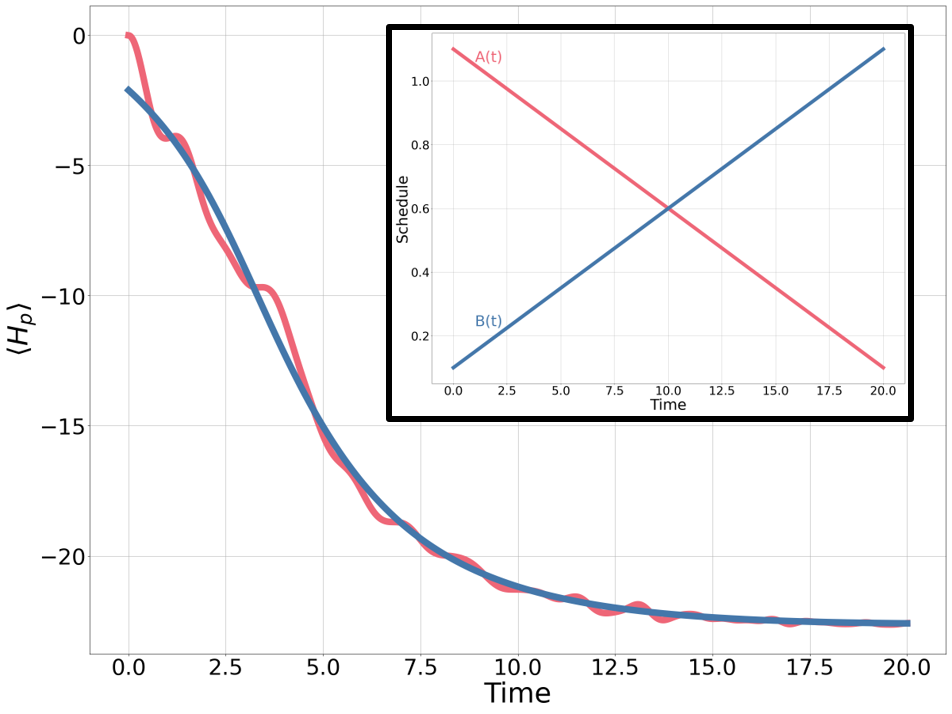}
        \caption{A comparison of $\langle H_p(t)\rangle$ calculated from the Schr\"odinger equation (the solid red line) and the PSTQA equations (the solid blue line).}
        \label{fig:sk_pstqa_inst}
    \end{subfigure}
    \hfill
    \begin{subfigure}[l]{0.48\textwidth}
        \centering
        \includegraphics[width=\textwidth]{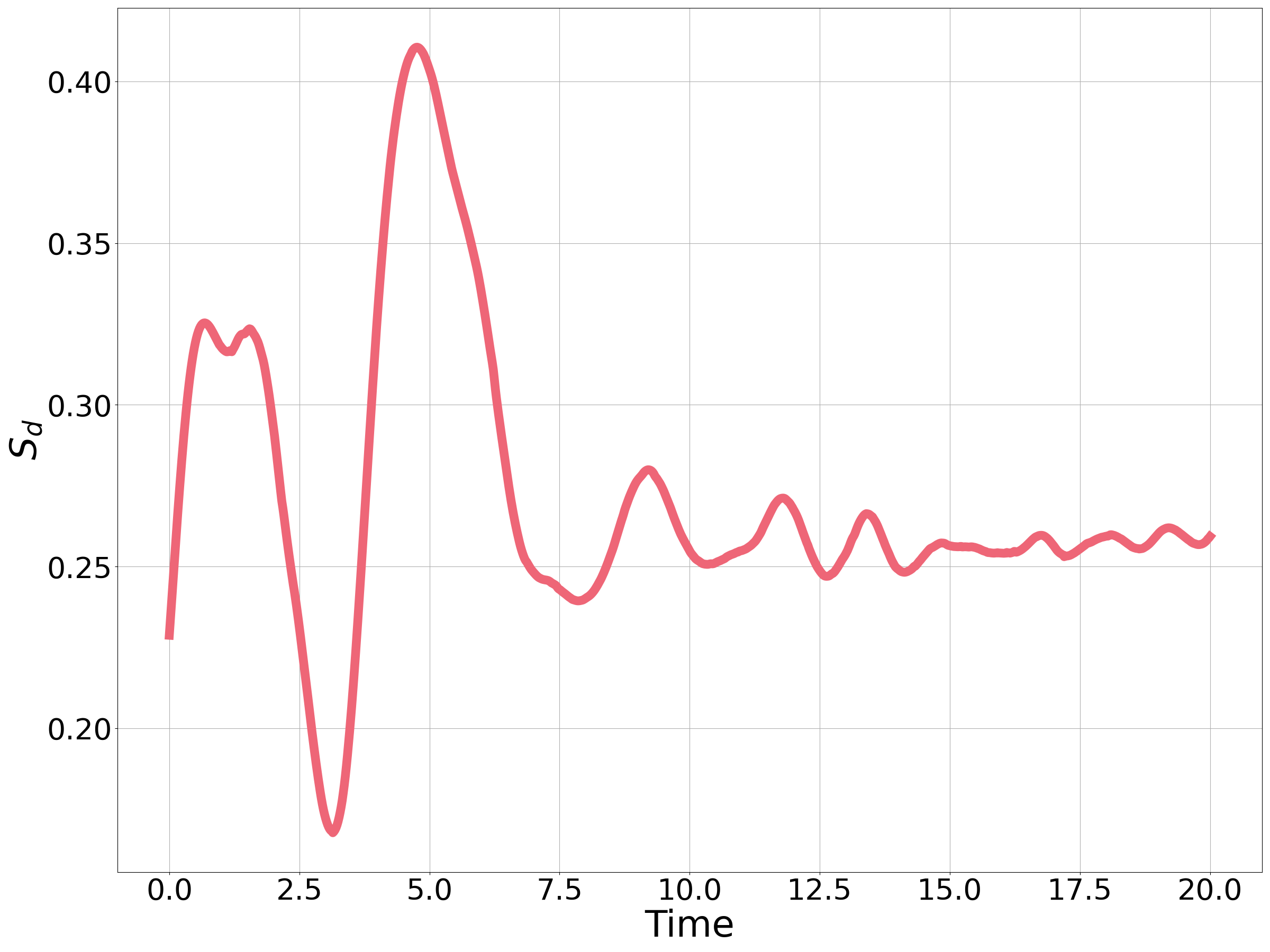}
        \caption{The diagonal entropy calculated from the Schr\"odinger equation.}
        \label{fig:sk_pstqa_sd}
    \end{subfigure}
    \hfill
    \begin{subfigure}[l]{0.48\textwidth}
        \centering
        \includegraphics[width=\textwidth]{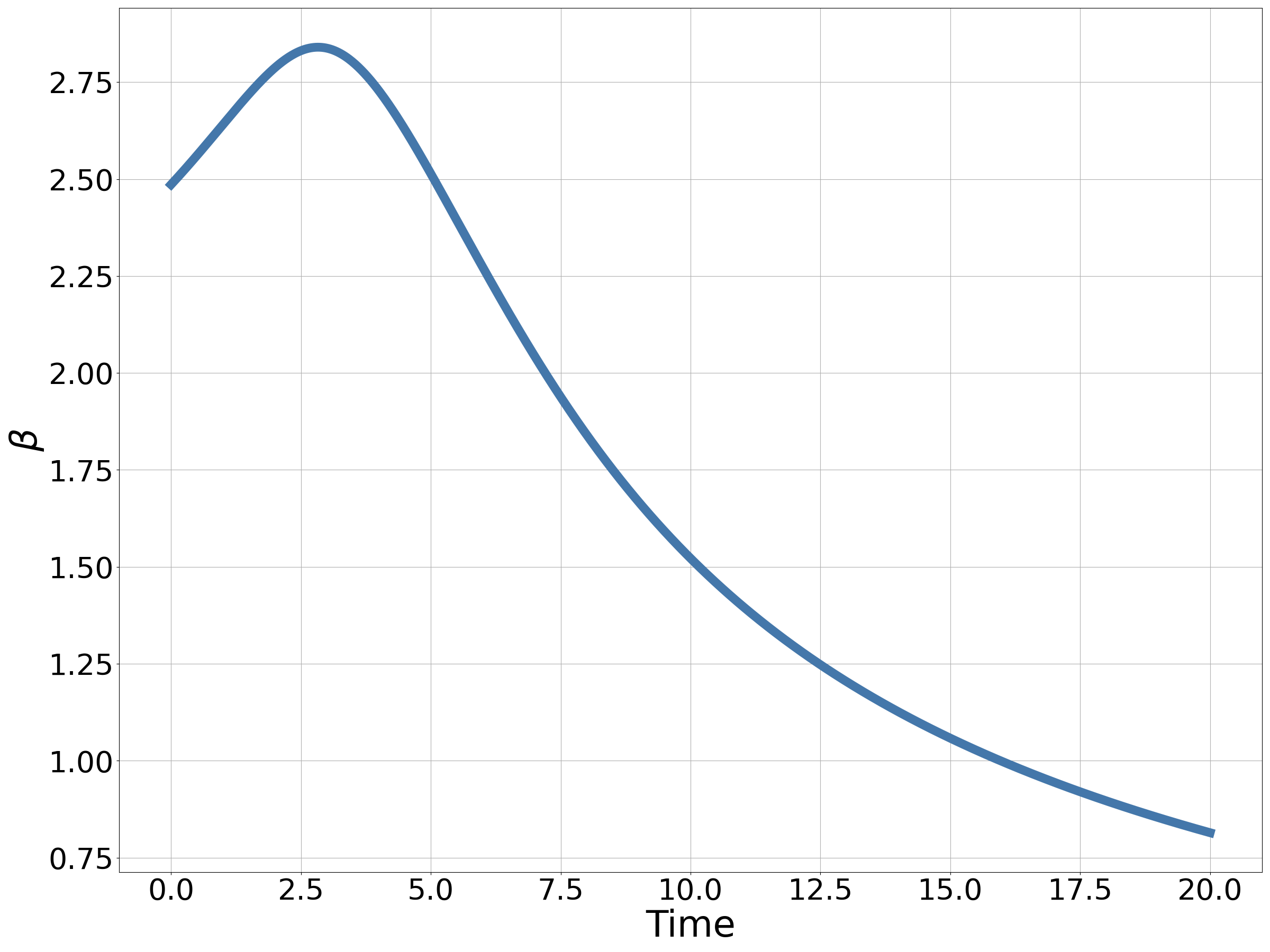}
        \caption{The inverse temperature calculated from the PSTQA equations.}
        \label{fig:sk_pstqa_beta}
    \end{subfigure}
    \caption{A comparison of the PSTQA equations and the Schr\"odinger equations for a 10 qubit SK instance. The schedule is shown in the inset of Fig.~\ref{fig:sk_pstqa_inst} and consists of linear ramps between 0.1 and 1.1.}
    \label{fig:enter-label}
\end{figure}

To show that PSTQA is not limited to MAX-CUT, we consider a Sherrington-Kirkpatrick (SK) inspired problem, with Ising Hamiltonian:
\begin{equation}
    H_p^{(SK)}=\sum_{i>j} J_{i,j} Z_i Z_j +\sum_i^nh_i Z_i,
\end{equation}
where $J_{i,j}$ and $h_i$ are drawn from a normal distribution with mean 0 and variance 1. For the driver Hamiltonian, we take the transverse-field, shown in Eq.~\ref{eq:TF}. The schedule under consideration is shown in the inset of Fig.~\ref{fig:sk_pstqa_inst}. Fig.~\ref{fig:sk_pstqa_inst} compares the value of $\langle H_p(t) \rangle$ according to the Schr\"odinger equation (red line) to the PSTQA equations (blue line) for a 10 qubit example. There is remarkably good agreement. As in the MAX-CUT case, Fig.~\ref{fig:sk_pstqa_sd} shows that the diagonal entropy is changing, so the system has not reached the adiabatic limit. Fig.~\ref{fig:sk_pstqa_beta} shows the inverse temperature, which is non-monotonic with time. 

\begin{figure}
    \centering
    \includegraphics[width=0.48\textwidth]{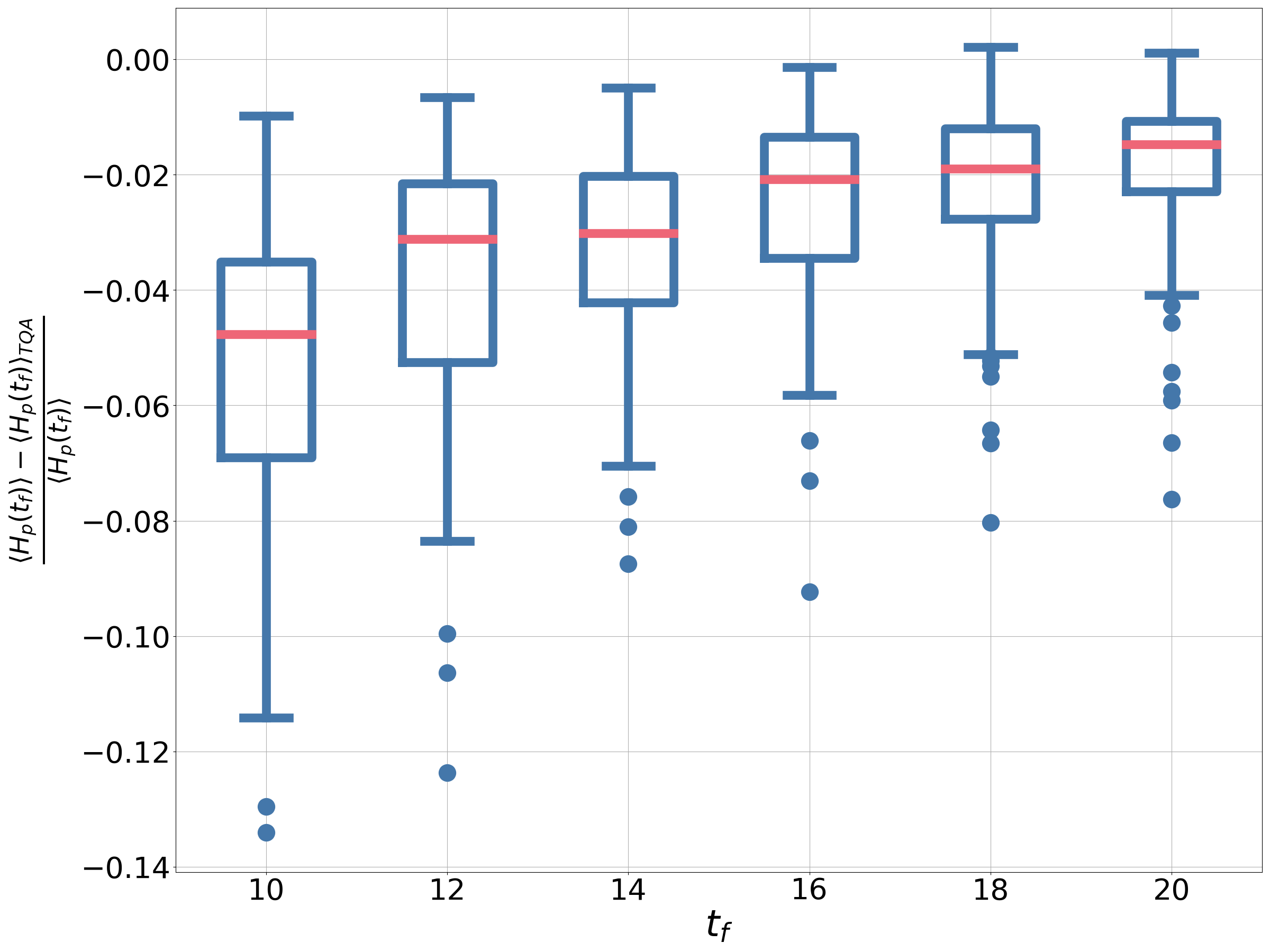}
    \caption{A box-plot showing $(\langle H_p(t_f) \rangle-\langle H_p(t_f) \rangle_{TQA})/\langle H_p(t_f) \rangle$ for  10 qubit SK instances. The final value of the Schr\"odinger evolution is denoted by $\langle H_p(t_f) \rangle$. The value predicted by the PSTQA equations is denoted by $\langle H_p(t_f) \rangle_{TQA}$. In each case a linear schedule is used with $A(0)=B(t_f)=1.1$ and $A(t_f)=B(0)=0.1$. At each value of $t_f$ 90 instances are considered.}
    \label{fig:sk_pstqa_many_hp}
\end{figure}

\begin{figure}
    \centering
    \includegraphics[width=0.48\textwidth]{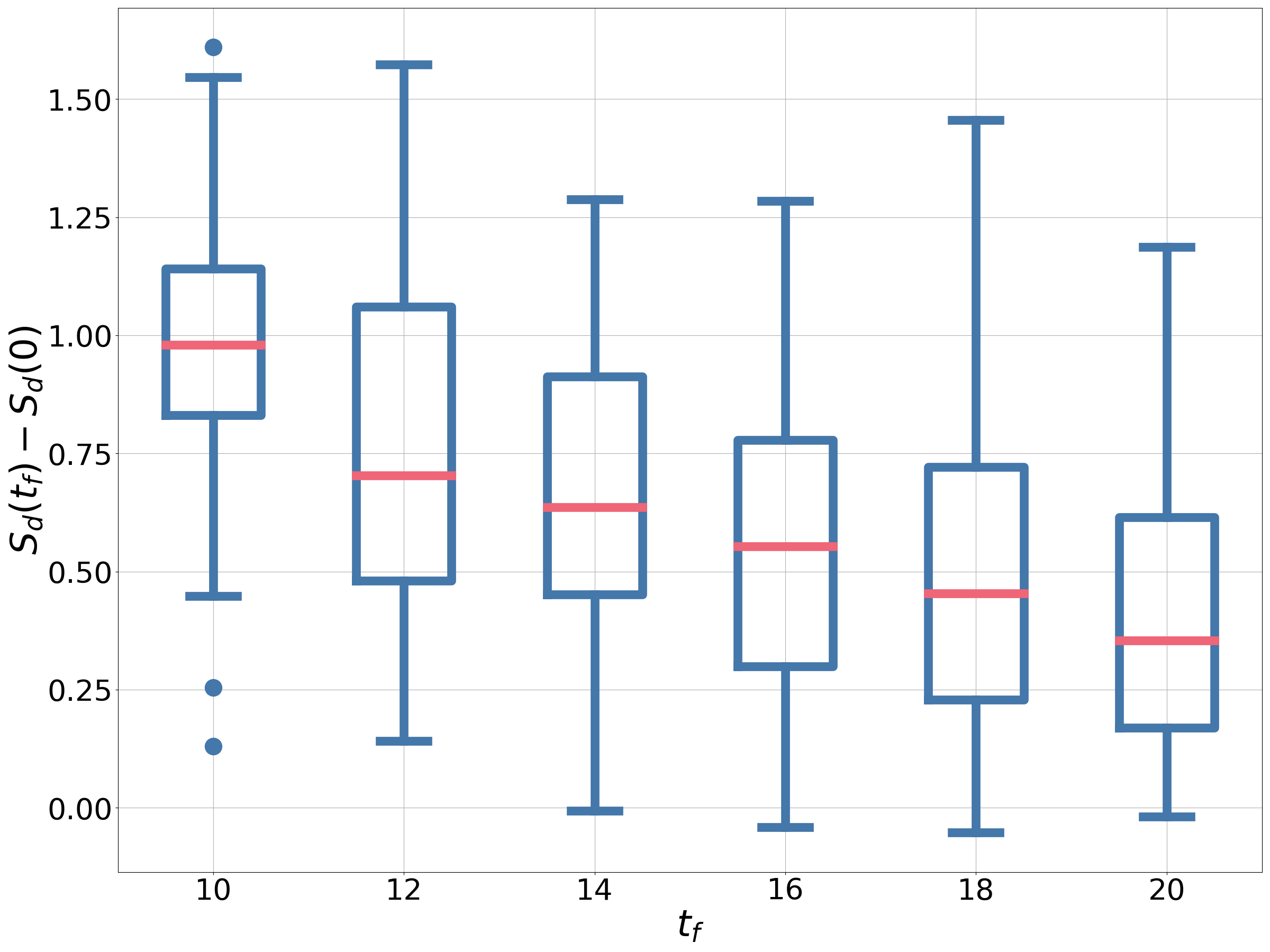}
    \caption{A box-plot showing the change in diagonal entropy of the 10-qubit SK instances shown in Fig.~\ref{fig:sk_pstqa_many_hp}.}
    \label{fig:sk_pstqa_many_sd}
\end{figure}

Again, we consider 90 instances of the SK model at various $t_f$. A similar trend for the error in $\langle H_p(t_f)\rangle$ as the MAX-CUT instances can be found in Fig.~\ref{fig:sk_pstqa_many_hp}. The change in diagonal entropy can be found in Fig.~\ref{fig:sk_pstqa_many_sd}, which decreases as the run time increases.

\FloatBarrier
\subsection{Ansatz based approach}

\begin{figure}
    \centering
    \includegraphics[width=0.48\textwidth]{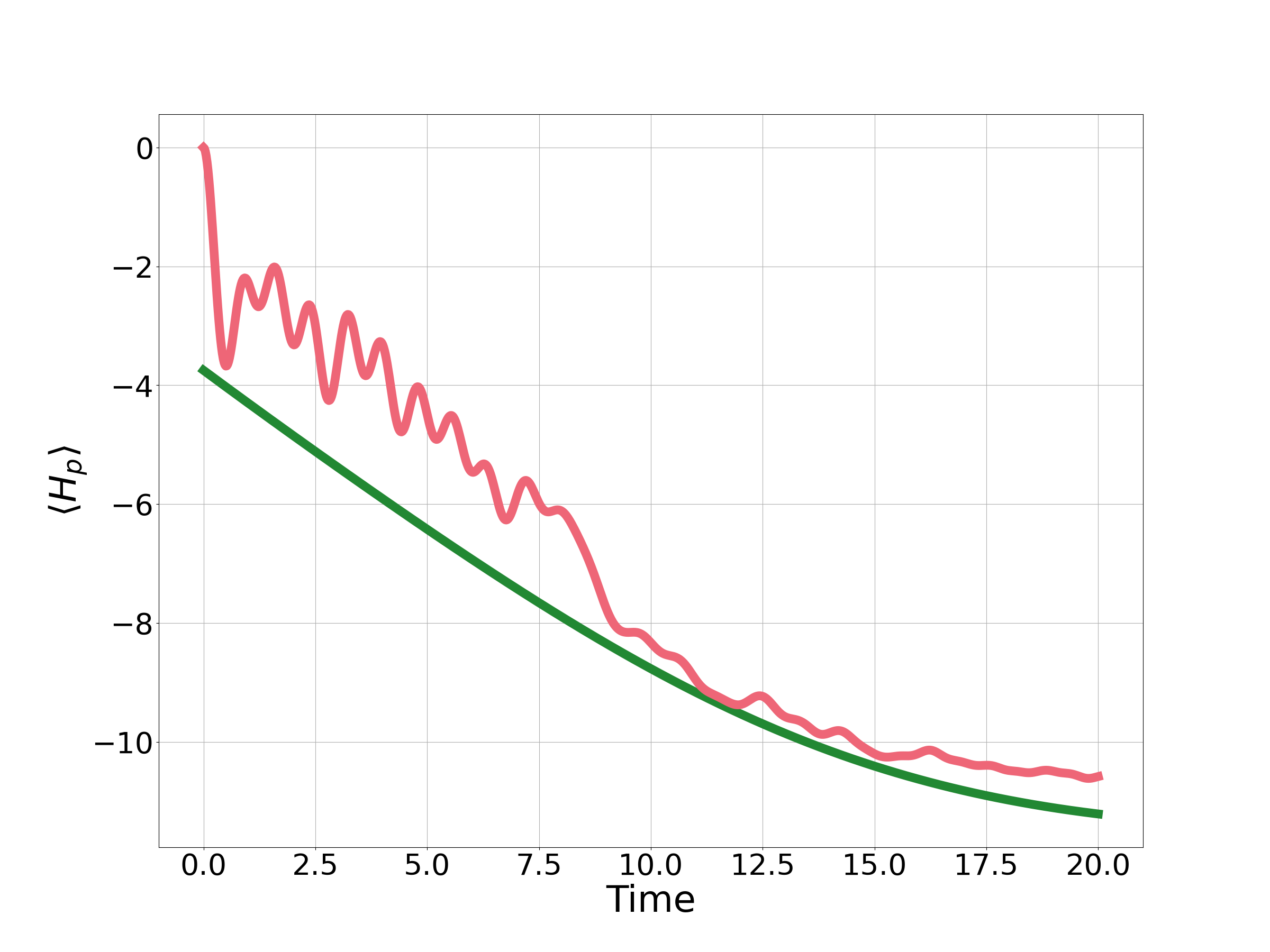}
    \caption{PSTQA for the 10-qubit MAX-CUT graph shown in the inset of Fig.~\ref{fig:mc_pstqa_inst}. The red line shows the Schr\"odinger evolution. The green line shows the prediction from the PSTQA equations when an exponentially modified Gaussian ansatz is made for the density-of-states.}
    \label{fig:emg_ex}
\end{figure}

Directly numerically solving the PSTQA equations (i.e.\ Eqs.~\ref{eq:dHdqa}-\ref{eq:hp_th}) is difficult, requiring repeated matrix exponentiation to find $\beta$. However, since the PSTQA equations depend only on the partition function, the equations can be tackled by making a good choice of ansatz for $\mathcal{Z}$. Ans\"atze for the partition function based on models of the density-of-states for MAX-CUT have been explored in \cite{Ban24} for the time-independent setting. In Fig.~\ref{fig:emg_ex} we show a specific ansatz, based on an exponentially modified Gaussian distribution for the density-of-states, for the time-dependent setting. The green line in Fig.~\ref{fig:emg_ex} shows the ansatz-based approach, and the red line again shows the Schr\"odinger evolution of $\langle H_p(t)\rangle$ for the 10-qubit instance considered in the previous section.  Although the agreement is not as good as exactly solving the PSTQA equations, this approach removes the need for matrix exponentiation and still provides a good approximation. The rest of this section details the ansatz based calculation, along with further numerical examples.

We consider two simple models for the density-of-states, first introduced for CTQWs for the combinatorial optimisation problem MAX-CUT in \cite{Ban24}. The two models are a Gaussian and an exponentially modified Gaussian. In \cite{Ban24} these models of the density-of-states were shown to make reasonable approximations of observables for CTQWs. We expect these models to be suitable when the state-vector has significant overlap with energy eigenstates in the middle of the spectrum. These models have no energy cut-off, so become less suitable at low (or high) energies. However, they do allow for some analytic analysis. 

\subsection{The Gaussian model}

First, we assume the density-of-states associated with Hamiltonian \ref{eq:hamtqa} can be well modelled by a Gaussian distribution:
\begin{equation}
    \Omega(\varepsilon,t) \, \dd\varepsilon=\frac{1}{\sqrt{2\pi\sigma^2(t)}}e^{\frac{-(\varepsilon -\mu(t))^2}{2 \sigma^2(t)}} \, \dd \varepsilon,
\end{equation}
where $\mu(t)$ is the mean and $\sigma^2(t)$ the variance of the density-of-states. A Gaussian density-of-states has been observed to be a good approximation for the density-of-states in the non-integrable setting for a number of models \cite{Ban24, Pro99, Mon17, Beu15}. The moments of the density-of-states can be calculated directly from the eigenvalues of $H_{TQA}(t)$, denoted be $E_k$. Again, let $D$ be the dimension of the state space. And $\Tr'=1/\mathcal{D} \Tr$ be the normalised Trace, with the scaled operators $\Tilde{H}_p=H_p-\Tr' H_p$ and $\Tilde{H}_d=H_d-\Tr' H_d$. The mean and variance of the density-of-states are then
\begin{align}
    \mu(t)&=\frac{1}{\mathcal{D}}\sum_k E_k(t) \nonumber\\
    &= \Tr' H_{TQA}(t) \nonumber\\
    &=A(t)\Tr' H_d+B(t) \Tr' H_d\\
    \sigma^2(t)&=\frac{1}{\mathcal{D}}\sum_k E_k^2(t) \nonumber\\
    &=\frac{1}{D} \Tr H_{TQA}^2(t) -\mu(t)^2 \nonumber\\
    &=A^2(t)\Tr'\Tilde{H}_d^2+B^2(t) \Tr'\Tilde{H}_p^2 \nonumber\\
    &\hspace{3 cm}+2A(t)B(t)\Tr'\Tilde{H}_d \Tilde{H}_p.
\end{align}

Evaluating the partition function gives:
\begin{align}
    \mathcal{Z}(t)&=\int_{-\infty}^\infty e^{-\beta(t) \varepsilon}\Omega(\varepsilon,t) \dd\varepsilon \nonumber\\
    &=\frac{1}{\sqrt{2\pi\sigma^2(t)}}\int_{-\infty}^\infty e^{-\beta(t) \varepsilon}e^{\frac{-\left(\varepsilon-\mu(t)\right)^2}{2 \sigma^2(t)}} \, \dd \varepsilon \nonumber\\
    &=e^{-\beta(t)\mu(t)+\frac{\beta(t)^2\sigma^2(t)}{2}}
\end{align}

Evaluating $\langle H_{TQA} (t)\rangle$ gives:
\begin{align}
    \langle H_{TQA} (t)\rangle &=-\frac{\partial \ln \mathcal{Z}(t)}{\partial \beta} \nonumber\\
    &=\mu(t)-\beta \sigma^2(t)
\end{align}

Hence:
\begin{equation}
    \beta(t)=\frac{\mu(t)-\langle H_{TQA} (t)\rangle}{\sigma^2}.
\end{equation}
Evaluating $\langle H_{d} (t)\rangle$ and $\langle H_{p} (t)\rangle$ gives:
\begin{align}
    \langle H_{d}(t) \rangle=&-\frac{1}{\beta(t)} \frac{\partial \ln \mathcal{Z}(t)}{\partial A} \nonumber\\
    &=\Tr'H_d-\beta(t) \left(A(t) \Tr'\Tilde{H}_d^2+ B(t) \Tr'\Tilde{H}_d \Tilde{H}_p \right)\\
    \langle H_{p}(t) \rangle=&-\frac{1}{\beta(t)} \frac{\partial \ln \mathcal{Z}(t)}{\partial B} \nonumber\\
    &=\Tr'H_p-\beta(t)\left(B(t) \Tr'\Tilde{H}_p^2+ A(t) \Tr'\Tilde{H}_d\Tilde{H}_p\right).
\end{align}
Substituting $\beta(t)$, $\langle H_{d} (t)\rangle$ and $\langle H_{p} (t)\rangle$ into Eq.\ \ref{eq:dHdqa} gives:
\begin{align}
    \frac{\dd \langle H_{TQA}(t) \rangle }{\dd t}&=\dot{A}(t)\langle H_{d}(t) \rangle+\dot{B}(t)\langle H_{p}(t) \rangle \nonumber\\
    &=\frac{\dd \mu}{\dd t}-\frac{\mu-\langle H_{TQA} (t)\rangle}{2\sigma^2} \frac{\dd \sigma^2}{\dd t}
\end{align}
Integrating the above expression gives:
\begin{equation}
    \langle H_{TQA}(t) \rangle= \mu(t)+c\sigma(t),
\end{equation}
where $c$ is the constant of integration, which can be fixed using the boundary condition $\langle H_{TQA}(0) \rangle=A(0) \bra{\psi_i} H_d \ket{\psi_i}+B(0) \bra{\psi_i} H_p \ket{\psi_i}$. Therefore:
\begin{equation}
    c=\frac{\langle H_{TQA}(0) \rangle-\mu(0)}{\sigma(0)}.
\end{equation}
With an expression for $\langle H_{TQA}(t)\rangle$, evaluating $\beta(t)$ and $\langle H_{p} (t)\rangle$ becomes trivial:
\begin{equation}
      \beta(t)=\frac{-c}{\sigma(t)}  
\end{equation}
and
\begin{multline}
    \langle H_p \rangle= \Tr'H_p\\
    +\frac{c}{\sigma(t)}\left(B(t) \Tr'\Tilde{H}_p^2+ A(t) \Tr'\Tilde{H}_d\Tilde{H}_p\right)
\end{multline}

\subsection{The Gaussian model applied to MAX-CUT}

The simple structure of the MAX-CUT problem and the transverse field allow for further simplification. Evaluating the moments gives:
\begin{align*}
    \mu(t)&=0\\
    \sigma^2(t)&=A^2(t)n+B^2(t)\kappa_2,
\end{align*}
where $\kappa_2$ is the number of edges in the MAX-CUT graph and $n$ is the number of nodes.
Hence:
\begin{align*}
     \beta&=\frac{n A(0)}{\sqrt{A^2(0)n+ B^2(0)\kappa_2}\sqrt{A^2(t)n+ B^2(t)\kappa_2}}\\
    \langle H_{TQA} (t)\rangle &=\frac{-n A(0) \sqrt{A^2(t)n+ B^2(t)\kappa_2}}{\sqrt{A^2(0)n+ B^2(0)\kappa_2}}\\
    \langle H_p \rangle&=\frac{-n A(0) B(t) \kappa_2}{\sqrt{A^2(0)n+ B^2(0)\kappa_2}\sqrt{A^2(t)n+ B^2(t)\kappa_2}}.
\end{align*}

The Gaussian case is primarily of interest as it can be handled analytically. It does not take into account any frustration in the system. In the next section we explore a model that begins to take this into account.

\subsection{The exponentially modified Gaussian model}

The above can be repeated for an exponentially modified Gaussian density-of-states. This model incorporates skewness into the density-of-states model, but is less tractable. The partition function for this model is given by:
\begin{equation}
    \mathcal{Z}(t)=\left(1+\frac{\beta(t)}{\lambda(t)}\right)^{-1}e^{-\nu(t) \beta(t)+\frac{1}{2}{\beta(t)^2 s^2}},
\end{equation}
where $\nu(t)$, $s(t)$ and $\lambda(t)$ are fitting parameters related to the mean ($\mu(t)$), variance ($\sigma^2(t)$) and skewness $\gamma(t)$ of the distribution:
\begin{align}
    \nu(t)&=\mu(t)-\sigma(t)\left(\frac{\gamma(t)}{2}\right)^{\frac{1}{3}}=\mu(t)-\Delta(t),\\
    s^2(t)&=\sigma^2\left(1-\left(\frac{\gamma(t)}{2}\right)^{\frac{2}{3}}\right)=\sigma^2(t)-\Delta^2(t),\\
    \lambda(t)&=\frac{1}{\sigma(t)}\left(\frac{\gamma(t)}{2}\right)^{-\frac{1}{3}}=\frac{1}{\Delta(t)},
\end{align}
where:
\begin{align}
    \mu(t)&=\Tr' H_{TQA}\\
    \sigma^2(t)&=\Tr'\left(H_{TQA}-\Tr'H_{TQA}\right)^2,\\
    \Delta(t)&=\frac{1}{2}\left(\Tr'\left(H_{TQA}-\Tr'H_{TQA}\right)^3\right)^{1/3}.
\end{align}

From the partition function, it is straight forward to estimate the relevant expectation values:
\begin{equation}
    \langle H_{TQA}(t) \rangle=\frac{\beta^2\Delta^3(t)}{1+\beta(t) \Delta(t) }+\mu(t)-\beta(t) \sigma^2(t).
\end{equation}
Inverting this equation gives:
\begin{equation}
    \beta(t)=\frac{-\sigma^2-\Delta \left(\langle H_{TQA}(t)\rangle -\mu\right)+\omega}{2 \Delta(t)\left(\sigma^2(t)-\Delta^2(t)\right)},
\end{equation}
where
\begin{multline}
    \omega=\biggl\{ \left[\sigma^2(t)+\Delta(t)\left( \langle H_{TQA}(t)\rangle-\mu(t)\right) \right]^2\\+4 \Delta \left(\langle H_{TQA}(t)\rangle -\mu(t)\right) \left(\Delta^2(t)-\sigma^2(t)\right)\biggr\}^{\! 1/2}.
\end{multline}
Calculating $\langle H_p \rangle$ gives: 
\begin{multline}
   \langle H_p(t) \rangle=\partial_B \mu-\frac{\Delta(t) \beta(t)}{1+\Delta \beta(t)}\partial_B \Delta\\
   -\beta(t)\left(\sigma(t) \partial_B \sigma(t)-\Delta(t) \partial_B \Delta(t)\right).
\end{multline}
The expression for $\langle H_d(t) \rangle$ is the same but with $B$ swapped for $A$.  Combining the above expressions, the energy of the system (Eq.~\ref{eq:dhdqa_s}) can be found using numerical integration, without resorting to full numerical integration of the state-vector.

\subsection{The exponentially modified Gaussian model applied to MAX-CUT}
\begin{figure}
    \centering
    \includegraphics[width=0.48\textwidth]{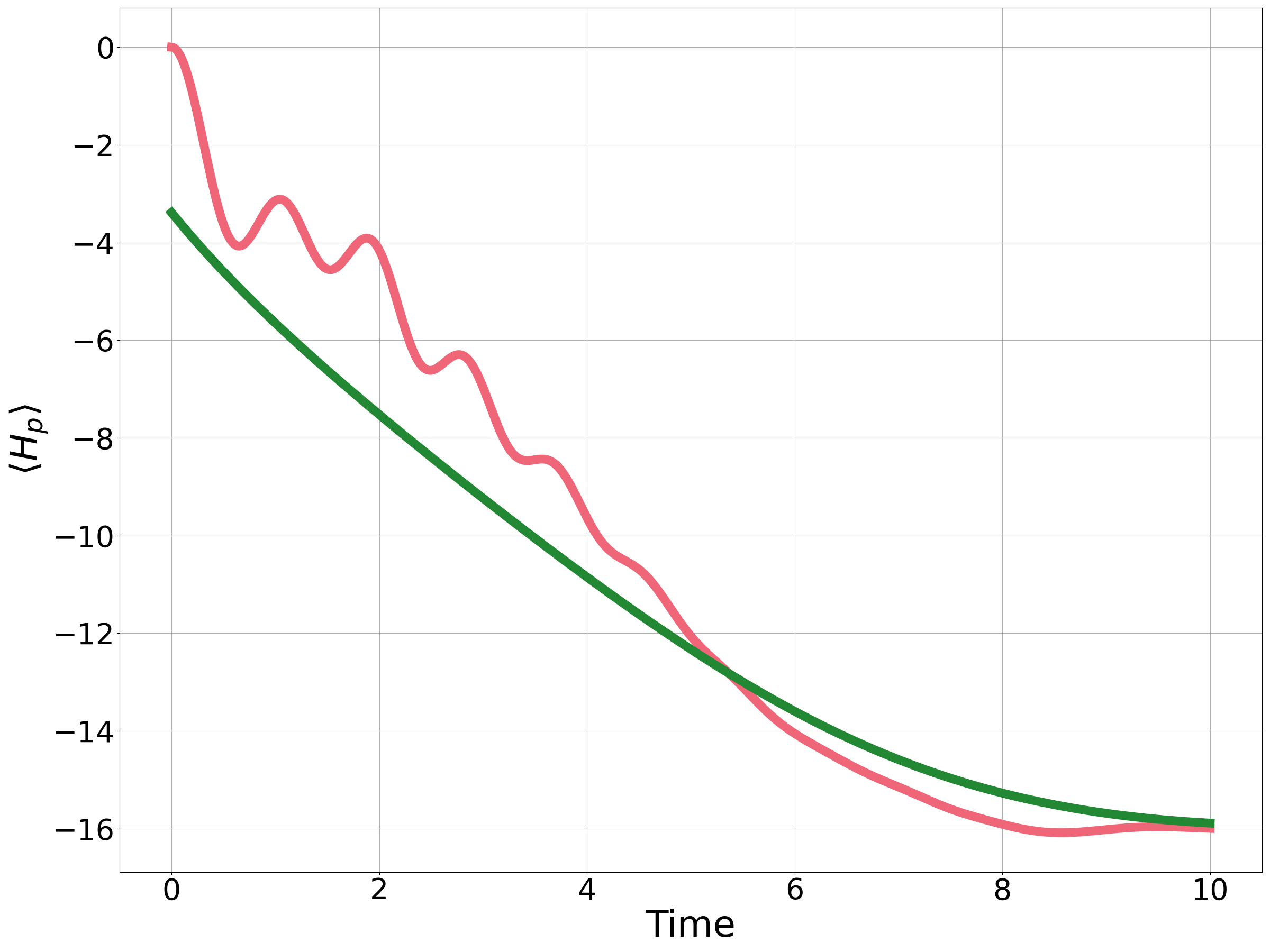}
    \caption{The evolution of $\langle H_p (t)\rangle $ for a MAX-CUT instance on 13 qubit binomial graph. The schedule is linear with $A(0)=B(t_f=10)=1.1$ and $A(t_f=10)=B(0)=0.1$. The red line shows the Schr\"odinger evolution, and the green line the prediction using an exponentially modified Gaussian ansatz for the density-of-states.}
    \label{fig:pstqa_emg_13}
\end{figure}

\begin{figure}
    \centering
    \includegraphics[width=0.48\textwidth]{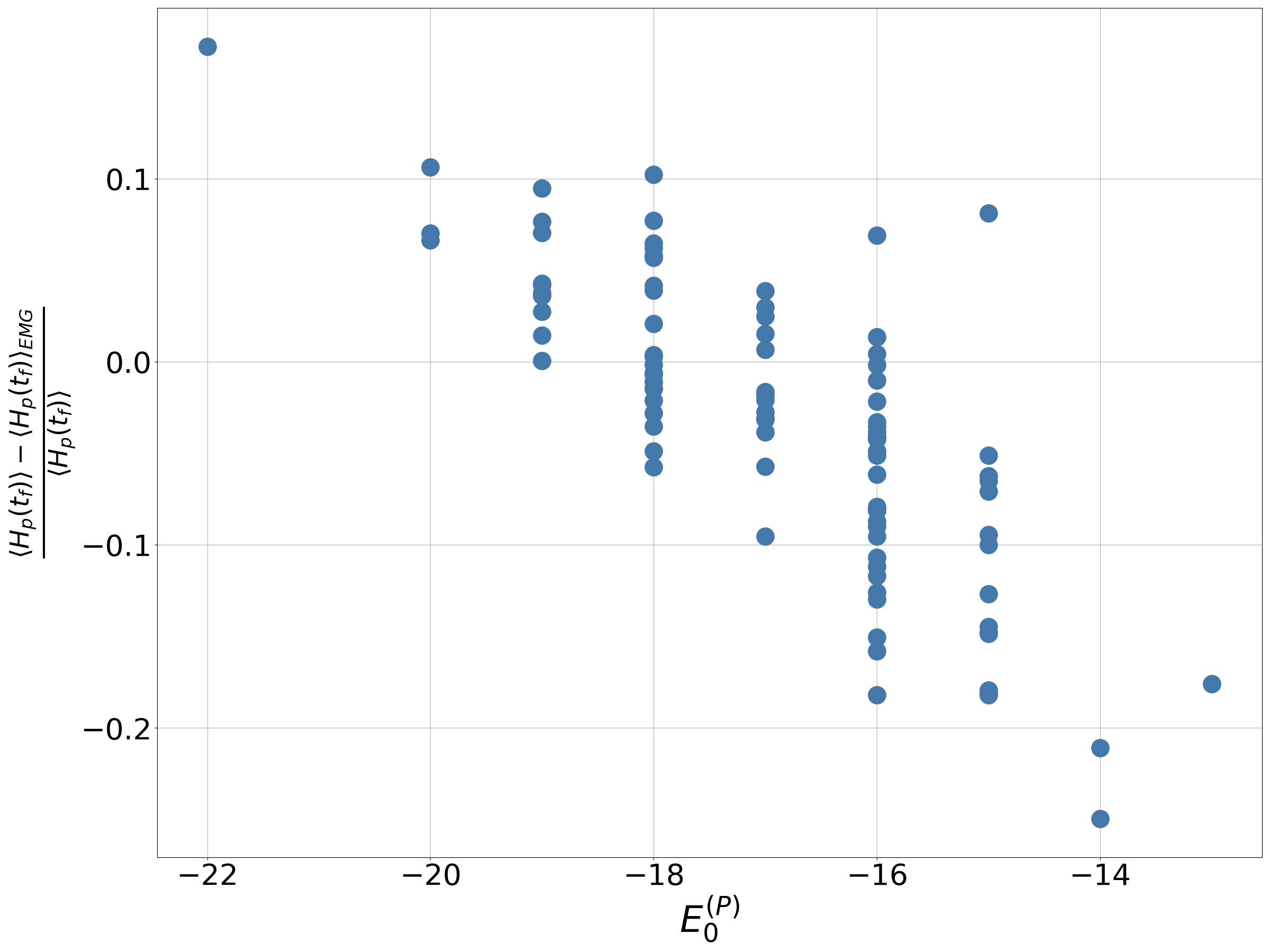}
    \caption{The error between $\langle H_p (t_f)\rangle$ from the Schr\"odinger equation and the exponentially modified Gaussian model, $\langle H_p (t_f)\rangle_{EMG}$. The figure shows 100 MAX-CUT instances on 13 qubit binomial graphs. The schedule is linear with $A(0)=B(t_f=10)=1.1$ and $A(t_f=10)=B(0)=0.1$. The prediction from the exponentially modified Gaussian is denoted by $\langle H_p(t_f)\rangle_{EMG}$, and the true value by $\langle H_p(t_f)\rangle$. The x-axis shows $E_{(0)}^p$, the ground state energy of $H_p$.}
    \label{fig:pstqa_emg_many}
\end{figure}

For MAX-CUT with a transverse field (and no terms proportional to the identity):
\begin{align}
    \mu &=0\\
    \sigma^2(t)&=A^2(t)n+B^2(t)\kappa_2\\
    \Delta^3(t)&=(3 \kappa_3)B^3(t).
\end{align}
Here $\kappa_2$ is the number of edges and $\kappa_3$ is the number of triangles in the MAX-CUT graph. Fig.~\ref{fig:pstqa_emg_13} shows the Schr\"odinger evolution (red line) and the prediction from using an exponentially modified Gaussian ansatz (green line). The instance considered is a 13-qubit binomial graph with a linear schedule with $A(0)=B(t_f=10)=1.1$ and $A(t_f=10)=B(0)=0.1$. There is good agreement, especially at the end of the evolution. Fig.~\ref{fig:pstqa_emg_many} shows the error between the true value of $\langle H_p (t_f)\rangle$ and the prediction from the ansatz $\langle H_p(t_f)\rangle_{EMG}$ for 100 13 qubit instances. The schedule is the same as before. The x-axis shows the ground state energy of the problem Hamiltonian, which is correlated with the error.

\section{Biased Quantum Annealing}
\label{app:bqa}
As discussed in Sec.~\ref{sec:notPP}, biased quantum annealing (BQA) is not isolated and therefore does not satisfy Planck's principle. In this section, we explore how this might lead to cooling. 

Since the energy of the system tends to increase a third term needs to be added to compensate for the global increase in energy,
\begin{equation}
    \label{eq:BQA}
    H_{cyc}(t)=H_p+H_b+G(t) H_d.
\end{equation}
Provided that the increase in energy associated with the bias Hamiltonian ($H_b$) is greater than the energy increase as a result of the cyclic process, the energy associated with $H_p$ will decrease.

The following protocol is inspired by works such as \cite{Wan22} and \cite{Zha24}.  In these works, the many-body localised phase transition in spin glasses are used to cyclically cool the system. We argue that this approach is likely to hold for a broad range of settings. Initially, we assume that we start in an eigenstate of $H_p$, denoted by $\ket{z^*}$ and $H_b=-\alpha \ket{z^*}\bra{z^*}$ with $\alpha>0$. This lowers the energy of the initial state, while all other eigenstates of $H_p$ remain unchanged. As long as the energy increase $\Delta W$ from the cyclic process is smaller than the energy shift from the bias, we will have reduced $\langle H_p \rangle$. This is sketched out in Fig.~\ref{fig:spec}. Note that for this choice of biasing Planck's principle no longer holds as a map between classical probability distributions, this is discussed in Sec.~\ref{sec:notPP}. However, we will assume that the net result of a typical cyclic process with sufficient dynamics is to move towards the bulk of the spectrum, increasing the energy. For the process to work, non-trivial dynamics need to take place. Given some starting state, the process could work as follows:
\begin{enumerate}
    \item Run the cyclic process
    \item If the resulting string corresponds to a better solution, update the initial state to be this state.
    \item Otherwise, keep the initial starting state but increase $\alpha$.
\end{enumerate}

\begin{figure}
    \centering
    \begin{tikzpicture}[domain=0:5]

        \draw[->] (0,0) -- (0,4.5) node[above] {Energy} ;
        \draw[-,ultra thick, color=myred]  (0.2,4) -- (1.2,4);
        \draw[-,ultra thick, color=myred]  (0.2,3.8) -- (1.2,3.8);
        \draw[-,ultra thick, color=myred]  (0.2,3.6) -- (1.2,3.6);
        \draw[-,ultra thick, color=myred]  (0.2,3.2) -- (1.2,3.2);
        \draw[-,ultra thick, color=myred]  (0.2,2.8) -- (1.2,2.8);
        \draw[-,ultra thick, color=myred]  (0.2,2.6) -- (1.2,2.6);
        \draw[-,ultra thick, color=mygreen]  (0.2,2.2) --  (1.2,2.2);
        \draw[-,ultra thick, color=myblue]  (0.2,1.8) -- (1.2,1.8);
        \draw[-,ultra thick, color=myblue]  (0.2,1.4) -- (1.2,1.4);
        \draw[-,ultra thick, color=myblue]  (0.2,0.8) -- (1.2,0.8);
        \draw[-,ultra thick, color=myblue]  (0.2,0) -- (1.2,0);
        \node at (0.7,4.5)  (hp) {$H_p$};

        \draw[->] (0,0) -- (0,4.5) node[above] {Energy} ;
        \draw[-,ultra thick, color=myred]  (1.7,4) -- (2.7,4);
        \draw[-,ultra thick, color=myred]  (1.7,3.8) -- (2.7,3.8);
        \draw[-,ultra thick, color=myred]  (1.7,3.6) -- (2.7,3.6);
        \draw[-,ultra thick, color=myred]  (1.7,3.2) -- (2.7,3.2);
        \draw[-,ultra thick, color=myred]  (1.7,2.8) -- (2.7,2.8);
        \draw[-,ultra thick, color=myred]  (1.7,2.6) -- (2.7,2.6);
        \draw[-, color=mygreen, dashed]  (0,2.2) -- (4.5,2.2);
        \draw[-, color=mygreen, dashed]  (2,1) -- (4.5,1);
        \draw[->, color=mygreen, ultra thick]  (4,2.2) -- (4,1);
        \node at (4.2,1.6)  (A) {$\alpha$};
        \draw[-,ultra thick, color=mygreen]  (1.7,1) -- (2.7,1);
        \draw[-,ultra thick, color=myblue]  (1.7,1.8) -- (2.7,1.8);
        \draw[-,ultra thick, color=myblue]  (1.7,1.4) -- (2.7,1.4);
        \draw[-,ultra thick, color=myblue]  (1.7,0.8) -- (2.7,0.8);
        \draw[-,ultra thick, color=myblue]  (1.7,0) -- (2.7,0);
        \node at (2.2,4.5)  (hphb) {$H_p+H_b$};
        

    \end{tikzpicture}
    \caption{The effect on the problem Hamiltonian from introducing $H_b$. The initial state is highlighted in green. states lower in energy are coloured blue, while states higher in energy are coloured red. As long as $\abs{\Delta W}<\alpha$, the expectation value of $\langle H_p \rangle$ will decrease.}
    \label{fig:spec}
\end{figure}

In practice, achieving a bias like $H_b=-\alpha \ket{z^*}\bra{z^*}$ will be infeasible. Such a bias would involve all-to-all interactions. A much more feasible driver is 
\begin{equation}
    \label{eq:lbias}
    H_b^{(l)}=-\alpha \sum_{i=1}^n (-1)^{z^*[i]} Z_i,
\end{equation}
where $z^*[i]$ is the $i^{\text{th}}$ bit in the $n$-bit string $z^*$, and $Z_i$ is the Pauli Z matrix acting on the $i^{\text{th}}$ qubit. This Hamiltonian has the ground state $\ket{z^*}$ with energy $-n \alpha$, and consists of one-local terms. However, this Hamiltonian will alter the energy of all the problem states. In the worst case, the energy gap opened up by $H_b^{(l)}$ becomes primarily occupied by high energy states of $H_p$ and no cooling of $\langle H_p\rangle$ occurs.

If, however, we assume that the ordering of energy eigenstates is uncorrelated between $H_b^{(l)}$ and $H_p$, then the average energy shift from $H_b^{(l)}$ is 0 with standard deviation $\alpha \sqrt{n}$. So the typical energy shift of the problem eigenstates is $\sqrt{n}$ times smaller than the shift on the initial state. Therefore, for large $n$ in the typical case, we would expect this local bias to mimic $-\ket{z^*}\bra{z^*}$. For this approach, if $\alpha$ is too big $H_b^{(l)}$ will dominate and the approach will not involve any information from $H_p$.  We numerically explore the performance of this approach in Appendix \ref{app:cyc_num}.

\subsection{Computational mechanism}

In the previous section, we discussed that if the system is likely to increase in energy, then the introduction of a bias can help improve $\langle H_p \rangle$. The idea that a system is more likely to move towards the bulk of the spectrum than away is physically reasonable given a sufficient level of dynamics. Here we elucidate the computational mechanism.

To make the system isolated, consider the introduction of a second register of qubits. The first register $Q$ contains the qubits used in the BQA process and contains $n$ qubits. The second register $C$ contains the same number of bits as $Q$, and each qubit in the second register is prepared in $\ket{0}$. The joint initial state is now:
\begin{equation}
    \rho_0^{(Q/C)}=\sum_z p_z \ket{z}\bra{z}_{Q}  \bigotimes_{i=0}^{n-1} \ket{0} \bra{0}_{i,C}.
\end{equation}
Each qubit in $Q$ is paired with a qubit in $C$. A controlled-NOT gate is applied between each pair of qubits such that the resulting state is:
\begin{equation}
    \rho_1^{(Q/C)}=\sum_z p_z \ket{z}\bra{z}_{Q} \otimes \ket{z} \bra{z}_{C}.
\end{equation}
The unitary, 
\begin{equation}
    U_t=\sum_z U_z \otimes \ket{z} \bra{z},
\end{equation}
where $U_z$ is a unitary dependent on the string $z$, is applied to $\rho_1^{(Q/C)}$. In the context of the previous section, this could be the unitary generated by Eq.~\ref{eq:BQA} with $H_b=-\alpha \ket{z}\bra{z}$ or $H_b^{(l)}$  with ground state $\ket{z}$. The resulting state after application of $U_t$ is
\begin{equation}
    \rho_2^{(Q/C)}=\sum_z  p_z U_z \ket{z}\bra{z}_Q U_z^{\dag} \otimes \ket{z} \bra{z}_C .
\end{equation}
This process is now unitary and therefore isolated. The diagonal entropy of the initial state is $S_0$. At the end of the process, the diagonal entropy cannot decrease as a result. The diagonal entropy of the $C$ register (i.e. the diagonal entropy of $\Tr_Q \rho_2^{(Q/C)}$) is exactly $S_0$. Hence, we conclude that the diagonal entropy of the $Q$ register must be greater than or equal to zero. For the adiabatic cycle discussed in the beginning of the section, the second law of thermodynamics is not violated, since all the entropy has been moved from $Q$ to $C$. Only by neglecting $C$ does it appear that the entropy of the system is decreasing. BQA is making use of a bath of classical bits $C$ prepared in a low entropy configuration to hopefully reduce the entropy of $Q$. Any statement that relies on the system being isolated requires the inclusion of $C$. We discuss the introduction of other baths further in Sec.~\ref{sec:oqs}.

\subsection{Numerically observing cyclic cooling}
\label{app:cyc_num}

Sec.~\ref{sec:cyc} made two predictions:
\begin{enumerate}
    \item Cyclic processes lead to heating, so RQA should lead to a greater value of $\langle H_p \rangle$.
    \item A cyclic process might achieve cooling of $\langle H_p \rangle$ with the introduction of a third term in the Hamiltonian.
\end{enumerate}
In this section, we numerically investigate this for the MAX-CUT problem and SK problem described in Appendix \ref{sec:pstqa}. 

\begin{figure}
    \centering
    \includegraphics[width=0.48\textwidth]{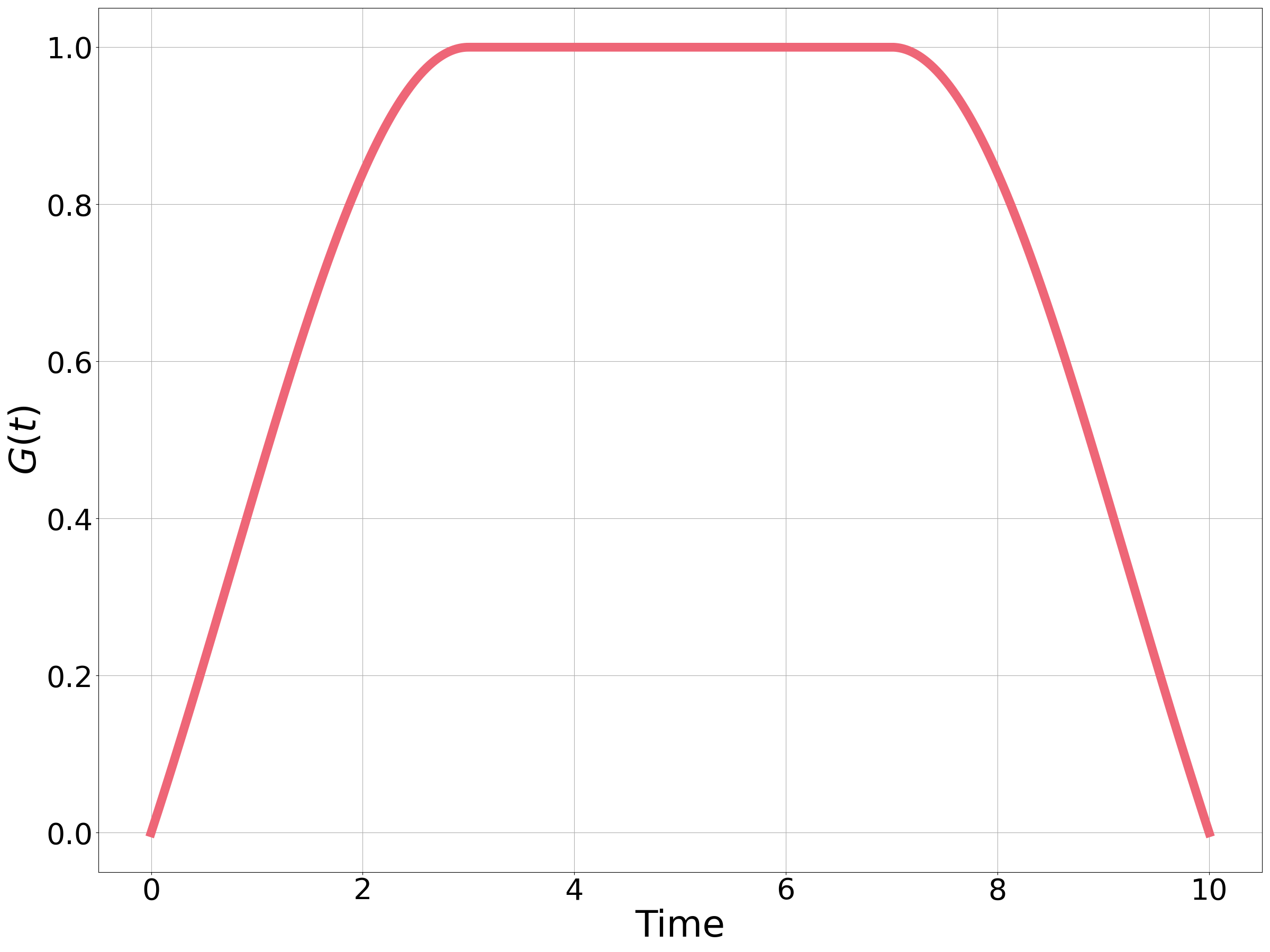}
    \caption{The schedule $G(t)$ appended to the driver Hamiltonian, used for the cyclic processes described in Appendix \ref{app:cyc_num}.}
    \label{fig:RQA_sc}
\end{figure}

\begin{figure}
    \centering
    \includegraphics[width=0.48\textwidth]{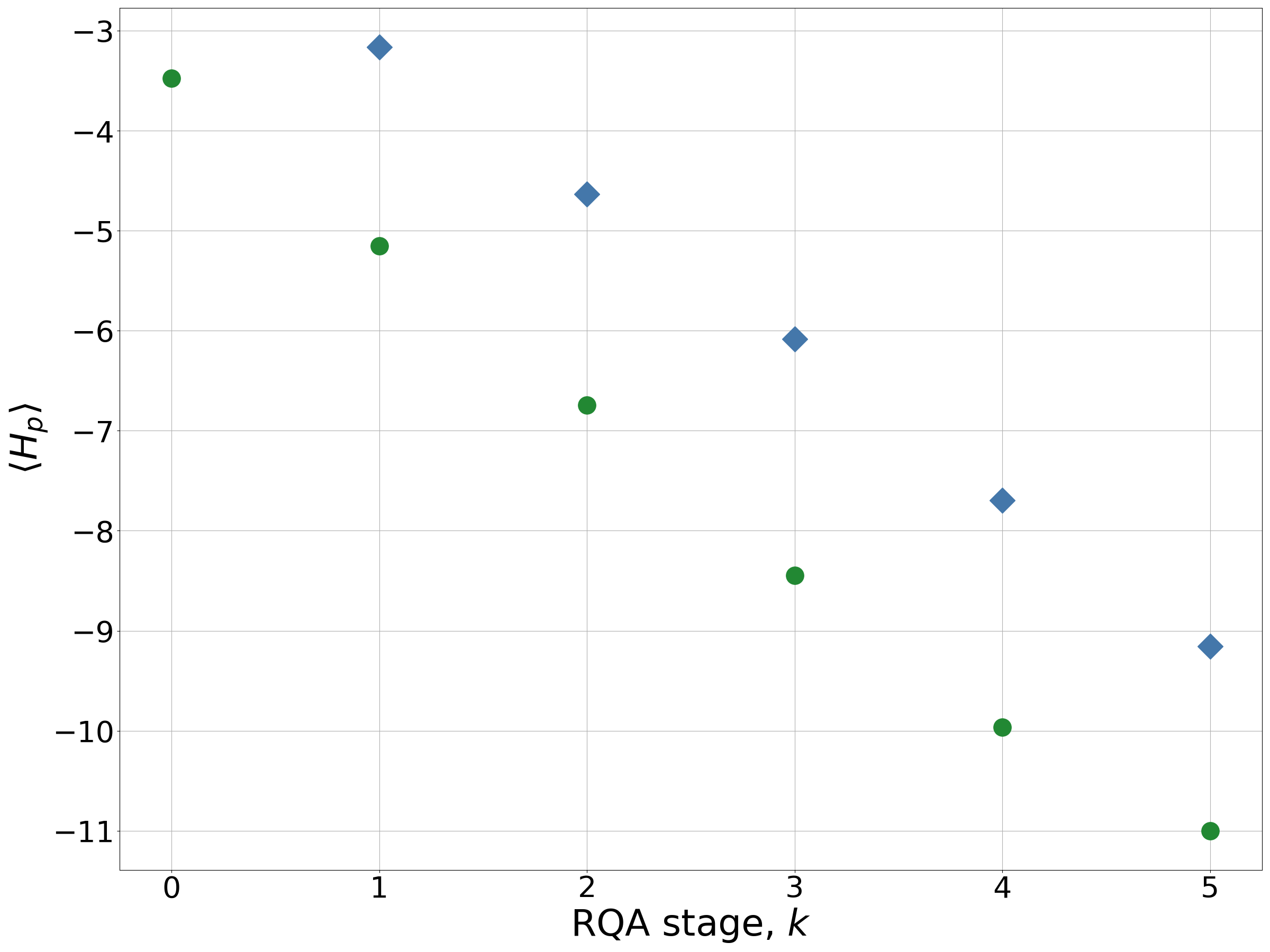}
    \caption{RQA for a 10 node MAX-CUT graph. The green dots show $\langle H_p \rangle$ for the post-selected distribution. The blue diamonds show $\langle H_p \rangle$ as sampled from application of $U_{\text{cyc}}$.}
    \label{fig:RQA_hp}
\end{figure}

\begin{figure}
    \centering
    \includegraphics[width=0.48\textwidth]{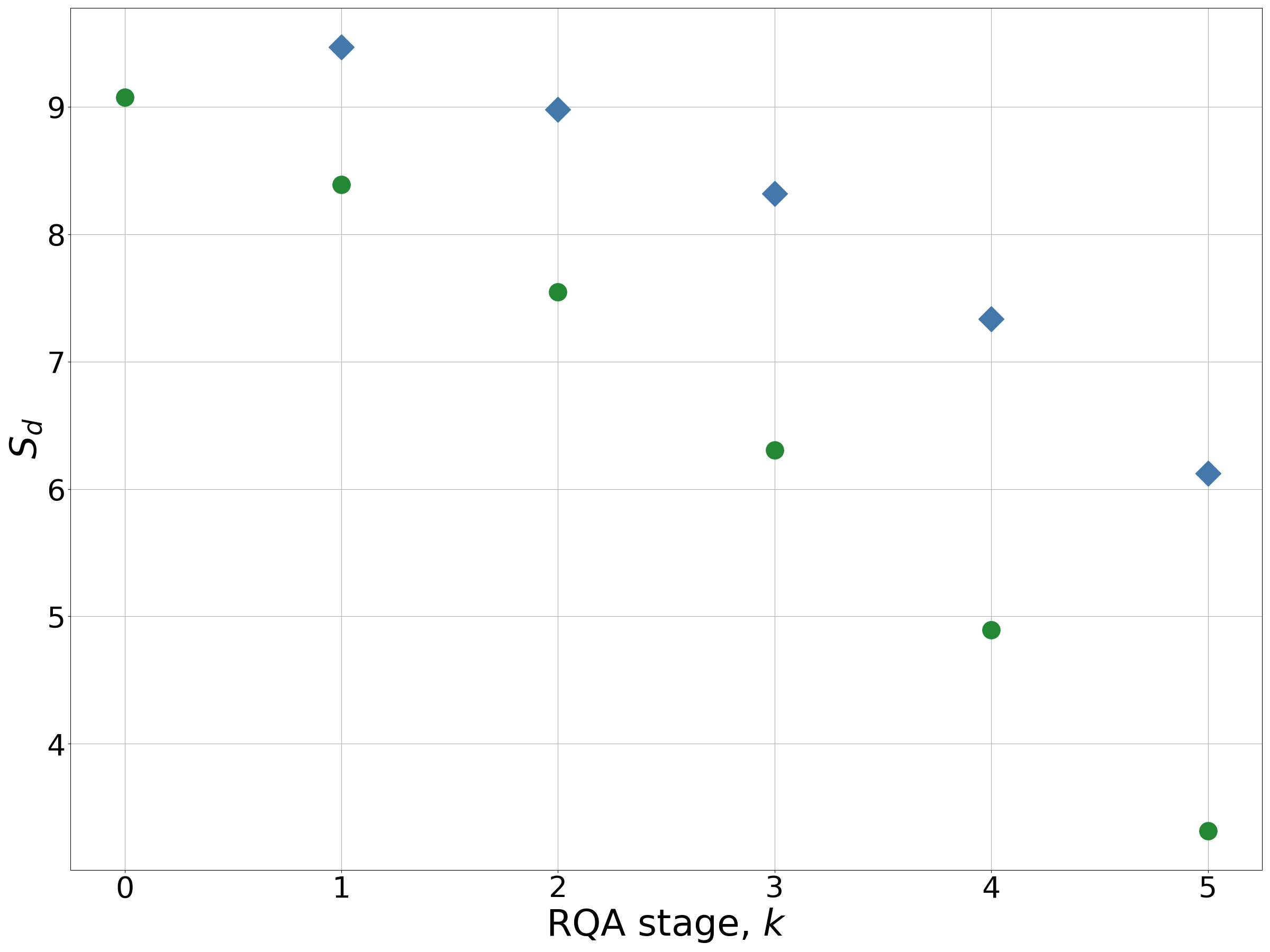}
    \caption{The diagonal entropy $S_d$ for RQA applied to a 10-node MAX-CUT graph. The green dots show $S_d$ for the post-selected distribution. The blue diamonds show $S_d$ as sampled after application of $U_{\text{cyc}}$. The logarithm used to calculate the diagonal entropy is taken to be base 2.}
    \label{fig:RQA_sd}
\end{figure}

\begin{figure}
    \centering
    \includegraphics[width=0.48\textwidth]{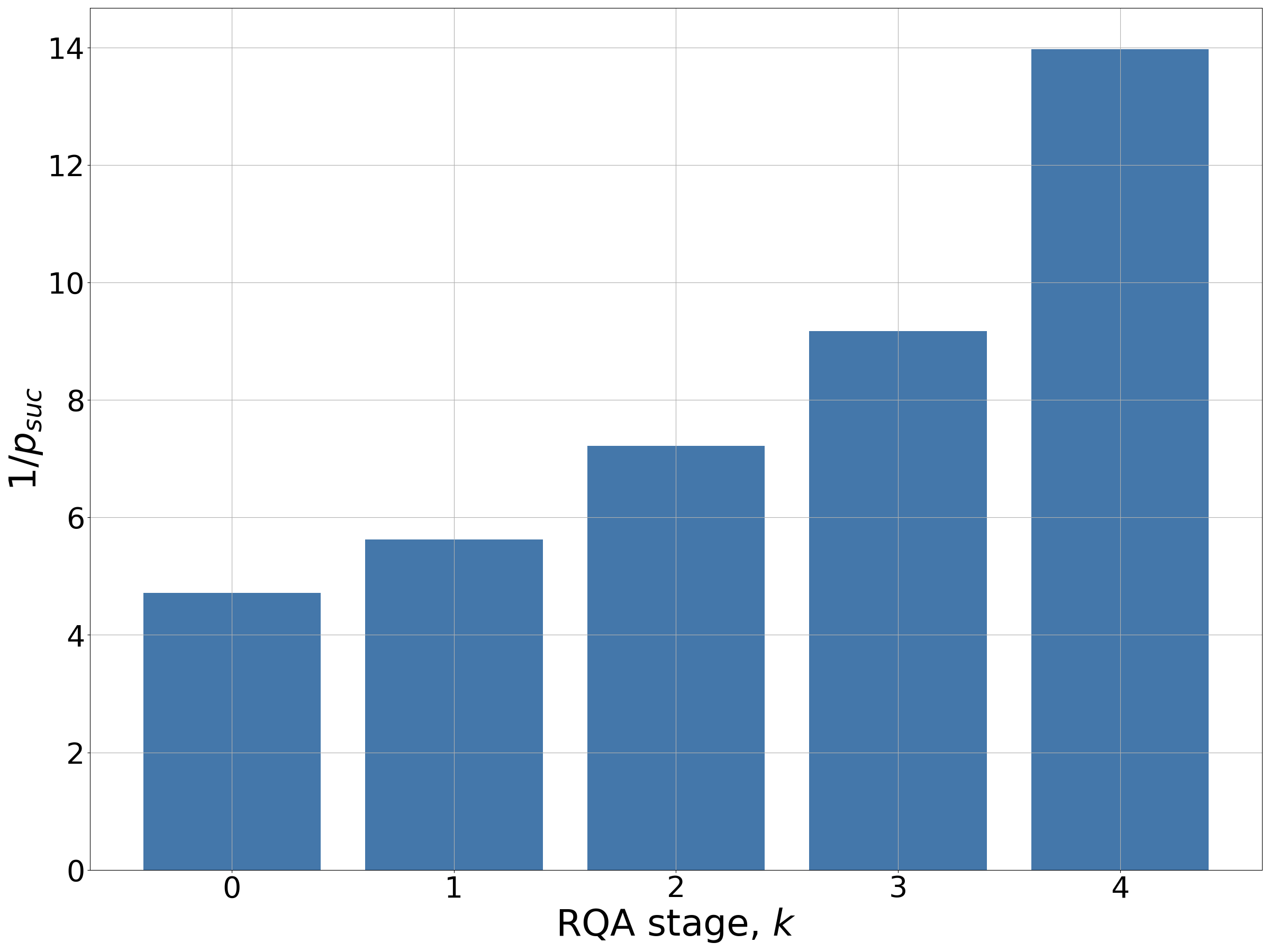}
    \caption{The inverse probability (Eq.~\ref{eq:psuc}) for each stage of RQA on a 10-vertex MAX-CUT graph.}
    \label{fig:RQA_psuc}
\end{figure}

Initially, we focus on reverse quantum annealing (RQA). With each cycle generated by the Hamiltonian:
\begin{equation}
    H_{cyc}=H_p+G(t)H_d.
\end{equation}
As in Sec.~\ref{sec:cyc_pass}, the initial state is the ensemble 
\begin{equation}
    \rho_0= \sum_{s} p(s) \ket{s}\bra{s},
\end{equation}
where $\ket{s}$ is an eigenstate of $H_p$ with eigenvalue $s$. Denoting the unitary associated with one cycle to be $U_{\text{cyc}}$, then the transition probability between states $\ket{s}$ and $\ket{j}$ is given by
\begin{equation}
    p(j|s)=\abs{\bra{j} U_{\text{cyc}} \ket{s}}^2.
\end{equation}

After each stage, there is some selection criterion to determine if a string is kept or if the RQA cycle is repeated with the initial string. The selection criterion could be the Metropolis-Hastings update rule \cite{Has70} or a more complicated update \cite{Cha17,Cha23}. In this work, we only keep states that lower the energy. 

The state after $k$ iterations and measurement but before post-selection is given by
\begin{multline}
    \rho_k=\frac{1}{\mathcal{N}_k}\sum_{s_k} \sum_{s_{k-1}<\dots<s_1<s_0}p(s_k|s_{k-1})p(s_{k-1}|s_{k-2})\\\dots p(s_{1}|s_{0})p(s_0) \ket{s_k}\bra{s_k},
\end{multline}
where $\mathcal{N}_k$ normalises the state. The probability of finding a lower state from this ensemble is 
\begin{multline}
    \label{eq:psuc}
    p_{\text{suc}}(k)=\frac{1}{\mathcal{N}_k}\sum_{s_k<s_{k-1}<\dots<s_1<s_0}p(s_k|s_{k-1})p(s_{k-1}|s_{k-2})\\\dots p(s_{1}|s_{0})p(s_0).
\end{multline}
This sets a rough estimate for the number of iterations to lower the energy. If $1/p_{\text{suc}}(k)$ becomes greater than some cut-off, we terminate the RQA process. If the approach is not terminated, the state that is fed into the next cyclic iteration is 
\begin{multline}
    \rho_k^{(ps)}=\frac{1}{\mathcal{N}_k'} \sum_{s_k<s_{k-1}<\dots<s_1<s_0}p(s_k|s_{k-1})p(s_{k-1}|s_{k-2})\\\dots p(s_{1}|s_{0})p(s_0) \ket{s_k}\bra{s_k}.
\end{multline}

To illustrate the analysis found in Sec.~\ref{sec:cyc}, we consider a single 10-vertex MAX-CUT instance, take $p(s)$ to be a uniform distribution on the interval $s<\Tr'H_p$. The driver Hamiltonian is taken to be $H_{TF}$. The schedule for the drive $G(t)$ is taken to be a square Gaussian, shown in Fig.~\ref{fig:RQA_sc}. For this instance $\langle H_p \rangle$ can be seen in Fig.~\ref{fig:RQA_hp}. The green circles show $\langle H_p \rangle$ for the state after post-selection. The blue diamonds show $\langle H_p \rangle$ sampled after each single application of $U_{\text{cyc}}$. At each stage $\langle H_p \rangle$ after the cyclic quantum process is greater than the post-selected state, therefore the cyclic quantum process is producing on average worse quality states. This is numeric evidence of heating at each stage. For this instance, RQA reaches the ground-state.

The diagonal entropy for the same process is shown in Fig.~\ref{fig:RQA_sd}. Again, the blue diamonds show the diagonal entropy sampling directly after an application of $U_{\text{cyc}}$. The green dots show the diagonal entropy of the post selected state. As is clear at each RQA leads to an increase of diagonal entropy corresponding to a broadening of the distribution.

Finally, Fig.~\ref{fig:RQA_psuc} shows $1/p_{\text{suc}}$, which gives a sense of the average number of shots required at each stage. The randomness $U_{cyc}$ introduces allows RQA to find the ground state with fewer classical evaluations of $H_p$. This is in spite of the fact, that $U_{\text{cyc}}$ results in a larger value of $\langle H_p \rangle$, as we have argued.

\begin{figure}
    \centering
    \begin{subfigure}[l]{0.48\textwidth}
        \includegraphics[width=\textwidth]{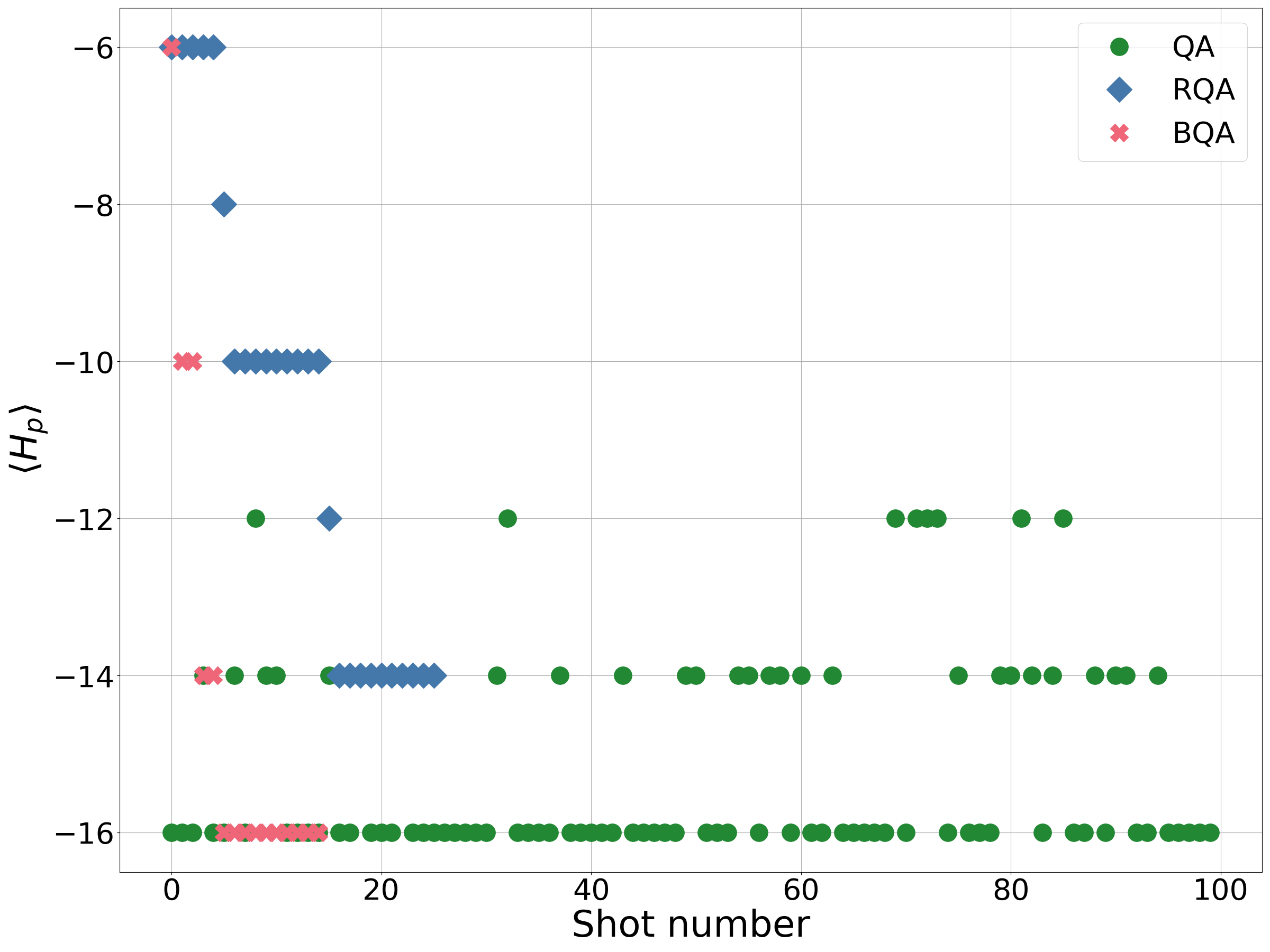}
        \caption{MAX-CUT}
        \label{fig:cyc_mc_ex}
    \end{subfigure}
    \vfill
    \begin{subfigure}[l]{0.48\textwidth}
        \includegraphics[width=\textwidth]{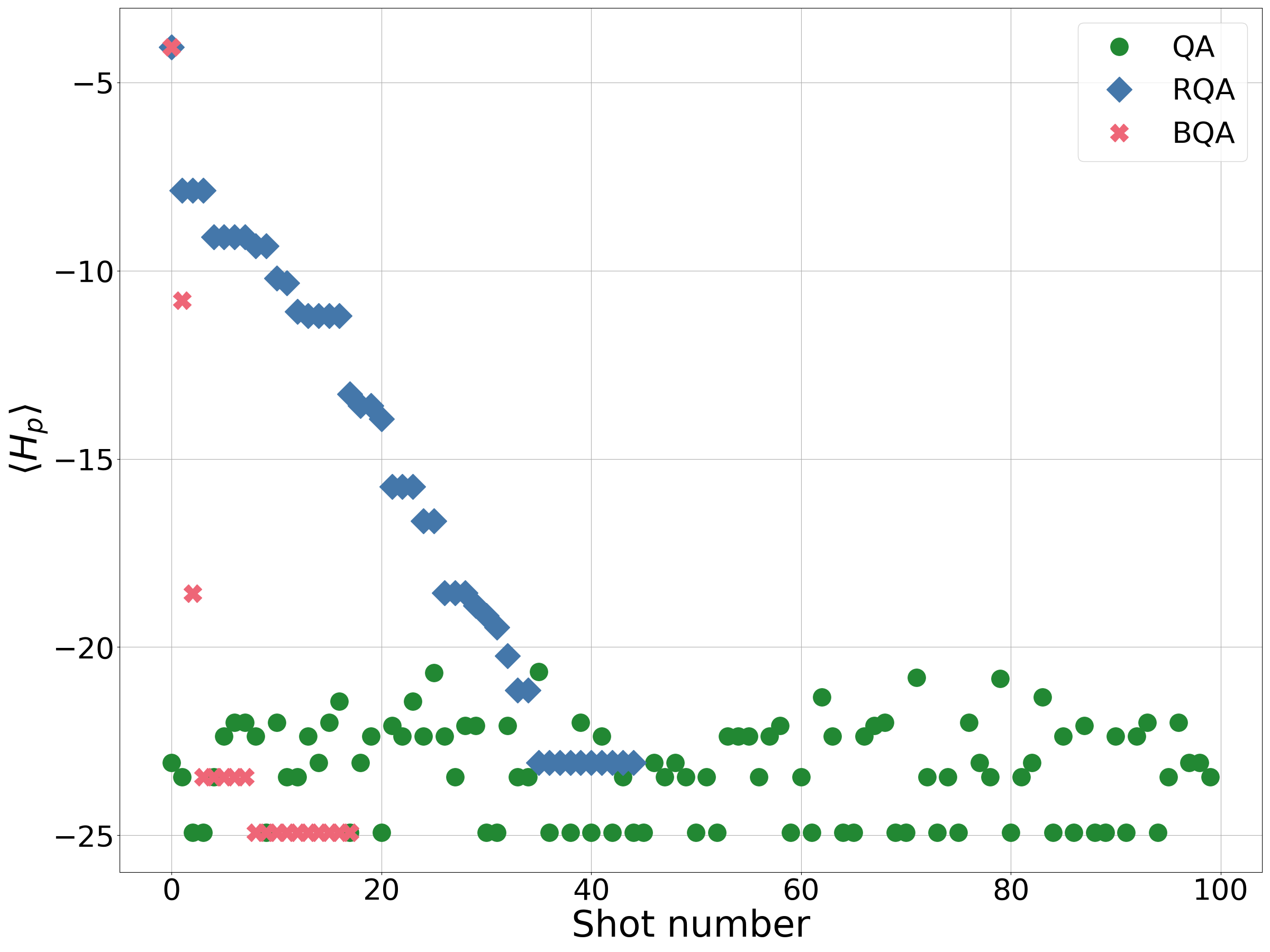}
        \caption{SK}
        \label{fig:cyc_sk_ex}
    \end{subfigure}
    \caption{The performance of QA (green circles), RQA (blue diamonds) and BQA (red crosses) for a 12 qubit example.}
    \label{fig:cyc_ex}
\end{figure}

\begin{figure}
    \centering
    \begin{subfigure}[l]{0.48\textwidth}
        \includegraphics[width=\textwidth]{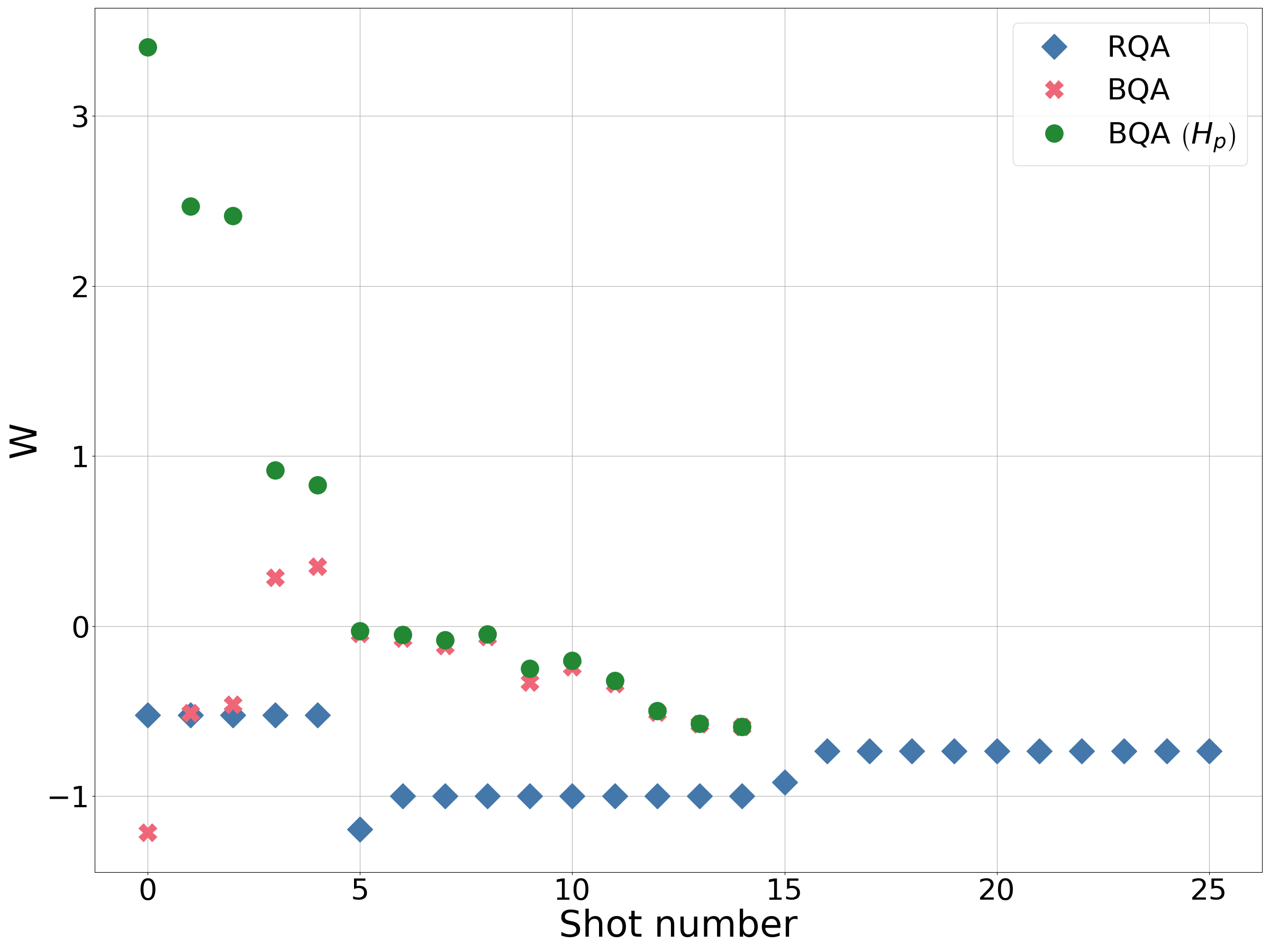}
        \caption{MAX-CUT}
        \label{fig:cyc_heat_mc_ex}
    \end{subfigure}
    \vfill
    \begin{subfigure}[l]{0.48\textwidth}
        \includegraphics[width=\textwidth]{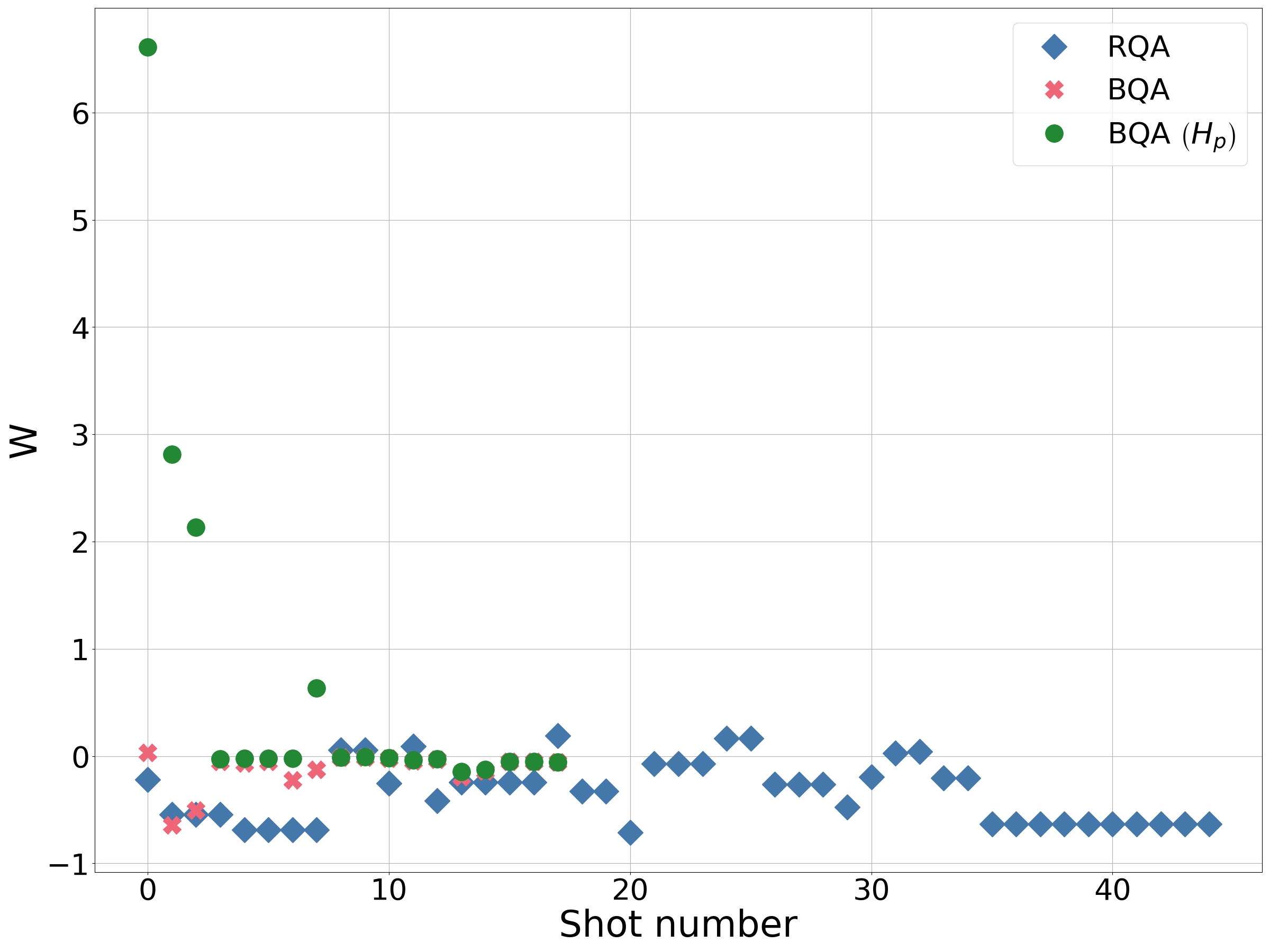}
        \caption{SK}
        \label{fig:cyc_heat_sk_ex}
    \end{subfigure}
     \caption{The extractable work, $W$, for the 12-qubit instances considered in Fig.~\ref{fig:cyc_ex}. The blue diamonds correspond to RQA. The red crosses BQA. The green circles show $W$ for BQA neglecting the change in energy of $H_b^{(l)}$, i.e, the change in $\langle H_p\rangle $.}
    \label{fig:cyc_heat_ex}
\end{figure}

\begin{figure}
    \centering
    \begin{subfigure}[l]{0.48\textwidth}
        \includegraphics[width=\textwidth]{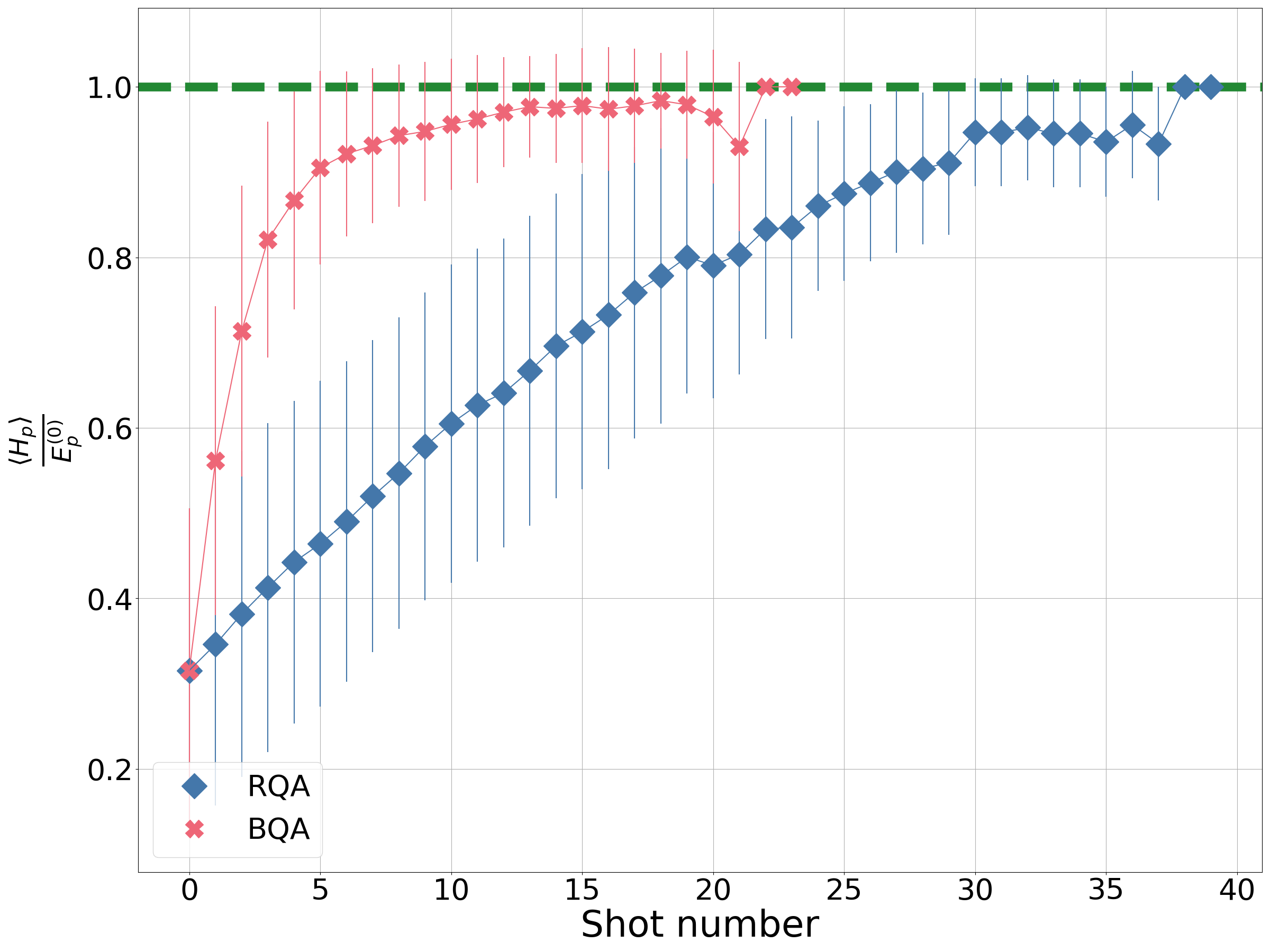}
        \caption{MAX-CUT}
        \label{fig:cyc_approx_mc}
    \end{subfigure}
    \vfill
    \begin{subfigure}[l]{0.48\textwidth}
        \includegraphics[width=\textwidth]{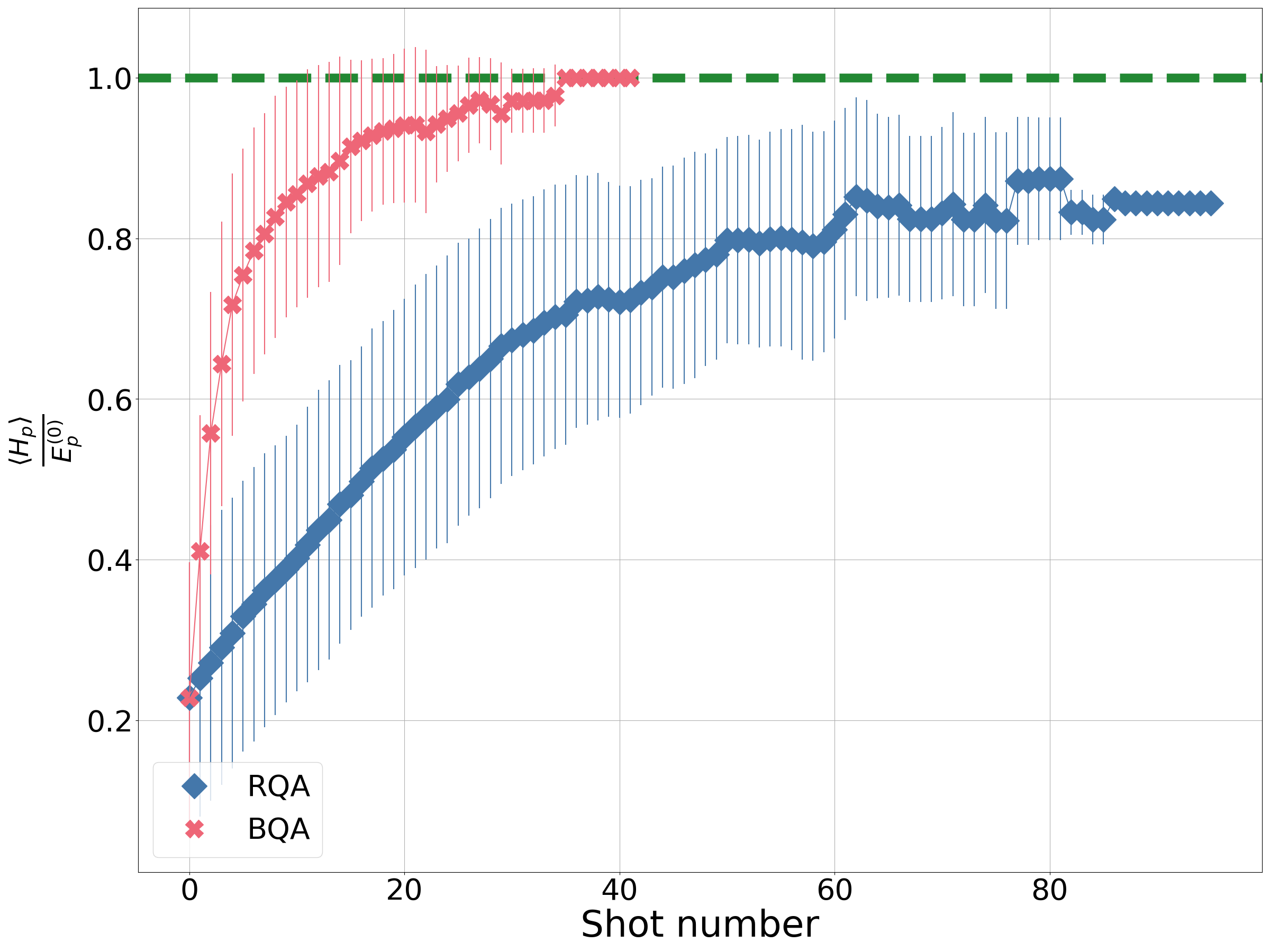}
        \caption{SK}
        \label{fig:cyc_approx_sk}
    \end{subfigure}
    \caption{The average approximation ratio for the 12 qubit instances at each shot number. Since the approaches terminate at different shot numbers, the number of instances decreases. The pink crosses show the average approximation ratio for BQA. The blue diamonds show the average approximation ratio for RQA. The lines show one standard deviation. The dashed green line shows an approximation ratio of 1. The final decrease in approximation ratio in BQA reflects more difficult problems for BQA, since it has not terminated after numerous shots. }
    \label{fig:cyc_approx}
\end{figure}

\begin{figure}
    \centering
    \begin{subfigure}[l]{0.48\textwidth}
        \includegraphics[width=\textwidth]{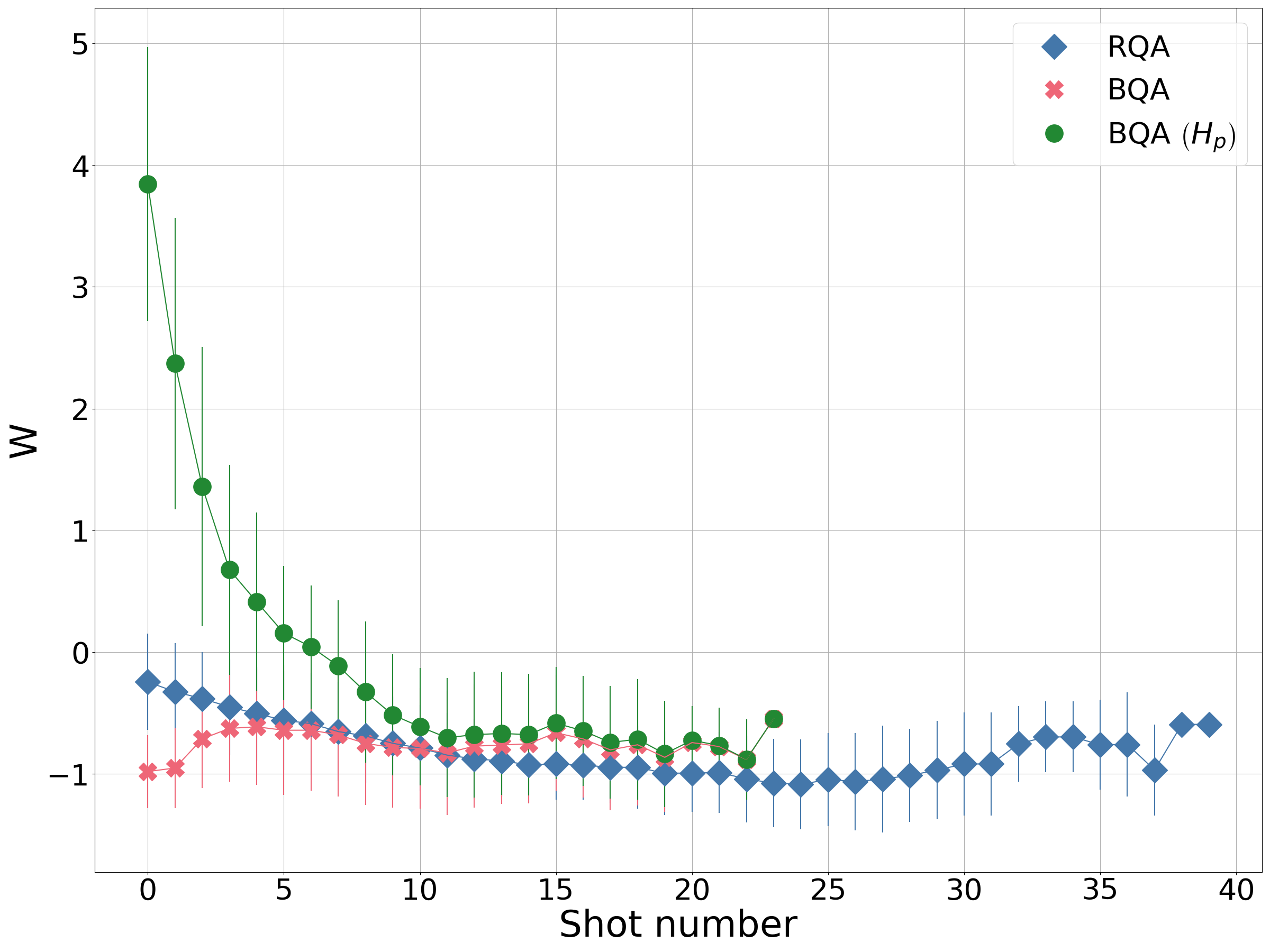}
        \caption{MAX-CUT}
        \label{fig:cyc_heat_mc}
    \end{subfigure}
    \vfill
    \begin{subfigure}[l]{0.48\textwidth}
        \includegraphics[width=\textwidth]{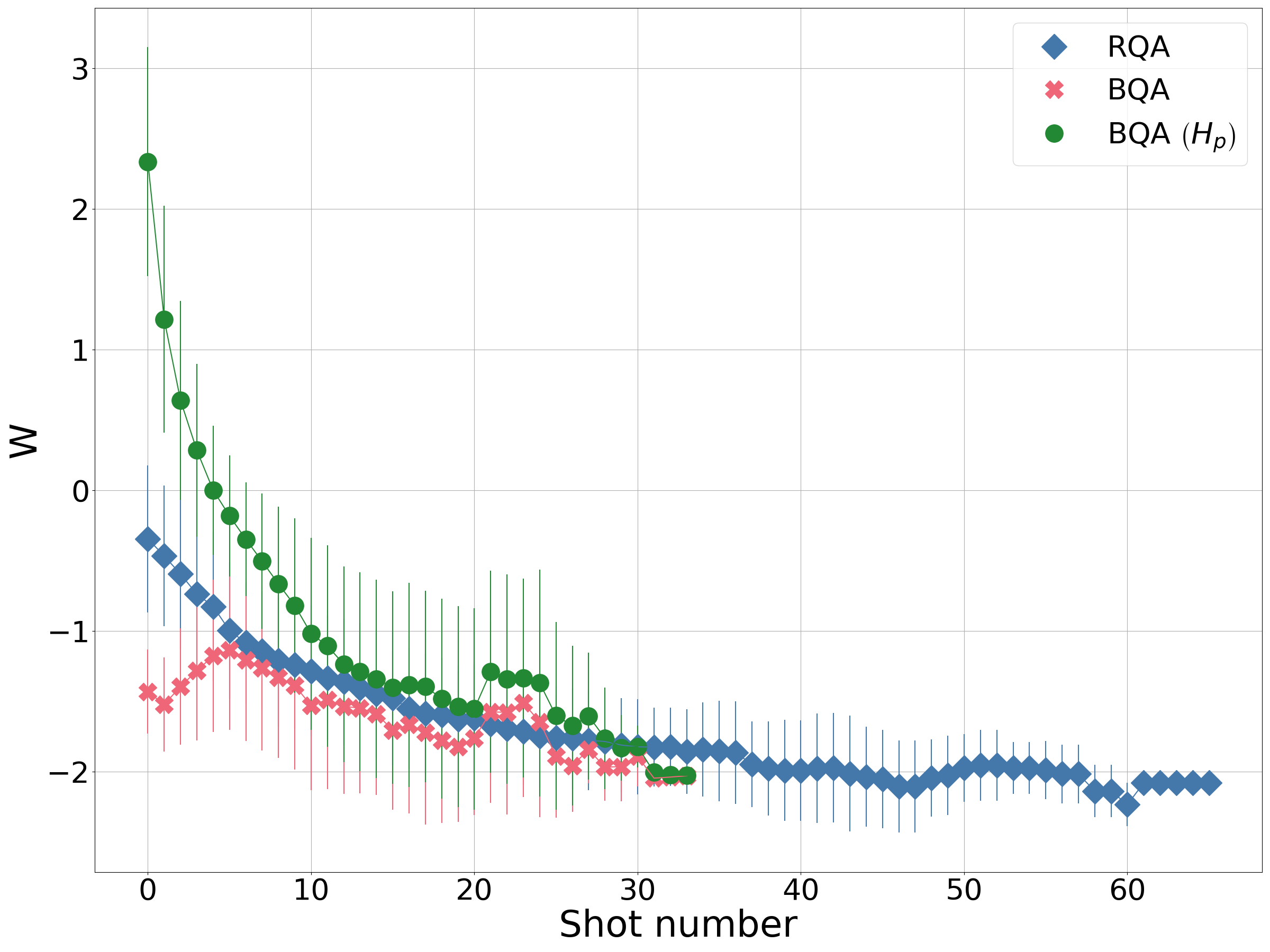}
        \caption{SK}
        \label{fig:cyc_heat_sk}
    \end{subfigure}
    \caption{The extractable work for the 12 qubit instances at each shot number. Since the approaches terminate at different shot numbers, the number of instances decreases. The pink crosses show the average extractable work for the BQA protocol. The blue diamonds show the average extractable work for RQA. The lines show one standard deviation. The dashed green circles shows the change in $\langle H_p \rangle$ for BQA. }
    \label{fig:cyc_work}
\end{figure}

Having explored RQA without a bias, we now introduce biased quantum annealing (BQA). Simulating the full ensemble at each stage is computationally more expensive than a single state-vector, so from here on the simulations use sampling as if the approach was being evaluated on an actual quantum device. The problems considered consist of 12 qubits. For each instance, up to $k_{\text{max}}$ shots are taken. The initial state is chosen by randomly selecting states until condition Eq.~\ref{eq:init_cond} is met. If the result of a run produces a better quality solution than the initial state, the initial state is updated to be this state. If the initial state is not updated after $k$ runs, the algorithm is assumed to have converged. For the numerics, we take $k_{max}=100$ and $k=10$. The schedule for the drive $G(t)$ is taken to be a square Gaussian, shown in Fig.~\ref{fig:RQA_sc}.

For BQA the biasing term is taken to be $H_{b}^{(l)}$ (i.e. Eq.~\ref{eq:lbias}). The initial value of $\alpha$ is taken to be $\alpha_0=\sqrt{\Tr'\left(H_p^2\right)}$. This is assumed to be typically an overshoot, so $\alpha$ is decreased with each shot if $\langle H_p \rangle$ does not decrease. We take $\alpha$ to decrease linearly by $\alpha_0/k$ each time.

With an annealing time of $t_a=10$, the problems are typically easy for QA. We consider 100 instances of MAX-CUT and SK. QA found the ground state with 100 shots or fewer for 99 of the MAX-CUT instances and all the SK instances. RQA found the ground state for 22 of the MAX-CUT instances and 12 of the SK instances. BQA found the ground state for 78 of the MAX-CUT instances and 48 of the SK instances. 

Fig.~\ref{fig:cyc_ex} shows a specific MAX-CUT and SK instance. The green circles show the result from QA. Though clearly not adiabatic, it finds the ground state in both cases. The blue diamonds show the initial value of $\langle H_p \rangle$ for RQA. In both cases, the termination condition is reached before the algorithm finds the ground state. The red crosses show BQA. BQA managed to find the ground state in both cases. BQA rapidly converges towards the ground state compared to RQA and terminates within 20 shots.

The above shows that BQA and RQA can tackle combinatorial optimisation problems. The aim of this section is to discuss heating. Fig.~\ref{fig:cyc_heat_ex} shows the extractable work for each shot for the instances considered in Fig.~\ref{fig:cyc_ex}. The RQA examples generally show a negative value of extractable work, meaning that $\langle H_p \rangle$ decreases — the blue diamonds in Fig.~\ref{fig:cyc_heat_ex}. BQA also shows heating (i.e, $W<0$), the pink crosses in Fig.~\ref{fig:cyc_heat_ex}. There are some instances of cooling in the MAX-CUT instance for BQA for two shots. The green circles show the change in $\langle H_p \rangle$ for the BQA protocol. We see $\langle H_p \rangle$ is decreasing as a result of BQA, i.e. cooling of $\langle H_p \rangle$. 

Finally, we consider the numerics for all the 100 MAX-CUT instances and 100 SK instances. Fig.~\ref{fig:cyc_approx} shows the approximation ration averaged over all instances. The approximation ratio is defined as $\langle H_p \rangle$ divided by the ground state energy of $H_p$. If the approach finds the ground state, the approximation ratio is 1 (and cannot exceed 1). In both cases, BQA converges much faster than RQA.

Fig.~\ref{fig:cyc_work} shows the extractable work averaged over all the instances. We observe heating for both protocols as predicted by Assumption \ref{ass:work}. However, at least for the first few shots, BQA is able to achieve cooling of $\langle H_p \rangle$. 

We have numerically demonstrated, with this set-up, that RQA leads to heating. This does not rule out RQA finding better solutions, as evidenced by Fig.~\ref{fig:cyc_approx}.  It does suggest there is scope for improvement and cooling $\langle H_p \rangle$ by the addition of a third term.

\end{document}